# Retail Pricing Format and Rigidity of Regular Prices

Sourav Ray
Department of Marketing and Consumer Science, G.S. Lang School of Business and Economics
University of Guelph
Guelph, ON N1G-1M8, Canada
s_ray@uoguelph.ca

Avichai Snir
Department of Economics, Bar-Ilan University
Ramat-Gan 5290002, Israel
Avichai.Snir@gmail.com

Daniel Levy**
Department of Economics, Bar-Ilan University
Ramat-Gan 5290002, Israel,
Department of Economics, Emory University
Atlanta, GA 30322, USA,
International Centre for Economic Analysis,
ISET at Tbilisi State University
Tbilisi 0108, Georgia, and
Rimini Center for Economic Analysis
Daniel.Levy@biu.ac.il

Revised: April 29, 2023

* We thank two anonymous reviewers for their constructive comments which led to a significantly improved paper. We very much appreciate the encouragement and guidance of the editor Daniel Gottlieb in the manuscript revision process. We thank the participants of the 2022 annual conference of the Royal Economic Society, and the 2022 Inflation Conference of the International Centre for Economic Analysis, especially Jurek Konieczny, Sebastian Link, Oleksiy Kryvtsov, Teresa Messner, and Danila Smirnov, for their helpful comments and suggestions. Ryan Chahrour kindly answered our questions concerning the implementation of his (Chahrour 2011) sales filter. We are grateful to Vinay Kanetkar for sharing with us his data on Canadian retail stores, and to Xiaogong Huang for his assistance with data gathering. We acknowledge the financial support of the SSHRC (Canada) and the IIMF (McMaster University). All errors are ours.

** Corresponding author: Daniel Levy, Daniel.Levy@biu.ac.il

# Retail Pricing Format and Rigidity of Regular Prices

*Abstract*

We study the price rigidity of regular and sale prices, and how it is affected by pricing formats (i.e., pricing strategies). We use data from three large Canadian stores with different pricing formats (Every-Day-Low-Price, Hi-Lo, and Hybrid) that are located within a 1 km radius of each other. Our data contains both the actual transaction prices and actual regular prices as displayed on the store shelves. We combine these data with two "generated" regular price series (filtered prices and reference prices) and study their rigidity. Regular price rigidity varies with store formats because different format stores treat sale prices differently, and consequently define regular prices differently. Correspondingly, the meanings of price cuts and sale prices vary across store formats. To interpret the findings, we consider the store pricing format distribution across the US.

# INTRODUCTION

"A central question in macroeconomics is whether nominal rigidities are important" (Eichenbaum et al. 2011, p. 234). A large literature, starting with Barro (1972), Mankiw (1985), Kashyap (1995), Carlton (1986), Cecchetti (1986), Lach and Tsiddon (1992, 1996), Levy et al. (1997, 1998), Dutta et al. (1999, 2002), Bils and Klenow (2004), and Konieczny and Skrzypacz (2005), assesses the importance of nominal rigidities by measuring how often prices change.[1]

An important distinction in this literature is between regular and sale prices. As Nakamura and Steinsson (2008) note, price changes associated with sale prices might have different macroeconomic effects than regular price changes. That is because sale prices are transient and, consequently, they do not generate as much accumulated effect on the aggregate price level as regular price changes. In addition, sale prices are less correlated with economic shocks than regular prices, suggesting that the adjustment to aggregate shocks takes place, mostly, via regular price changes.[2] Recent studies of price-setting models, therefore, distinguish between regular and transaction prices.[3]

In this paper, we argue that the distinction between regular and sale prices does not depend only on whether the price change is temporary or not. It also depends on the store pricing format (i.e., pricing strategy): stores that follow different pricing formats have different notions of temporary price changes, which can potentially lead to different patterns of regular price rigidity. This aspect of retail pricing practice, however, has not received much attention in the context of price rigidity, although it might affect the transmission of supply shocks to consumer prices (He et al., 2023). Our goal is to fill this gap in the literature.

---

[1] The empirical literature on nominal price rigidity has expanded dramatically since then. For older surveys, see Gordon (1981, 1990) Romer (1993), Weiss (1993), Taylor (1999), Willis (2003), and Wolman (2007). More recent surveys include Klenow and Malin (2011), Leahy (2011), and Nakamura and Steinsson (2013).

[2] The literature uses different terms. In general, "transaction prices" refer to "final prices" or "posted prices," which is the same as the "discounted prices" in case there are discounts or "regular prices" in case there are no discounts.

[3] Examples include Nakamura (2008), Nakamura and Steinsson (2008, 2013), Eichenbaum et al. (2011), Guimaraes and Sheedy (2011), Midrigan (2011), Klenow and Malin (2011), Campbell and Eden (2014), Beradi et al. (2015), Coibion et al. (2015), Kehoe and Midrigan (2015), Anderson et al. (2015, 2017), Eden (2018), Nakamura et al. (2018), DellaVigna and Gentzkow (2019), Levy et al. (2020), Wu (2022), and the studies cited therein.

Whereas Hi-Lo stores run frequent sales, which they promote using sale signs, leaflets, etc., Every-Day-Low-Price (EDLP) stores run fewer temporary price cuts, and when they do, the price cuts are rarely promoted. Hybrid (HYB) stores combine some features of the Hi-Lo and EDLP formats. We find that the number of temporary price cuts at the HYB store in our data is similar to the number of temporary price cuts at the EDLP store. Most of the temporary price cuts at the HYB stores, however, are promoted, as in the Hi-Lo store.

We use a unique dataset from three large Canadian food stores. The dataset has three features that are particularly important for the questions we ask. First, it includes *both* the actual regular and the actual transaction prices as posted on the stores' shelves. Second, the stores differ in their pricing format: one follows Hi-Lo, the second follows EDLP, and the third is a HYB. Third, the stores are located within a 1 km radius, serving the same pool of clientele.

To assess how the treatment of sale prices affects price rigidity, we study the rigidity of four price series at each store. One is the *transaction price* series, which includes the sale prices as defined by the store, and three are regular price series: *regular prices* as defined by the stores, *filtered prices* which we generate from the transaction price series by filtering out temporary price cuts using Nakamura and Steinsson's (2008) Sales Filter A, and *reference prices*, which we generate from the transaction prices by removing all short-lived prices using Chahrour's (2011) algorithm, building on Eichenbaum et al. (2011).

We find that the pricing format has a large effect on regular price rigidity. If we follow the stores' notion of regular prices, then regular prices at the EDLP store are more flexible than at the Hi-Lo or at the HYB stores: the likelihood that the EDLP store changes a regular price on a given week is 13.38%, in contrast to the HYB store—5.34%, and the Hi-Lo store—4.06%. If we treat filtered prices as regular prices, then regular prices at the HYB and EDLP stores are more flexible than at the Hi-Lo store: the likelihood that the HYB store changes a regular price on a given week is 4.50%, similar to the EDLP store—4.25%, while the likelihood at the Hi-Lo store is 3.55%. If we treat reference prices as regular prices, then regular prices at the HYB store are more flexible than at the EDLP or at the Hi-Lo stores: the likelihood that a HYB store changes a regular price on a given week is 3.95%, the EDLP store—2.70%, and the Hi-Lo store—2.23%.

We recognize that the empirical studies in this literature usually report their results for filtered (or for reference) price series because they are primarily interested in identifying specific patterns of price changes to match and/or replicate data parameters for fitting structural models. For that purpose, knowing how retailers label their regular prices is less consequential.

However, recognizing that stores that follow different pricing formats treat temporary price cuts differently is likely to matter for macro-level price rigidity for several reasons. First, the macroeconomic literature often treats temporary price cuts as pre-planned events, designed to maintain a brand's image and/or market share (Anderson et al. 2017, Warner and Barsky 1995). Our finding that stores that follow different formats treat temporary price cuts differently suggests that they might be driven by different motivations when they make decisions on temporary price cuts.

Second, as Chevalier and Kashyap (2019) note, sales have a large effect on sales volumes. Indeed, the effective price level at Hi-Lo and HYB stores is strongly affected by temporary price cuts (Glandon, 2018). However, the effect of temporary price cuts might be different at an EDLP store, where the price cuts are not promoted as sales. Therefore, the role that temporary price cuts play in the transmission of monetary shocks likely depends on the distribution of store pricing formats.

Third, although the stores are located within 1 km of each other and serve the same clientele, we find that the stores' regular prices exhibit different degrees of rigidity, depending on the definition of regular prices. If we consider regular prices as defined by the store, then the most flexible prices are 3.6 times as flexible as the least flexible prices. If we consider the filtered prices, then the most flexible prices are only 1.3 times as flexible as the least flexible prices. If we consider the reference prices, then the most flexible prices are 1.7 times as flexible as the least flexible prices. Thus, the choice of a definition of regular prices can have a large effect on the differences found in the measured price rigidity across stores.[4]

Fourth, Carvalho (2006), Konieczny and Rumler (2006), Nakamura and Steinsson (2008), Álvarez et al. (2016), and Baley and Blanco (2021) emphasize the role that heterogeneity in price rigidity plays in determining the market responsiveness to monetary shocks. We find that regardless of the definition of "regular" prices, there is a significant heterogeneity of price rigidity across stores that follow different pricing formats. This suggests that the heterogeneity of pricing formats might play a role in determining the macro level price rigidity.

Fifth, Ellickson and Misra (2008) find regional differences in the geographical distribution of Hi-Lo, EDLP, and HYB stores across the US. This, along with our finding that stores with different pricing formats have different regular price rigidities, suggests that the heterogeneity in the distribution of pricing formats across the US might play a part in the variability of the effects of monetary policy by regions or states, as reported by Angeloni and Ehrmann (2007) and Francis et al. (2012).

In addition to discussing the frequency of price changes, we study the size of price changes. Firms can respond to shocks by either changing the frequency of price changes or by changing the size of price changes (Sheshinski and Weiss, 1977). Álvarez et al. (2016) show that when all firms respond to a common shock, the variance of the shock equals the number of price changes times the average size of price changes. They also show that for a large class of sticky price models that assume zero inflation and symmetric menu costs, the responsiveness of a market to shocks is captured by a sufficient statistic, which is proportional to the kurtosis of the size of price changes divided by the number of price changes. We find that if we look at the transaction prices, or at the regular prices as defined by the stores, there are significant differences across stores between the product of the number of price changes and the average size of price changes, and between the sufficient statistics of the three stores. If we look at the filtered or at the reference price, there are still differences across stores in the product of the number of price changes and the average size of price changes, but the differences in the sufficient statistics are no longer statistically significant.

---

[4] This also raises a question of whether the term "sales filter" is appropriate, because a large share of the prices that a sales filter tags as sale prices might not be perceived as such by shoppers.

Further, it is possible that the differences that we find in the product of the number of price changes and the average size of price changes are an artifact of the short time series that we have. Therefore, it seems that the filtered and reference price series are capturing the stores' responsiveness to shocks better than the transaction prices or the regular prices as defined by the stores. However, the differences in the price rigidity across stores likely affect the aggregate price rigidity, especially because we find evidence of synchronization across stores. As Carvalho (2006) shows, strategic complementarities across stores with different levels of price rigidity tend to slow down the market responsiveness to aggregate shocks.

An important limitation of our data is its shortness – we have data for only 52 weeks. This might lead to two types of errors. First, for many goods, the filtered and reference price series have no price changes, which may bias downwards our estimates of the duration of prices. Second, sales filters are inherently less accurate near the endpoints (Nakamura and Steinsson, 2008), and this might also affect the accuracy of our estimates. To overcome this limitation, we obtained data for a supermarket in Israel that belongs to a chain that follows a HYB strategy. The data, which covers 171 weeks for 447 products, includes both the regular and the transaction prices as posted by the store. We augment these price series with filtered and reference price series. We find that the frequency of price changes at the Israeli supermarket is similar to that of the Canadian HYB store. We also conclude that short time series leads to underestimation of the price duration. The effect of the imprecision due to endpoints, however, is quite small.

The paper is organized as follows. In section I, we discuss the pricing format landscape in the US and Canada. In section II, we describe the data. In section III, we highlight key characteristics of the data. In section IV, we discuss temporary price cuts and generated regular prices. In section V, we assess the rigidity of the transaction, regular, filtered, and reference prices. In section VI, we discuss the role that the size of the price changes plays in the adjustment process. In section VII, we discuss the synchronization of price changes across stores. In section VIII, we describe the various robustness checks that we have conducted and discuss the results of the analyses we have conducted using Israeli data. In section IX, we address data representativeness and limitations. We end the paper in section X with a summary of key findings, possible macroeconomic implications of our findings, and avenues for future research.

## I. PRICING FORMAT LANDSCAPE IN THE US AND CANADA

Studying price rigidity at stores with EDLP, Hi-Lo, and HYB pricing formats is important because these are the most common pricing formats in North America, Canada, and other countries. Indeed, according to Ellickson and Misra (2008, p. 813), "The majority of both marketers and practitioners frame a store's pricing decision as a choice between offering everyday low prices [EDLP] or deep but temporary discounts [Hi-Lo]."

EDLP retailers guarantee "every-day-low-price," and thus they rarely offer discounts (He et al., 2023). Hi-Lo (or PROMO) retailers charge higher prices but offer frequent promotions by temporarily cutting prices below the EDLP prices (Hoch et al. 1994; Lal and Rao 1997; Gauri et al.

2008).

As Ellickson and Misra (2008) note, however, the EDLP/Hi-Lo dichotomy is too narrow, because firms can choose a hybrid (HYB) format by adopting a mixture of EDLP and Hi-Lo, for example, by varying the product categories on sale, or by changing the frequency of sales across the categories (Bolton and Shankar 2003). Thus, the HYB format can take various forms, combining some features of the EDLP and Hi-Lo formats, while adapting the particulars to the relevant settings, depending on the overall market positioning, local market structure, etc. The specific features of an HYB format can therefore vary by area, competitive landscape, etc.

In other words, the pricing format space in practice is a continuum along the entire spectrum between the EDLP and Hi-Lo. For example, Walmart, Costco, and Food Lion are EDLP retailers, Target follows a Hi-Lo format, while Publix is in-between, following mostly a HYB format.[5]

**Table 1 – About Here**

Table 1 shows the distribution of food store pricing formats by store type in the US. According to the table, the share of the three pricing formats among large stores is about 33% EDLP, 30% Hi-Lo, and 37% HYB, and among small stores—about 22% EDLP, 50% Hi-Lo, and 28% HYB.

Figure 1 shows the distribution of food store pricing formats across the US. Although all three formats are present in all parts of the US, Ellickson and Misra (2008) find regional variations as follows: EDLP format stores are particularly popular in the South, South-East, Southern Central, and the South-West; Hi-Lo format stores are particularly popular in the Great Lakes, Southern Central, North-East, and West Coast; and HYB format stores are particularly popular in the North

---
[5] See Table J1 in the Online Supplementary Appendix J for details.

West, South-West, West Coast, North-East, and South-East.[6]

Unfortunately, we know far less about the pricing format landscape of the Canadian retail food market. In particular, we do not have information on the pricing format of Canadian supermarket chains, or on their geographical distribution. We can offer several observations, nevertheless.

First, based on the population size difference between US and Canada, the Canadian market size is about 1/10 of the US market. Second, four Canadian provinces populate most of Canada's large retail food stores. Ontario, the most populous province of Canada, currently has 5,583 supermarket stores. Quebec—4,405 stores, British Columbia—1,681 stores, and Alberta—1,645 stores. Yukon has just 16 supermarket stores.[7] Third, unlike the US where the stores are spread across the entire country, in Canada, the majority of the retail supermarket stores are located in the southern part of the country along the US border, which is a populated part of Canada.

Perhaps most importantly, however, the pricing practices of Canadian retailers are similar to the pricing practices of American retailers. Indeed, some of the US-based large retail food chains operate stores in Canada as well. These include Shop n'Save, SuperValu, Safeway, Lucky, Costco, Walmart, Whole Foods, A&P, Price Chopper, etc.[8]

A retail food chain might operate stores with different pricing formats.[9] Pricing format, however, is one of the key components of the stores' strategic positioning. A decision about the pricing format, therefore, has long-term consequences, since the cost of changing a pricing format is likely to be prohibitively expensive. Indeed, there are many examples, including J.C. Penny, Sears, Montgomery World, etc., that tried to re-position themselves by changing their pricing format, but failed, often dramatically, many of them going bankrupt.[10] Consequently, stores do not change pricing formats in response to small or temporary shocks, but only as part of long-term strategic decisions.

---

[6] See Table J2 in the Online Supplementary Appendix J for details.
[7] Source: https://www.statista.com/topics/2874/supermarkets-and-grocery-stores-in-canada/#topicOverview, accessed March 29, 2023.
[8] Appendix I offers more details about the retail supermarket landscape in Canada.
[9] For example, among Kroger's stores, the largest US food retailer, 13% are EDLP, 47% Hi-Lo, and 40% HYB (Ellickson and Misra 2008). The data in Table 1 and Figure 1 are based on the 1998 Trade Dimensions survey.

## II. DATA: TRANSACTION PRICES AND REGULAR PRICES

Our data come from three large Canadian supermarket stores that belong to three large chains: Loblaw's, Provigo, and Super-C. The stores are located in Montreal's Notre-Dame-de-Grâce neighborhood, a middle-class residential district.[11] Loblaw's is the largest Canadian retailer, operating 400 stores.[12] Super-C operates 97 stores and Provigo 300 stores.[13]

**Table 2 – About Here**

Table 2 reports the stores' size, parking area, annual sales, number of products, and number of employees. Based on sales (equivalent to $16–$23 million) and the number of products, these are large superstores, comparable to some of the largest US chain stores, such as those studied by Levy et al. (1997) for measuring menu costs.

---

[10] The Warehouse Group of New Zealand switched from Hi-Lo to EDLP in 2017. In their 2018 annual report, the Group explicitly acknowledges how costly it is to switch pricing strategies. The costs include expected sales and margin declines in the near term (an expected 3-year turnaround), the need to restructure the supply chain, building up the private label lines, and even renegotiating trading cycles with business partners. Fast forward to 2020 - the effectiveness of the EDLP move is still a "work in progress" with a focus on identifying the right product portfolio and leveraging better forecasting abilities. These suggest that changing a pricing strategy is a long-term decision. See:
https://www.thewarehousegroup.co.nz/application/files/3815/3775/0583/2018_Annual_Report_EDLP_Case_Study.pdf, https://www.thewarehousegroup.co.nz/download_file/force/1583/174, and
https://www.thewarehousegroup.co.nz/download_file/force/2600/174, all accessed November 22, 2021. Iceland, an UK-based frozen goods retailer, had ditched its EDLP strategy in 1997 and went back to Hi-Lo. See: https://www.campaignlive.co.uk/article/iceland-freezes-edlp-policy/63585, accessed November 22, 2021.

[11] The district population is 104,974, average age 41.2, 42.8% with academic degrees, a median gross household income of $58,178, and 8.6% unemployment rate. As a comparison, the median gross household income in Canada is $70,336, 54% with academic degrees, and 6.8% unemployment rate (Statistics Canada 2016).

[12] Sources: https://mtltimes.ca/Montreal/local-businesses/loblaws-cavendish-store-in-notre-dame-de-grace-becomes-a-provigo/, and https://www.theglobeandmail.com/business/article-loblaw-profit-revenue-gain-as-bigger-baskets-help-offset-slower/, both accessed May 13, 2020.

[13] In November 1998, Provigo was purchased by the Loblaw's Group. Loblaw's and Provigo, however, are run independently of each other, each with its own market positioning, format, and identity. In January 2016, the Loblaw's store in our sample was turned into a Provigo store. We collected the data from July 2003–July 2004, long before that happened. See:
https://www.wsj.com/articles/SB909782024300867500, accessed November 18, 2021.

The three stores differ in their pricing format: Provigo is a Hi-Lo store, Loblaw's is an EDLP store, and Super-C is an HYB store.[14] Figure 2 shows the stores' locations.

In each store, we hand-collected weekly price data during July 30, 2003–July 23, 2004. Every week, for each good, we manually recorded either one price or two prices, shown on the shelf price tags. If a good was not on sale, then we recorded the regular price, which was also the transaction price. If a good was on sale, then the retailer posted *the regular price next to the transaction price*, and we recorded both. Figure 3 shows actual price tag examples from the three stores.

We thus have two weekly price series for each good at each store: *transaction prices* and *regular prices*. Both are classified as such by the store managers. That is, our regular and transaction prices are *regular prices* and *transaction prices* as viewed by the store management, and as communicated to the shoppers through shelf price tags. On any week, the two prices of a good differ from each other if the good is on sale on that week. Otherwise, the two prices coincide.

From each store, we have observations for 89 national brand (*NB*) goods in 11 product categories. In addition, we have price data for 39 private label (*PL*) goods (10 at the EDLP store, 10 at the Hi-Lo store, and 19 at the HYB store).[15] For each good, we have 52 weekly price observations. In other words, we have no missing observations, giving us a total of 15,912 weekly observations.[16] In addition to the prices, we recorded the products' location: back/middle/front aisle and bottom/eye-level/top shelf.[17] These serve as controls in the regression equations we estimate.

---

[14] The pricing format of each store was self-identified by the store managers when we interviewed them. Super-C follows a discount format, a type of HYB format. It offers low daily prices like EDLP stores, but with occasional discounts like Hi-Lo stores, to generate an image of "best deals," in addition to the image of everyday best prices.

[15] PL goods are specific to each chain/store and therefore they are not comparable across the stores.

[16] The total number of observations $n = (89 \times 52 \times 3) + (39 \times 52) = 15,912$. Appendix D lists the products included in our sample, and the corresponding regular and transaction prices.

[17] Manufacturers compete for eye-level shelf spaces by paying the supermarkets various slotting and display fees.

Our dataset is small because it was hand-collected, but it has three unique features. First, we have *both* the actual regular prices and the actual transaction prices of each product, each week, as posted on store shelves. Thus, we can match the regular prices with the corresponding transaction prices (if they differ), as viewed by the stores' management and the shoppers. Second, the stores in our sample represent three pricing strategies (EDLP, Hi-Lo, and HYB). Third, the stores are located close to each other (Figure 2), catering to consumers from the same geographical area.

**Table 3 – About Here**

Table 3 gives the average regular and transaction prices at each of the stores, along with the results of Wilcoxon rank-sum tests for pairwise comparisons. We find that all pairwise comparisons are statistically significant for both the regular and the transaction prices. The Hi-Lo store has the highest average regular (transaction) prices, C$4.58 (C$4.47), the HYB store has the lowest average regular (transaction prices), C$3.98 (C$3.94), and the EDLP has average regular (transaction) price in-between C$4.12 (C$4.11).

**III. SAMPLE PRICE SERIES: A GENERAL PICTURE AND SOME CHARACTERISTICS**

To illustrate price behavior and its effects on price rigidity/flexibility, we depict in Figure 4 the regular and transaction price series at the three stores for five products: (1) Dishwashing Liquid, (2) Perrier, (3) Frozen Dessert, (4) Eggs, and (5) Cheerios. A visual examination of the plots leads to the following general observations:

a) Average prices at the Hi-Lo store are higher than at the EDLP and HYB stores.
b) The Hi-Lo store offers frequent and deep temporary price cuts, where its price falls below the EDLP and HYB prices.

c) The price gaps between the EDLP and HYB stores are small. For three products (Dishwashing Liquid, Frozen Dessert, and Cheerios), the average EDLP store prices are below the HYB store prices, and for two products (Perrier and Eggs) it is the other way around.

d) The transaction prices at the Hi-Lo store change more often than at the other stores.

e) The regular prices at the EDLP store change more frequently than at the other stores.

f) The total number of price changes at the EDLP and HYB stores are similar. But there is an important difference between them: at the EDLP store, the vast majority of these price changes are presented as changes in *regular prices*, whereas at the HYB store, most price changes are classified as *sale prices*.[18]

g) The EDLP store rarely presents temporary price cuts as sales. In our entire data, we find only 12 price cuts that the EDLP store classifies as "sales." Figure 4 shows one such case – the sale of eggs at the start of the sample period. This "sale" is characteristic of sales at the EDLP store: when the EDLP store defines a price cut as a "sale," it is usually an exceptionally deep price cut.

h) Temporary price cuts that are not sales occur also at the Hi-Lo and HYB stores. For example, in the price of Perrier at the Hi-Lo store, we see price cuts on the $8^{th}$ and $35^{th}$ weeks, which the store classifies and presents as a cut in the regular price. Such temporary regular price changes are rare in the Hi-Lo and HYB stores.

i) Sales do not always end with the pre-sale price. Sometimes the post-sale price is lower than the pre-sale price. For example, at the Hi-Lo store, on the $43^{rd}$ week, there is a sale of Perrier, but when the price returns to the regular price after the sale ends, the new regular price is below the previous regular price.

j) Consistent with Anderson et al. (2017), the post-sale price may exceed the pre-sale price as well.

---

[18] Consequently, there are many temporary price cuts at the EDLP store which visually resemble "sales" (i.e., deep and temporary price cuts), but which the store classifies and presents as changes in *regular prices*.

For example, the transaction price of Frozen Dessert at the HYB store drops on the 31st week because of a sale, and then it goes back up, but to a higher level than before the sale. We see a similar event in the case of Eggs at the HYB store on the 21st week.

k) Prices sometimes go up for very short periods, consistent with Chahrour (2011) and Syed (2015). For example, the price of Dishwashing Liquid at the EDLP store, on the 47th week, or the price of Frozen Dessert at the EDLP store, on the 33rd week.

To summarize, the Hi-Lo store offers far more temporary price cuts than the other two stores. In addition, at the Hi-Lo and HYB stores, when the transaction prices are temporarily reduced, the regular prices usually remain unchanged. At the EDLP store, in contrast, the temporary price cuts are treated by the store as cuts in regular prices.

Thus, at the Hi-Lo and HYB stores, when a price is temporarily cut, buyers observe the reduced price along with the unchanged regular price, allowing them to assess the gains from buying at the reduced price. This also alerts them that if they do not buy now, they will likely face higher prices next time. Consumers, facing such situations, are likely to buy more than they would normally do, especially if the good is storable (Hendel and Nevo, 2013; Fox and Syed, 2016; Glandon 2018).

When an EDLP store temporarily reduces the price, however, the consumers likely treat the reduced price as a regular price because that is how the store presents it. In such situations, they have no incentive to buy more than they would normally do, as they do not see any sign that hints at the possibility of higher prices in the weeks ahead.

## IV. TEMPORARY PRICE CUTS AND GENERATED REGULAR PRICES

To assess the effect of the store-pricing format on price rigidity/flexibility, we consider the behavior of four price series for each good, at each store. The first two price series are the transaction and regular price series as defined by the store. These are the actual prices we hand-collected from the store shelves. The third is a filtered price series which we generated by employing Nakamura and Steinsson's (2008) sales filter A. The filter identifies temporary V-shaped (not necessarily symmetric) price cut patterns as sale prices. Following Chahrour (2011), we assume that sales last 6 weeks or less. The fourth price series is a *reference* price series which we generate based on Eichenbaum et al. (2011), using Chahrour's (2011) algorithm, which employs a 13-week rolling window.[19] Chahrour's (2011) algorithm is designed to remove temporary price changes, regardless of their direction. It sets the reference price equal to the modal price in a rolling window and then makes some modifications to smoothen the resulting reference price series.

Figure 5 demonstrates the properties of the four price series. The figure shows the transaction, regular, filtered, and reference prices of a dishwashing liquid at the EDLP, Hi-Lo, and HYB stores.[20] At the EDLP store, there is a small number of V-shaped price cuts that look like sale prices, but the EDLP store views them as regular price changes, and thus the transaction and regular prices at the EDLP store coincide. The filtered prices are smoother because the filter removes all V-shaped price cuts. The reference prices are similar to the filtered prices, with the exception that Chahrour's (2011) filter also removes the one-week price hike that occurs on the $48^{th}$ week.

At the Hi-Lo store, regular prices last long periods, with frequent V-shaped price cuts. Consequently, transaction prices are more volatile than regular prices. The filtered series resemble the regular series, demonstrating that the filter performs well with the Hi-Lo store data. The reference price series is a smoothed transaction price series, and thus, in this case, it remains unchanged during the sample period.

---

[19] Eichenbaum et al. (2011) define a reference price as the modal price in a quarter, but we have only 52 weeks of data, and thus with their algorithm we would have a maximum of four different price changes per good. Klenow and Malin (2011) define a reference price based on the modal price in a 13-observation rolling window. Chahrour (2011) and Kehoe and Midrigan (2015) suggest that such an algorithm might result in the reference price changing either too early or too late and offer procedures for mitigating this problem.

[20] The regular and transaction prices of this product are also shown in the top row of Figure 4.

The HYB store offers few V-shaped transaction price cuts, and thus its regular prices are smoother than the transaction prices. The filtered prices are similar to regular prices, suggesting that the sales filter performs well also at the HYB store. The only place where it misses a sale is at the end of the sample period, where the price cut is treated by the filter as a regular price change because it "cannot find" a price increase that must follow the price cut.

Figure 5 illustrates well the general pattern found in our data. We summarize the findings on temporary price cuts in Table 4, which shows the number of sale events according to the various price series. Column 1 reports the number of promoted sales, defined as periods when a store's transaction price was below the store's regular price, thus informing the consumers that a product is on sale. Column 2 reports the number of filtered sale events, defined as periods when the filtered price is above the transaction price. Column 3 reports the number of reference sale events, defined as periods when the reference price is above the transaction price. Columns 4–6 give the results of Pearson $\chi^2$ tests for comparing the shares of price changes between each pair of stores.

**Table 4 – About Here**

The results underscore the differences in the way the stores use and treat temporary price cuts. According to all definitions, the Hi-Lo store uses more temporary price cuts than the other stores. It had 508 promoted sale events, 509 price cuts that Nakamura and Steinsson's (2008) sales filter A defined as filtered sale events, and 497 price cuts that Chahrour's (2011) algorithm defined as reference sale events. All these numbers are significantly larger than at either of the other stores.

The EDLP store offers 12 promoted sales, significantly less than at either the Hi-Lo store (508 promoted sales) or the HYB store (265 promoted sales). However, the number of temporary price cuts at the EDLP and HYB stores is similar. According to the sales filter, there are 265 filtered sale events at the EDLP store and 280 in the HYB store. According to the reference prices, there are 261 sale events at the EDLP store and 262 at the HYB store. In both cases, the differences are not statistically significant.

Thus, according to all price series, the Hi-Lo store has significantly more temporary price cuts than any of the other stores. The EDLP and HYB stores have a similar number of temporary price cuts, but the HYB store promotes them as sales, while the EDLP store does not. Instead, the EDLP store treats most of these price cuts as regular price changes.

## V. VARIATION IN PRICE RIGIDITY: PRICE CHANGE FREQUENCY AND PRICE SPELL DURATION

*Summary statistics*

Given the emphasis in the literature on the different effects that regular and sale prices have on the aggregate price level, it is of interest to study how the stores' treatment of temporary price cuts affects their price rigidity. In Panel A of Table 5, we present category-level average weekly price change frequencies at each store for the 11 product categories, for each of the four price series (transaction, regular, filtered, and reference). The averages are computed over all goods in each category.

**Table 5 – About Here**

The figures in the table indicate substantial heterogeneity in the average frequency of weekly transaction price changes across categories, consistent with Bils and Klenow (2004) and Nakamura and Steinsson (2008). Except for the EDLP store, which treats temporary price cuts as regular price changes, the price variability is smaller for regular prices than for transaction prices. In all stores, the variance is even smaller for the filtered and the reference prices.

Despite the heterogeneity we find across categories, however, when we compare across stores, we find a consistent pattern. We, therefore, compare the overall average frequencies across stores rather than across categories. Table 6 presents the results of Pearson $\chi^2$-tests of proportions of price changes for pairwise comparisons.

**Table 6 – About Here**

The Hi-Lo store has the highest frequency of weekly transaction price changes, 23.29%, compared to 13.83% and 13.76% at the EDLP and HYB stores, respectively. The differences between the Hi-Lo and the EDLP stores, and between the Hi-Lo and the HYB stores, are significant ($\chi^2 = 152.39$ and $\chi^2 = 162.89$, respectively, $p < 0.01$ in both cases). There is no statistically significant difference between the EDLP and HYB stores ($\chi^2 = 0.01$, $p > 0.92$).

If we look at the regular prices as defined by the stores, then the EDLP store has the highest frequency of weekly price changes with 13.38%, compared to 4.06% and 5.34% at the Hi-Lo and HYB stores, respectively. All pairwise differences are statistically significant, with the $\chi^2$ statistics being 281.09 and 208.19 for comparing the EDLP store with the Hi-Lo and HYB stores, respectively, and 9.80 for comparing the Hi-Lo store with the HYB store ($p < 0.01$ in all cases).

The results are somewhat different for the filtered series. The average frequency of filtered price changes at the EDLP and HYB stores, 4.25% and 4.50%, respectively, are about the same $\chi^2 = 0.40$, $p > 0.52$), and both exceed the corresponding figure at the Hi-Lo store, 3.55%. The $\chi^2$ statistics for comparing the Hi-Lo store with the EDLP and HYB stores, are 3.35 ($p < 0.07$) and 6.24 ($p < 0.02$), respectively.

Focusing on the reference prices, i.e., "long-lived" prices, the HYB store has the highest frequency of weekly price changes with 3.95%., exceeding the frequency at the EDLP, 2.70% ($\chi^2 = 13.01$, $p < 0.01$), and at the Hi-Lo store, 2.23% ($\chi^2 = 26.17$, $p < 0.01$). The gap between the EDLP and Hi-Lo stores is not statistically significant ($\chi^2 = 2.33$, $p > 0.12$).

Thus, if we consider the filtered prices, then the EDLP and HYB stores have more flexible prices than the Hi-Lo store. If we look at the reference prices, then the HYB store has the most flexible prices.

In Panel B of Table 5, we report the implied average price duration in weeks. To alleviate the effect of truncation on the measured duration, we calculate the duration as

$$(1) \quad -\left[ ln\,(1 - \underline{f}) \right]^{-1},$$

where $\underline{f}$ is the average weekly price change frequency (Levy et al. 1997; Nakamura and Steinsson, 2008). For the EDLP store, the average durations of price spells are 6.72, 6.96, 23.00, and 36.53 for transaction, regular, filtered, and reference prices, respectively. For the Hi-Lo (HYB) store, the corresponding values are 3.77 (6.75), 24.13 (18.22), 27.63 (21.69), and 44.26 (24.79).

As Carvalho (2006) shows, however, calibrating sticky price models using information on average frequencies, may underestimate price stickiness because of Jensen's inequality. We, therefore, calculate an alternative measure of price durations,

$$(2) \quad -\frac{1}{N_C}\Sigma_{i \in C} \quad [ln(1-f_i)]^{-1},$$

where $f_i$ is the weekly price change frequency of product $i$ in category $C$, and $N_C$ is the number of products in category $C$. Using our data to calculate Equation 2, however, has a drawback, because we have to drop products that have 0 price changes, biasing our estimates downwards. The results are summarized in Panel B of Table E1, in Appendix E.

For the EDLP store, the expected durations of price spells are 10.70, 10.94, 26.01, and 33.52 for transaction, regular, filtered, and reference prices, respectively. For the Hi-Lo (HYB) store, the corresponding values are 8.94 (10.55), 27.66 (21.96), 29.53 (24.44), and 36.23 (28.30).

Comparing the duration figures based on Equation 2 to those based on Equation 1, we find that for the transaction prices, Equation 1 underestimates the durations by 33.0%–137.1%. For the regular prices, as defined by the store, Equation 1 underestimates the durations by 14.6%–57.2%. For the filtered prices, Equation 1 underestimates the durations by 6.9%–13.1%. For the reference price, we find that for both the EDLP and Hi-Lo stores, the results based on Equation 1 are larger than the results based on Equation 2, implying that the bias due to the removal of products with 0 price changes is quite significant. This also suggests that the results reported in Panel B of Table E1 for the regular and the filtered price series, might be biased downwards.

Therefore, to better compare the two measures of price stickiness, we use Equation 1 again, but this time we apply it to the sample we used to calculate Equation 2. I.e., we use information only on products that had at least 1 price change. The results are reported in Table E2, in the Appendix. We find that when we use the same samples, the price stickiness based on Equation 2 is 19.5%–142.9% greater than when we use Equation 1. We also find that the differences depend on both the price flexibility of the products and on the variance in the flexibility across products. Using the average frequencies, therefore, leads to particularly severe underestimation of the price stickiness for transaction prices at the Hi-Lo store (142.9%) and at the EDLP store (70.1%). The underestimation is least severe for the filtered and reference prices at the HYB store, 19.5% and 22.2%, respectively.

*Econometric estimation*

The results above are suggestive, but they could be affected by heterogeneity across goods, categories, etc. To account for the effects of covariates, while controlling for cross-category heterogeneity in the average frequency of price changes and for truncation, we estimate a series of Cox semi-parametric hazard functions, one regression for each series of price changes:

(3) $\quad h(t) = h_0(t) \times exp\left(\beta_1 EDLP + \beta_2 HYB + X\gamma + Z\delta\right)$

where $h(t)$ is the hazard of a price change at time $t$, and $h_0(t)$ is the baseline hazard when all covariates equal 0. The main covariates are dummies for the EDLP and HYB stores. The elements of *X* are further covariates, which include the price level of the good, defined for each store as the average transaction price over the sample period, a dummy for private label (*PL*) product, a dummy for price changes made in January, and a dummy for price changes made in Christmas week. *Z* includes fixed effects for the product location in the stores—for the aisle (back/front/middle) and for the shelf (bottom/top/middle). We allow for recurrent price changes, and we stratify the data by categories to allow the hazard in different categories to be non-proportional.

Table 7 summarizes the estimation results. The values in the table are the proportional changes in the hazard in response to a one-unit change in each covariate. Robust standard errors, clustered at the good-store level, are reported in parentheses.

### **Table 7 – About Here**

We find that prices are more likely to change in January, consistent with Nakamura and Steinsson (2008). We also find that except for the regular price series, prices are more likely to change in the Christmas week than at other times.[21] More importantly, we find that the results of the hazard function estimation corroborate the findings we discuss above. Consider first the hazard function estimate for transaction prices. According to the first column, the hazard that a price will change at the EDLP (HYB) store is 0.66 (0.64) times the hazard at the Hi-Lo store, and the differences are statistically significant ($p < 0.01$ in both cases). There are no statistically significant

---

[21] Observing more price changes on Christmas week is consistent with Warner and Barsky (1995) and Levy et al. (2010), who find a higher frequency of price changes in the week prior to Christmas.

differences between the hazards at the EDLP and HYB stores ($\chi^2 = 0.09, \ p > 0.76$).

Next, consider the hazard function estimate for regular prices. Here we find the opposite. The hazard that a regular price will change at the EDLP store is 3.62 times the hazard at the Hi-Lo store ($p < 0.01$). The hazard that a regular price will change at the HYB store is 1.41 times the hazard at the Hi-Lo store ($p < 0.01$). The difference between the EDLP and HYB stores is also significant ($\chi^2 = 116.87, \ p < 0.01$).

When we look at the filtered prices, we find that the hazard that a filtered price will change at the EDLP store is 1.26 higher than at the Hi-Lo store. The difference is not statistically significant ($p > 0.24$). The hazard that a filtered price will change at the HYB store is 1.28 times the hazard at the Hi-Lo store ($p < 0.02$). There is no significant difference between the hazards of the EDLP and HYB stores ($\chi^2 = 1.10, \ p > 0.29$).

Finally, considering the reference prices, we find that the hazard that a reference price will change at the EDLP store is 1.20 times the hazard at the Hi-Lo store, but the difference is not statistically significant ($p > 0.14$). The hazard that a reference price will change at the HYB store is 1.70 times the hazard at the Hi-Lo store ($p < 0.01$). The difference between the EDLP and HYB stores is statistically significant too ($\chi^2 = 10.81, \ p < 0.01$).

In sum, there are significant differences in the price rigidity between the stores, regardless of which price series we consider. If we consider transaction prices, then the Hi-Lo store has the most flexible prices. If we consider regular prices, then the Hi-Lo store has the least flexible prices whether as defined by the store, as defined by the sales filter A, or as defined by Chahrour's (2011) algorithm. The EDLP store has the most flexible prices if we consider regular prices as defined by the store. The HYB store has the most flexible prices if we consider reference prices. If we consider filtered price series, then the EDLP and HYB store have a similar level of price rigidity.

# VI. THE SIZE OF PRICE CHANGES

Retailers can respond to changes in economic conditions by varying either the frequency of price changes or the size of price changes (Sheshinski and Weiss, 1977, Caballero and Engel, 2007). Above we studied the frequency of price changes. In this section, we focus on the size of price changes.

Álvarez et al. (2016) show that for a large class of sticky price models, the volatility of the underlying shocks in the steady state should equal $N(\Delta p_{i,s,t}) \times var(\Delta p_{i,s,t})$, where $N(\Delta p_{i,s,t})$ is the average number of price changes per year, and $var(\Delta p_{i,s,t})$ is the variance of price changes (Baley and Blanco, 2021). Following Álvarez et al. (2016), we calculate the size of price changes as $\Delta p_{i,s,t} = ln(p_{i,s,t}) - ln(p_{i,s,t-1})$, where $p \in \{transaction\ price,\ regular\ price,\ filtered\ price,\ reference\ price\}$ is the price of product $i$ sold in store $s \in \{EDLP,\ Hi-Lo,\ HYB\}$ on week $t$, conditional on a price change.

Assuming that stores that are located in the same neighborhood face similar shocks, we expect that they would have similar values of $N(\Delta p_{i,s,t}) \times var(\Delta p_{i,s,t})$. Figure 6 depicts the values of $N(\Delta p_{i,s,t}) \times var(\Delta p_{i,s,t})$ for each store, for each of the four price series. We calculate $N(\Delta p_{i,s,t})$ using the values in Panel B of Table E1. The vertical lines in the figure represent 95% confidence intervals, which we calculate by bootstrapping the calculation 1,000 times.

For transaction prices, the results suggest that the Hi-Lo store responds to shocks more than the EDLP store, which responds more than the HYB store. This reflects the large number of sales at the Hi-Lo store, which are both common and large in percentage terms. Therefore, the responsiveness of the transaction prices likely over-estimates the response of the Hi-Lo store to shocks, as sales are often set in advance (Warner and Barsky 1995; Nakamura and Steinsson, 2008; Anderson et al., 2017).

For regular prices as defined by the stores, we find that the EDLP store responds much more than the Hi-Lo and HYB store. This reflects the EDLP store's practice of defining temporary price changes as regular price changes. It is therefore likely that focusing on regular prices as defined by the store would likely overestimate the responsiveness of the EDLP store to underlying shocks.

For filtered prices, we find that although the likelihood of a filtered price change at the EDLP and HYB stores are about the same, the HYB store has a lower measure of responsiveness than either of the other two stores. This reflects the smaller size of price changes at the HYB store.

The HYB store has a lower measure of responsiveness than either of the other two stores also for the reference price, again reflecting the small price changes that it makes. However, it is important to note that our measures of the average number of price changes are lower bounds on the true values. Therefore, it is possible that we underestimate the number of price changes that each of the stores makes in steady state. If we had a longer time series, we might have found that the HYB store's response is not different than that of the other stores.

To explore this possibility further, we employ an alternative measure of the responsiveness of stores to shocks. Álvarez et al. (2016) show that if we assume symmetric menu costs and zero inflation, then for a large class of sticky price models, the real effects of a small monetary shock depend on the ratio of two statistics: the kurtosis of the size of price changes, and the frequency of price changes. The frequency is important because when firms adjust prices more often, they can respond quickly to a monetary shock, dampening the real effects of the shock. The kurtosis is equally important because a small kurtosis implies a selection (Golosov and Lucas, 2007). I.e., price changes are made by the firms whose prices are in the greatest need of adjustment, which also reduces the real effects of a monetary shock.

Figure F1 in Appendix F depicts the histograms of the size of price changes. We find that there is a large variation in the kurtoses, both across stores and across price measures. If we look at the transaction price, we find that the kurtoses are between 3.48 at the Hi-Lo store and 4.63 at the HYB store. When we focus on regular prices, the kurtosis of the EDLP store remains almost unchanged (4.29), but the removal of sales, which are usually large in percentage terms, leads to an increase in the kurtoses of the Hi-Lo (8.52) and HYB (5.78) stores. For filtered prices, the kurtoses are more similar across the three stores: 8.64 at the EDLP store, 7.52 at the Hi-Lo store, and 7.48 at the HYB store. There is also a large variation in the kurtoses of the reference prices: 5.17 at the EDLP store, 7.32 at the Hi-Lo store, and 4.24 at the HYB store.

The large variation of the kurtoses across stores suggests that depending on the price series, variations in the size of price changes might play a different role in the transmission of monetary shocks. However, Álvarez et al. (2016) show that when we pool price changes of different products, we might bias the estimated kurtosis upwards (Midrigan, 2011). Therefore, following Álvarez et al. (2016), we standardize the price changes by computing $Z_{\Delta p_{i,s,t}} = \frac{(\Delta p_{i,s,t} - \overline{\Delta p_{i,s}})}{\sigma_{\Delta p_{i,s}}}$, where $\overline{\Delta p_{i,s}}$ is the average size of price changes in the category of product $i$ in store $s$, and $\sigma_{\Delta p_{i,s}}$ is the standard deviation of the size of price changes in the category of product $i$ in store $s$. We then pool the standardized price changes and draw the histograms again.

Figure 7, which depicts the resulting histograms, suggests that once we standardize the data, the kurtoses are more similar across stores and price measures. For the transaction prices, the kurtoses are in the range 3.77–4.30, for the regular prices 4.02–5.29, for the filtered prices 3.74–5.05, and for the reference prices 3.76–4.03. These results are in the same range as Midrigan (2011), Álvarez et al. (2016) and Cavallo (2018).

To offer an intuitive assessment of the significance of these results for the transmission of monetary shocks, we use Álvarez et al.'s (2016) sufficient statistics. According to Álvarez et al. (2016), the real effects of a monetary shock are proportional to $kur(\Delta p_{i,s,t})/N(\Delta p_{i,s,t})$, where $kur(\Delta p_{i,s,t})$ is the kurtosis of price changes in store $s$. Using information on the kurtoses from Figure 7, we calculate $kur(\Delta p_{i,s,t})/N(\Delta p_{i,s,,t})$ for each of the four price series, in each of the three stores. To obtain 95% confidence intervals we bootstrap the calculations 1,000 times.

Figure 8 depicts the values of $kur(\Delta p_{i,s,t})/N(\Delta p_{i,s,,t})$ for each of the four price measures and for each of the three stores. We find that for transaction prices, the Hi-Lo store has a significantly lower value of $kur(\Delta p_{i,s,t})/N(\Delta p_{i,s,,t})$ than the other two stores. For regular prices as defined by the stores, the EDLP store has a ratio that is significantly lower than at the other two stores. If we focus on the filtered or reference prices, however, the ratios are not statistically different across the three stores. Therefore, if we believe that the regular prices, as defined by the filtered or by the reference prices, capture the responsiveness of the firms to shocks, then for the class of models studied by Álvarez et al. (2016), an economy populated by any of the three stores would have a similar response to a monetary shock.

## VII. SYNCHRONIZATION OF PRICE CHANGES

Carvalho (2006) finds that when an economy is populated by firms with different price rigidities, then the macro level price rigidity would be exacerbated relative to an economy in which all firms have the same level of price rigidity. This is particularly true in the presence of strategic complementarities, because when stores keep their prices close to the average price of their competitors, stores with rigid prices slow down the market response to shocks.

To obtain an estimate for strategic complementarities, we study the synchronization of price changes across stores by calculating for each product the Fisher-Konieczny's (2000) index of synchronization. The index attains values between 0 and 1, where 0 indicates perfect staggering (firms change prices at constant intervals), and 1 indicates perfect synchronization (prices change simultaneously in all stores). We focus on the 89 national brand products that are sold in all three stores.

Figure 9 depicts the histogram of the Fisher-Konieczny's (2000) indices of the 89 products. Focusing on the transaction prices, the indices are in the range 0–0.61, with an average of 0.26. This value is higher, indicating more synchronization, than reported by Fisher and Konieczny (2000), and similar to Lach and Tsiddon (1996).

In other words, the stores in our sample do not perfectly synchronize price changes, but they are also far from perfect staggering. It, therefore, seems that the stores in our sample respond to price changes made by other stores, which seems reasonable since they are located in the same neighborhood.

Looking at regular prices, as defined by the store, by Nakamura and Steinsson's (2008) sales filter and by Chahrour's (2011) reference price series, we find that they are less synchronized than changes in the transaction prices. For regular prices, the average Fisher-Konieczny index is 0.11, for the filtered prices, 0.09, and for the reference prices, 0.10. In other words, the stores in our sample seem to change their transaction prices in response to changes in the transaction prices of the other stores. Changes in the regular prices, however, are less synchronized. A possible explanation is that when a store changes a price, the other stores might not be certain whether the change is temporary or more permanent (Levy et al. 2002). Therefore, they might change their transaction price in response, but change the regular price only after they identify the competitor's price change as a regular price change. This type of behavior will result in a more staggered pattern of regular price changes.

## VIII. ROBUSTNESS TESTS

*Robustness check with Canadian data*

We run four robustness tests, which we briefly summarize below. The Online Supplementary Appendix contains a detailed description of these tests. It also contains a discussion of how our data compares to similar but larger datasets in terms of the distribution of the last digit and the last two digits of the price (Appendix C), and a detailed list of all the products included in our sample (Appendix D).

First, in the paper, we study the price level at each store. We find that the Hi-Lo store has the highest regular and transaction prices and that the prices at the HYB store are lower than at the EDLP store. Repeating the analyses at the category level yields similar results. See Appendix A for details.

Second, we assess the extent of price rigidity at the level of product categories. Above, we compare the weekly frequencies of price changes at the store level. We find similar results for category-level data. See Appendix B for more details.

Third, in Appendix E, we discuss alternative measures of the average price durations to gauge the effect of Jansen's inequality on the measured price durations.

Fourth, one disadvantage of sales filters such as Nakamura and Steinsson (2008), is that they are less precise near the endpoints. Because we have only 52 weeks of data, we cannot completely overcome this problem. As a partial solution, we exclude observations close to the endpoints and recalculate the frequency of price changes and implied price durations. Under this assumption, we find that the average frequencies of price changes, which are lower bounds on the real average frequencies, are 3.69%, 2.91%, and 4.20% for the EDLP, Hi-Lo, and HYB stores, respectively. This is compared to 4.25%, 3.55%, and 4.50% when we use the filter as designed.[22] See Appendix G for more details.

*Robustness check with Israeli data*

---

[22] When we exclude the final weeks from the sample, we risk biasing our estimates if the probability of sales is higher in the excluded weeks than in other weeks. However, to the best of our knowledge, neither the academic nor the practitioners' literature suggests that sales of the products that we study, are particularly common in June–July (Volpe and Lee, 2012). See also: https://maplemoney.com/canadas-yearly-sales-cycle/, accessed June 20, 2023.

The short span of our data, only 52 weeks, creates two problems. First, for many products, the regular, filtered and reference prices do not change at all, which leads to a downward bias in duration estimates. Second, the reduced precision of sales filters near the endpoints is more severe when the data is short. We, therefore, employ data for a large retail food store in Israel that follows the HYB format. We have 171 weeks of data on 447 products. For each product, we have data on both the regular and the transaction prices.

We use this data to estimate the frequency and duration of prices. We find that for all 4 data series, the frequency of price changes in the Israeli store is similar to the frequency of price changes we find in the Canadian HYB store. The likelihood that a transaction, regular, filtered and reference price will change in a given week are 13.62%, 4.68%, 5.79%, and 4.47%, respectively.[23] When we use Equation (2) to estimate the implied durations, we find that the transaction, regular, filtered and reference prices change every 18.34, 69.95, 43.97, and 47.85 weeks, respectively.

This suggests that the durations we obtain using Equation (2) for the Canadian HYB store, 10.55 weeks, 21.96 weeks, 24.44 weeks, and 28.30 weeks, for the transaction, regular, filtered and reference prices, are perhaps biased downwards.

Assessing the extent of price rigidity by excluding the first and the final weeks, however, has only a small effect on the estimated durations. If we remove observations less than 6 weeks from the beginning or the end of each series, we find that the implied durations are shorter by 3.0%–6.4%.

## IX. DATA REPRESENTATIVENESS AND DATA LIMITATIONS

An important caveat that we shall note concerns our dataset's limitations, which is due to the hand-collection process we had to employ to collect it. We only have one year of data from three stores, and for a limited number of products, raising a question over the representativity of our data. Also, we do not have quantity data, and the dataset itself is dated. We recognize these shortcomings. An important question, therefore, is whether we can treat the stores we sampled and their price data as reasonably representative of similar store formats, their price-setting practices, etc.

---

[23] The likelihood of a regular price change, as defined by the store, is lower than the likelihood of a filtered price change, and it is similar to that of the reference price. It turns out that the reason is that the chain tends to keep a high regular price and a low transaction price for long periods. Occasionally, it changes the transaction ("sale") price. These changes are then recorded as changes in the filtered price and, consequently, there are more filtered price changes than regular price changes.

To answer this question, we offer four observations. First, we interviewed the managers of the three stores, and they self-identified and confirmed our information about their store formats, which was consistent also with the general knowledge at the time. They also offered details of how they manage prices in their stores, etc., and these details are consistent with the typical characteristics of their store formats.[24]

Second, we looked at the existing empirical studies that use retail price datasets, focusing on their descriptive statistics. We identified eight studies (all in marketing journals), that use comparable but larger datasets from EDLP and Hi-Lo stores.[25] The studies use prices of different products, from different locations, and address different questions, yet overall they report that (1) Hi-Lo prices are higher than EDLP prices, (2) the average change in Hi-Lo prices is higher than in EDLP prices, (3) the variance of the change in Hi-Lo prices is higher than in EDLP prices, and (4) Hi-Lo stores offer more deals than EDLP stores. We find similar attributes in our data. The average weekly frequencies of price changes reported by Levy et al. (1997) for Hi-Lo and EDLP chains are also consistent with the behavior we document here.

Third, the price behaviors we find at the three stores are typical and consistent with textbook descriptions of similar-format stores. For example, the descriptions of price setting and price adjustment practices of EDLP, Hi-Lo, and HYB stores, found in textbooks on retail pricing, are consistent with the pricing behavior found in our data (e.g., Monroe 2002, Nagle and Müller 2017).

---

[24] During the data collection period, one of the coauthors of this paper as well as our research assistant lived in Montreal, and as part of the general knowledge, they both knew quite well the pricing formats of the three stores.

[25] These are Hoch et al. (1994), Shankar (1996), Bell and Lattin (1998), Galata et al. (1999), Bolton and Shankar (2003), Voss and Seiders (2003), Rondan-Cataluña et al. (2005), and He et al. (2023). We were unable to find a study that considers an HYB store similar to ours, that also reports the store's regular and transaction price statistics.

Fourth, in Appendix C, we compare the price ending distribution in our data to those found in large scanner price datasets and find that 9- and 99-ending prices are a dominant feature in our data, consistent with the findings in the literature (Levy et al. 2011, Anderson et al. 2015, Knotek 2019, Snir and Levy 2021).

Thus, the descriptive statistics, and other attributes of the price behavior we report, as well as the pricing practices of the stores in our sample, are all in line with comparable figures and information reported in the literature for larger datasets. We believe, therefore, that our dataset, although small, is still a good representative of the price-setting practices of the retailers that follow EDLP, Hi-Lo, and HYB pricing formats.

## X. CONCLUSIONS, MACROECONOMIC IMPLICATIONS, AND FUTURE RESEARCH

We use a hand-collected dataset to study different notions of sale and regular prices, and their variability with stores' pricing format (i.e., pricing strategy). The dataset is unique for three key reasons. First, the stores in our sample follow different pricing formats—one is EDLP, the second is Hi-Lo, and the third is HYB. Second, we have both the actual transaction prices and the actual regular prices. Third, the stores are located within a 1 km radius.

We study four price series at each store— the actual transaction prices, the actual regular prices, filtered prices, and reference prices. We find substantial variability in the extent of price rigidity across the three store formats and the four price series.

Irrespective of the measure of the regular price we use, we find that the Hi-Lo store has the lowest frequency of regular price changes. If we use the stores' notion of regular prices, then the likelihood that the Hi-Lo store changes a regular price on a given week is 4.06%, compared to 13.38% at the EDLP store, and 5.34% at the HYB store. If we use the filtered series, then we find that the likelihood that the Hi-Lo store changes a price is 3.55%, compared to 4.25% at the EDLP store and 4.50% weeks at the HYB store. If we focus on reference prices, we find that the likelihood that the Hi-Lo store changes a price is 2.23%, compared to 2.70% at the EDLP store, and 3.95% at the HYB store. Thus, in our data, the most flexible regular prices are 1.3–3.3 times as flexible as the least flexible regular prices, depending on the store format and the definition of the regular price.

Several conclusions follow. First, when using data that comes from stores that follow different pricing formats, the choice of the definition of regular prices can have significant effects on the measured variation in regular price rigidity: our results suggest that it will be the lowest when regular prices are generated using a sales filter, and the highest when we adopt the stores' own notion of regular prices. However, the label "sales filter" might be misleading in this context, because the filter removes temporary price cuts, which, especially at EDLP stores, are not necessarily promoted as sales.

Second, our results suggest that the behavior of filtered and reference price series is consistent with the predictions of sticky price models with no inflation and symmetric menu costs as discussed by Álvarez et al. (2016). We find, for example, that if we use either the filter or the reference prices, then the sufficient statistics score (Álvarez et al., 2016) of the three stores is similar.

Third, Stores with different pricing formats treat temporary price cuts differently and, therefore, they also have different frequencies of regular price changes, regardless of how we define regular prices. As Carvalho (2006) notes, firm-level heterogeneity in price rigidity can exacerbate market-level price rigidity. In particular, if there are strategic complementarities, then firms with rigid prices can slow down the response of the economy to a shock, increasing the real effects. We find evidence for some synchronization of price changes across stores, suggesting that strategic complementarities might indeed play a role in the market-level response to shocks.

Fourth, while it is hard to gauge the macroeconomic implications of our findings because we do not have information on quantities sold, we can expect that the geographic distribution of store pricing formats will likely affect the economy's response to monetary shocks if stores tend to respond to macroeconomic shocks through regular price changes (Coibion et al., 2015; Kehoe and Midrigan, 2015; Anderson et al., 2017; DellaVigna and Gentzkow, 2019). In Table J2 in the Online Supplementary Appendix J, we present the geographic distribution of retail store formats across the US. Based on the figures in the table, we anticipate that the effect of sales on the aggregate price level will be higher in the Great Lakes region, which has a high share of Hi-Lo stores, than in the South-East region, which has a high share of EDLP stores. This can be important for assessing the variability in the local effects of monetary policy by regions and/or states as in Angeloni and Ehrmann (2007) or Francis et al. (2012), for example. Thus, our findings may have implications for the aggregate empirics and macroeconomic models of price-setting.[26]

Fifth, some studies suggest that temporary price cuts might have large effects on the aggregate effective price level (i.e., the price level that accounts for quantities purchased).[27] These studies are partly based on the premise that price cuts are promoted by stores, increasing the quantities purchased. However, because EDLP stores tend not to promote temporary price cuts, such price cuts are likely to have only a modest effect on sales volumes.[28] Therefore, temporary price cuts are likely to have only a small effect on the aggregate price level in markets that are dominated by EDLP stores than by Hi-Lo stores. This underscores the value of having the actual store-level data on regular and sale prices, rather than using mechanical algorithms, such as sales filters, to distinguish between regular and sale prices.

---

[26] At the technical level, the geographical variability in the pricing format suggests that removing sale prices from the analysis might be more appropriate in the context of the price data from some regions than others.

[27] See Klenow and Willis (2007), Hendel and Nevo (2013), Fox and Syed (2016), Glandon (2018), Chevalier and Kashyap (2019), Kryvtsov and Vincent (2021), and Wu (2021).

[28] Note an important difference between temporary price cuts at Hi-Lo vs EDLP stores, as seen from the point of view of consumers. A shopper at a Hi-Lo store knows that the price cut is temporary, and thus s/he has an incentive to buy more than the usual quantity. At an EDLP store, however, even if the shopper knows that the price is low, s/he does not perceive it as temporary, and therefore s/he has no incentive to buy more than the usual quantity.

Sixth, to check the robustness of the results, we use data from a supermarket in Israel. We find that the frequency of price changes in the Israeli supermarket, which belongs to a chain that promotes itself as HYB, is similar to the frequency at the Canadian HYB store. Thus, although the distribution of pricing formats may differ across countries, heterogeneity due to pricing formats likely affects price rigidities also in countries outside of North America.

Future work could consider larger data sets that contain information on stores' pricing formats and quantities sold to explore the robustness of the results we report. Considering our findings, we believe it will be beneficial, when studying the behavior of temporary price cuts and their implications, to focus on the prices from the point of view of both buyers and sellers. Depending on store formats and the corresponding notions of regular and sale prices, store managers and shoppers do not necessarily interpret price cuts as "sales." Therefore, considering how they interpret price cuts, is important for accurately assessing the effects of micro-level price changes on the aggregate price level.

Another avenue for future research that is worth exploring should be a study of the aggregate implications of our findings. We have argued that the heterogeneity we document in pricing policies can affect the degree of nominal price rigidity. However, just because prices change more/less frequently or by smaller/larger amounts, does not necessarily imply that they are more/less responsive to aggregate shocks. Exploring this more methodically requires developing a theoretical framework for assessing how stores with different pricing formats may respond to aggregate shocks. These and similar questions could also be explored empirically with larger datasets that contain information about prices and quantities sold as well as about the stores' pricing format. We hope that the current manuscript offers a starting point for such explorations.
33

# REFERENCES

ÁLVAREZ, F., LE BIHAN, H. and LIPPI, F. (2016). The real effects of monetary shocks in sticky price models: a sufficient statistic approach. *American Economic Review*, **106** (10), 2817–2851.

ANDERSON, E., JAIMOVICH, N. and SIMESTER, D. (2015). Price stickiness: empirical evidence of the menu cost channel. *Review of Economics and Statistics*, **97** (4), 813–826.

ANDERSON, E., MALIN, B., NAKAMURA, E., SIMESTER, D. and STEINSSON, J. (2017). Informational rigidities and the stickiness of temporary sales. *Journal of Monetary Economics*, **90**, 64–83.

ANGELONI, I. and EHRMANN, M. (2007). Euro area inflation differentials. *B.E. Journal of Macroeconomics*, **7** (1).

BALEY, I. and BLANCO, A. (2021). Aggregate dynamics in lumpy economics. *Econometrica*, **89** (3), 1235–1264.

BARSKY, R., BERGEN, M., DUTTA, S. and LEVY, D. (2003). What can the price gap between branded and private label products tell us about markups? In R. Feenstra and M. Shapiro (eds), *Scanner Data and Price Indexes* (pp. 165–225). Chicago, IL: University of Chicago Press.

BARRO, R. (1972). A theory of monopolistic price adjustment. *Review of Economic Studies*, **39** (1), 17–26.

BELL, D.R. and LATTIN, J.M. (1998). Shopping behavior and consumer preference for store price format: why 'large basket' shoppers prefer EDLP. *Marketing Science*, **17** (1), 66–88.

BERADI, B., GAUTIER, E. and LE BIHAN, H. (2015). More facts about prices: France before and during the great recession. *Journal of Money, Credit and Banking*, **47** (8), 1465–1502.

BILS, M. and KLENOW, P. (2004). Some evidence on the importance of sticky prices. *Journal of Political Economy*, **112**, 947–985.

BOLTON, R. N. and SHANKAR, V. (2003). An empirically derived taxonomy of retailer pricing and promotion strategies. *Journal of Retailing*, **79** (4), 213–224.

CABALLERO, R.J. and ENGEL, E.M.R.A. (2007). Price stickiness in *s*, *S* models: new interpretations of old results. *Journal of Monetary Economics*, **54**, 100–121.

CAMPBELL, J. and EDEN, B. (2014). Rigid Prices: evidence from U.S. scanner data. *International Economic Review*, **55** (2), 423–442.

CARLTON, D. (1986). The rigidity of prices. *American Economic Review*, **76**, 637–658.

CARVALHO, C. (2006). Heterogeneity in price stickiness and the real effects of monetary shocks. *Frontiers of Macroeconomics*, **2** (1).

CAVALLO, A. (2018). Scraped data and sticky prices. *Review of Economics and Statistics*, **100** (1), 105–119.

CECCHETTI, S. (1986). The frequency of price adjustment: a study of the newsstand prices of magazines. *Journal of Econometrics*, **31**, 255–274.

CHAHROUR, R. (2011). Sales and price spikes in retail price data. *Economics Letters*, **110**, 143–146.

CHEN, H.A., LEVY, D., RAY, S. and BERGEN, M. (2008). Asymmetric price adjustment in the small. *Journal of Monetary Economics*, **55**, 728–737.

CHEVALIER, J. and KASHYAP, A. (2019). Best prices: price discrimination and consumer substitution. *American Economic Journal: Economic Policy*, **11** (1), 126–159.

COIBION, O., GORODNICHENKO, Y. and HONG, G.H. (2015). The cyclicality of sales, regular and effective prices: business cycle and policy implications. *American Economic Review*, **105** (3), 993–1029.

DELLAVIGNA, S. and GENTZKOW, M. (2019). Uniform pricing in US retail chains. *Quarterly Journal of Economics*, **134** (4), 2011–2084.

DUTTA, S., BERGEN, M., LEVY, D. and VENABLE, R. (1999). Menu costs, posted prices, and multi-product retailers. *Journal of Money, Credit, and Banking*, **31**, 683–703.

DUTTA, S., LEVY, D. and BERGEN, M. (2002). Price flexibility in channels of distribution. *Journal of Economic Dynamics and Control*, **26**, 1845–1900.

EDEN, B. (2018). Price dispersion and demand uncertainty: evidence from U.S. scanner data. *International Economic Review*, **59** (3), 1035–1075.

EICHENBAUM, M., JAIMOVICH, N. and REBELO, S. (2011). Reference prices, costs and nominal rigidities. *American Economic Review*, **101** (1), 234–262.

ELLICKSON, P. and MISRA, S. (2008). Supermarket pricing strategies. *Marketing Science*, **27** (5), 811–828.

FISHER, T.C.G and KONIECZNY, J.D. (2000). Synchronization of price changes by multiproduct firms: evidence from Canadian newspaper prices. *Economics Letters*, **68**, 271–277.

FOX, K.J and SYED, I.A. (2016). Price discounts and the measurement of inflation. *Journal of Econometrics*, **191**, 398–406.

FRANCIS, N., OWYANG, M. and SEKHPOSYAN, T. (2012). The local effects of monetary policy. *The B.E. Journal of Macroeconomics*, **12** (2), 1–38.

GALATA, G., RANDOLPH, E., BUCKLIN, R. and HANSSENS, D. (1999). On the stability of store format choice. Working Paper, Anderson School, UCLA.

GAURI, D.K., TRIVEDI, M. and GREWAL, D. (2008). Understanding the determinants of retail strategy: an empirical analysis. *Journal of Retailing*, **84** (3), 256–267.

GLANDON, P.J. (2018). Sales and the (mis)measurement of price level fluctuations. *Journal of Macroeconomics*, **58**, 60–77.

GOLOSOV, M. and LUCAS, R.E. (2007). Menu costs and Philips curves. *Journal of Political Economy*, **115** (2), 171–199.

GORDON, R.J. (1981). Output fluctuation and gradual price adjustment. *Journal of Economic*

*Literature*, **19**, 492–530.

GORDON, R.J. (1990). What is New-Keynesian economics? *Journal of Economic Literature*, **28**, 1115–1171.

GORODNICHENKO, Y. and TALAVERA, O. (2017). Price setting in online markets: basic facts, international comparisons, and cross-border integration. *American Economic Review*, **107** (1), 249–282.

GUIMARAES, B. and SHEEDY, K. (2011). Sales and monetary policy. *American Economic Review*, **101**, 844–876.

HE, G., LAFRANCE, J.T., PERLOFF, J.M. and VOLPE, R. (2023). How do every-day-low-price supermarkets adjust their prices? Manuscript.

HENDEL, I. and NEVO, A. (2013). intertemporal price discrimination in storable goods markets. *American Economic Review*, **103** (7), 2722–2751.

HOCH, S.J., DRÈZE, X. and PURK, M.E. (1994). EDLP, Hi-Lo, and margin arithmetic. *Journal of Marketing*, **50**, 16–27.

KASHYAP, A. (1995). Sticky prices: new evidence from retail catalogues. *Quarterly Journal of Economics*, **110**, 245–274.

KEHOE, P. and MIDRIGAN, V. (2015). Prices are sticky after all. *Journal of Monetary Economics*, **75**, 35–53.

KLENOW, P. and WILLIS, J. (2007). Sticky information and sticky prices. *Journal of Monetary Economics*, **54**, 79–99.

KLENOW, P. and MALIN, B. (2011). Microeconomic evidence on price setting. In B. Friedman and M. Woodford (eds.), *Handbook of Monetary Economics, Volume 3A*. New York, NY: North Holland.

KNOTEK, E. II (2019). The roles of price points and menu costs in price rigidity. Working Paper

No. 19-23, Federal Reserve Bank of Cleveland.

KONIECZNY, J. and RUMLER, F. (2006). Regular adjustment: theory and evidence. Working Paper No. 669, European Central Bank.

KONIECZNY, J. and SKRZYPACZ, A. (2005). Inflation and price setting: evidence from a natural experiment. *Journal of Monetary Economics*, **52**, 621–632.

KRYVTSOV, O. and VINCENT, N. (2021). The cyclicality of sales and aggregate price flexibility. *Review of Economic Studies*, **88**, 334–377.

LAL, R. and RAO, R. (1997). Supermarket competition: the case of everyday low pricing. *Marketing Science*, **16**, 60–80.

LACH, S. and TSIDDON, D. (1992). The behavior of prices and inflation: an empirical analysis of disaggregated data. *Journal of Political Economy*, **100**, 349–389.

LACH, S. and TSIDDON, D. (1996). Staggering and synchronization in price setting: evidence from multiproduct firms. *American Economic Review*, **86**, 1175–1196.

LEAHY, J. (2011). A survey of New Keynesian theories of aggregate supply and their relation to industrial organization. *Journal of Money Credit & Banking*, **43**, 87–110.

LEVY, D., BERGEN, M., DUTTA, S. and VENABLE, R. (1997). The magnitude of menu costs: direct evidence from large U.S. supermarket chains. *Quarterly Journal of Economics*, **112**, 791–825.

LEVY, D., DUTTA, S. and BERGEN, M. (2002). Heterogeneity in price rigidity: evidence from primary micro-level data. *Journal of Money, Credit, and Banking*, **34**, 197–220.

LEVY, D., DUTTA, S., BERGEN, M. and VENABLE, R. (1998). Price adjustment at multiproduct retailers. *Managerial and Decision Economics*, **19**, 81–120.

LEVY, D., MÜLLER, G., DUTTA, S. and BERGEN, M. (2010) Holiday price rigidity and the cost of price adjustment. *Economica*, **77** (305), 172–198.

LEVY, D., SNIR, A., GOTLER, A. and CHEN, H.A. (2020). Not all price endings are created equal: price points and asymmetric price rigidity. *Journal of Monetary Economics*, **110** (April), 33–49.

LEVY D., LEE, D., CHEN, H.A., KAUFFMAN, R.J. and BERGEN, M. (2011). Price points and price rigidity. *Review of Economics and Statistics*, **93**, 1417–1431.

MANKIW, N. G. (1985). Small menu costs and large business cycles: a macroeconomic model of monopoly. *Quarterly Journal of Economics*, **100**, 529–539.

MIDRIGAN, V. (2011). Menu costs, multiproduct firms, and aggregate fluctuations. *Econometrica*, **79** (4), 1139–1180.

MONROE, K. (2002). *Pricing: Making Profitable Decisions*. New York, NY: McGraw Hill.

NAGLE, T. and MÜLLER, G. (2017). *The Strategy and Tactics of Pricing: A Guide to Growing More Profitably*. New York, NY: Routledge.

NAKAMURA, E. (2008). Pass-through in retail and wholesale. *American Economic Review, Papers and Proceedings*, **98** (2), 430–437.

NAKAMURA, E. and STEINSSON, J. (2008). Five facts about prices: a reevaluation of menu cost models. *Quarterly Journal of Economics*, **123** (4), 1415–1464.

NAKAMURA, E. and STEINSSON, J. (2013). Price rigidity: microeconomic evidence and macroeconomic implications. *Annual Review of Economics*, **5** (1), 133–163.

NAKAMURA, E., STEINSSON, J., SUN, P. and VILLAR, D. (2018). The elusive costs of inflation: price dispersion during the US great inflation. *Quarterly Journal of Economics*, **133** (4), 1933–1980.

RAY, S., CHEN, H., BERGEN, M. and LEVY, D. (2006). Asymmetric wholesale pricing: theory and evidence. *Marketing Science*, **25**, 131–154.

ROMER, D. (1993). The New Keynesian Synthesis. *Journal of Economic Perspectives*, **7**, 5–22.

RONDAN-CATALUÑA, F.J., SÁNCHEZ-FRANCO, M.J. and VILLAREJO-RAMOS, Á.F. (2005). Are hypermarket prices different from discount store prices? *Journal of Product and Brand Management*, **14** (5), 330–337.

SHANKAR, V. (1996). Relating price sensitivity to retailer promotional variables and pricing policy: an empirical analysis. *Journal of Retailing*, **72** (3), 249–272.

SHESHINSKI, E. and WEISS, Y. (1977). Inflation and costs of price adjustment. *Review of Economic Studies*, **44** (2), 287–303.

SNIR, A., CHEN, H.A. and LEVY, D. (2022). Zero-ending prices, cognitive convenience, and price rigidity. *Journal of Economic Behavior and Organization*, **203**, 519–542.

SNIR, A. and LEVY, D. (2021). If you think 9-ending prices are low, think again. *Journal of the Association for Consumer Research*, **6** (1), 33–47.

STATISTICS CANADA (2016). *Census profile, 2016 census, Notre-Dame-de-Grâce-Westmount*, Quebec, https://goo.gl/HNoqnk, accessed January 27, 2018.

SYED, I. (2015). Sale spotter: an algorithm to identify sale prices in point-of-sale data. UNSW Business School Research Paper No. 2015 ECON 13.

TAYLOR, J. (1999). Staggered price and wage setting in macroeconomics. In: J. Taylor and M. Woodford (eds.), *Handbook of Macroeconomics*. New York, NY: Elsevier.

VOLPE, R.J. and LI, C. (2012). On the frequency, depth, and duration of sales at high-low pricing supermarkets. *Agribusiness*, **28** (2), 222–238.

VOSS, G. B. and SEIDERS, K. (2003). Exploring the effect of retail sector and firm characteristics on retail price promotion strategy. *Journal of Retailing*, **79** (1), 37–52.

WARNER, E. and BARSKY, R. (1995). The timing and magnitude of retail store markdowns: evidence from weekends and holidays. *Quarterly Journal of Economics*, **110** (2), 321–352.

WEISS, Y. (1993). Inflation and price adjustment: a survey of findings from micro-data. In E.

Sheshinski and Y. Weiss (eds.), *Optimal Pricing, Inflation, and the Cost of Price Adjustment*. Cambridge, MA: The MIT Press, pp. 3–17.

WILLIS, J. (2003). Implications of structural changes in the U.S. economy for pricing behavior and inflation dynamics. *Federal Reserve Bank of Kansas City Economic Review*, (1$^{st}$ quarter), 5–26.

WOLMAN, A. (2007). The frequency and costs of individual price adjustment: a survey. *Managerial and Decision Economics*, **28** (6), 531–552.

WU, W. (2022). Sales of durable goods and the real effects of monetary policy. *Review of Economic Dynamics*, **43**, 80–92.

TABLE 1

DISTRIBUTION OF STORE PRICING FORMATS BY STORE TYPE IN THE US: ALL FOOD RETAILERS

| Type of Chain/Store | Percentage of | | |
|---|---|---|---|
| | EDLP Stores (%) | Hi-Lo Stores (%) | HYB Stores (%) |
| **(a) Large Chains/Stores** | | | |
| Chain | 33 | 30 | 37 |
| Vertically Integrated | 35 | 29 | 36 |
| Large Store Size | 32 | 30 | 38 |
| Many Checkouts | 31 | 30 | 39 |
| **(b) Small Chains/Stores** | | | |
| Independent | 22 | 50 | 28 |
| Not Vertically Integrated | 21 | 47 | 32 |
| Small Store Size | 23 | 52 | 26 |
| Few Checkouts | 22 | 52 | 26 |

Notes:

The distinction between large vs small stores/chains is based on four criteria: chain/independent, vertically/not-vertically integrated, large/small store, and many/few checkouts. A "chain" has 11 or more stores, an "independent" has 10 or fewer. Vertically integrated firms operate their own distribution centers. Large vs. small store size and many vs. few checkouts are defined by the upper and lower quartiles of the full store level census. The figures are the averages for 17,388 stores in the US, with annual revenues of at least $2 million.

Source: Ellickson and Misra (2008).

TABLE 2

GENERAL INFORMATION ON THE STORES SAMPLED

|  | **EDLP** (Loblaw's) | **Hi-Lo** (Provigo) | **HYB** (Super-C) |
|---|---|---|---|
| Total Floor Area ($m^2$) | 7,695 | 2,969 | 7,133 |
| Total Parking Area ($m^2$) | 19,204 | 3,021 | 10,700 |
| Annual Sales (in Canadian $) | 30 million | 24 million | 21 million |
| Total Number of Products | 39,000 | 28,000 | 33,000 |
| Total Number of Employees | 235 | 175 | 180 |

TABLE 3

STATISTICAL SIGNIFICANCE OF THE AVERAGE PRICE DIFFERENCES BETWEEN THE STORES, FOR REGULAR AND TRANSACTION PRICES

|  | EDLP (Loblaw's) | Hi-Lo (Provigo) | HYB (Super-C) | EDLP vs. Hi-Lo | EDLP vs. HYB | Hi-Lo vs. HYB |
|---|---|---|---|---|---|---|
| Regular price | C$4.12 | C$4.58 | C$3.98 | $z = 8.66$*** | $z = 3.16$*** | $z = 11.42$*** |
| Transaction price | C$4.11 | C$4.47 | C$3.94 | $z = 6.60$*** | $z = 3.99$*** | $z = 10.18$*** |

Notes:

The table reports the average regular and transaction prices at each store, together with the results of a pairwise comparison of regular and sale prices across the three stores. The EDLP vs. Hi-Lo column reports the results of Wilcoxon rank-sum tests for the equality of the average price at the EDLP store and the average price at the Hi-Lo store. The EDLP vs. HYB column reports the results of Wilcoxon rank-sum tests for the equality of the average price at the EDLP store and the average price at the HYB store. The Hi-Lo vs. HYB column reports the $z$-values of Wilcoxon rank-sum tests for the equality of the average price at the Hi-Lo store and the average price at the EDLP store.

*** $p < 0.01$

TABLE 4

GENERATED REGULAR PRICES: PROMOTED SALE EVENTS VS FILTERED SALE EVENTS

|  | EDLP (Loblaw's) | Hi-Lo (Provigo) | HYB (Super-C) | EDLP vs. Hi-Lo | EDLP vs. HYB | Hi-Lo vs. HYB |
|---|---|---|---|---|---|---|
| Promoted sale events | 12 | 508 | 265 | $\chi^2 = 498.27^{***}$ | $\chi^2 = 215.55^{***}$ | $\chi^2 = 106.84^{***}$ |
| Filtered sale events | 265 | 509 | 280 | $\chi^2 = 83.17^{***}$ | $\chi^2 = 0.15$ | $\chi^2 = 95.00^{***}$ |
| Reference sale events | 261 | 497 | 262 | $\chi^2 = 79.32^{***}$ | $\chi^2 = 0.95$ | $\chi^2 = 102.00^{***}$ |

Notes:

The Promoted Sale Events are the number of promoted sales, i.e., the cases where the sale price displayed on the shelf price tag was lower than the regular price posted next to it. The Filtered Sale Events are the number of sale events identified as sales by Nakamura and Steinsson's (2008) sales filter A. The Reference Sale Events are sale events identified as sales by the Chahrour (2011) algorithm. The EDLP vs. Hi-Lo column reports the Pearson $\chi^2$-test statistics for the differences in the proportion of sale events between the EDLP store and the Hi-Lo store. The EDLP vs. HYB column reports the Pearson $\chi^2$-test statistics for the differences in the proportion of sale events between the EDLP store and the HYB store. The Hi-Lo vs. HYB column reports the Pearson $\chi^2$-test statistics for the differences in the proportion of sale events between the Hi-Lo store and the HYB store.

*** $p < 0.01$

TABLE 5
FREQUENCY OF PRICE CHANGES AND IMPLIED PRICE DURATION

| Product Category | EDLP (Loblaw's) | | | | Hi-Lo (Provigo) | | | | HYB (Super-C) | | | |
|---|---|---|---|---|---|---|---|---|---|---|---|---|
| | Transaction | Regular | Filtered | Reference | Transaction | Regular | Filtered | Reference | Transaction | Regular | Filtered | Reference |
| **A. Average Weekly Frequency of Price Changes** | | | | | | | | | | | | |
| Baby Products & Foods | 5.77% | 5.77% | 1.92% | 0.77% | 5.77% | 0.38% | 0.38% | 0.00% | 6.54% | 5.38% | 2.69% | 3.08% |
| Beverages | 23.08% | 22.69% | 6.92% | 2.88% | 35.58% | 4.42% | 5.77% | 1.92% | 17.83% | 5.59% | 3.50% | 2.97% |
| Breakfast/Cereals | 14.50% | 14.20% | 2.96% | 2.07% | 24.26% | 2.81% | 2.81% | 1.48% | 17.47% | 4.81% | 4.65% | 4.33% |
| Condiments, Sauces & Spread | 19.58% | 19.58% | 4.37% | 2.27% | 29.37% | 4.72% | 4.02% | 2.97% | 14.96% | 5.13% | 4.06% | 4.49% |
| Dairy Products | 14.81% | 14.42% | 5.77% | 4.42% | 15.58% | 5.38% | 4.04% | 3.46% | 15.38% | 6.25% | 4.97% | 3.85% |
| Frozen Food | 15.11% | 14.56% | 3.85% | 2.75% | 21.15% | 3.57% | 3.85% | 1.92% | 14.74% | 6.20% | 5.34% | 4.06% |
| Health & Beauty Aid | 12.31% | 9.81% | 5.58% | 4.23% | 15.00% | 7.50% | 5.77% | 4.42% | 10.96% | 5.19% | 4.42% | 4.42% |
| Households | 9.48% | 9.48% | 3.71% | 2.75% | 20.33% | 2.20% | 2.06% | 1.24% | 12.14% | 5.41% | 5.29% | 3.73% |
| Juices | 15.38% | 14.90% | 3.13% | 1.68% | 34.38% | 4.57% | 4.09% | 2.16% | 15.87% | 6.25% | 5.29% | 5.29% |
| Paper Towel, Tissue & Pet Supplies | 3.85% | 3.85% | 2.75% | 1.65% | 18.68% | 3.02% | 2.47% | 2.47% | 7.69% | 3.21% | 3.42% | 2.56% |
| Soups/Canned Foods | 11.54% | 11.54% | 4.81% | 3.37% | 34.62% | 6.25% | 1.92% | 1.44% | 13.74% | 5.22% | 4.67% | 4.95% |
| **Total** | **13.83%** | **13.38%** | **4.25%** | **2.70%** | **23.29%** | **4.06%** | **3.55%** | **2.23%** | **13.76%** | **5.34%** | **4.50%** | **3.95%** |
| **B. Implied Average Price Duration in Weeks** | | | | | | | | | | | | |
| Baby Products & Foods | 16.83 | 16.83 | 51.50 | 129.50 | 16.83 | 259.50 | 259.50 | N/A | 14.79 | 18.07 | 36.64 | 32.00 |
| Beverages | 3.81 | 3.89 | 13.94 | 34.16 | 2.27 | 22.10 | 16.83 | 51.50 | 5.09 | 17.37 | 28.10 | 33.14 |
| Breakfast/Cereals | 6.38 | 6.53 | 33.30 | 47.78 | 3.60 | 35.08 | 35.08 | 67.10 | 5.21 | 20.30 | 21.01 | 22.61 |
| Condiments, Sauces & Spread | 4.59 | 4.59 | 22.38 | 43.50 | 2.88 | 20.68 | 24.37 | 33.14 | 6.17 | 19.00 | 24.13 | 21.78 |
| Dairy Products | 6.24 | 6.42 | 16.83 | 22.10 | 5.91 | 18.07 | 24.26 | 28.39 | 5.99 | 15.49 | 19.62 | 25.50 |
| Frozen Food | 6.10 | 6.35 | 25.50 | 35.90 | 4.21 | 27.50 | 25.50 | 51.50 | 6.27 | 15.63 | 18.22 | 24.13 |
| Health & Beauty Aid | 7.61 | 9.69 | 17.43 | 23.13 | 6.15 | 12.83 | 16.83 | 22.10 | 8.61 | 18.75 | 22.10 | 22.10 |
| Households | 10.04 | 10.04 | 26.46 | 35.90 | 4.40 | 45.00 | 48.03 | 80.39 | 7.73 | 17.98 | 18.40 | 26.34 |
| Juices | 5.99 | 6.20 | 31.50 | 58.93 | 2.37 | 21.39 | 23.97 | 45.72 | 5.79 | 15.49 | 18.40 | 18.40 |
| Paper Towel, Tissue & Pet Supplies | 25.50 | 25.50 | 35.90 | 60.17 | 4.84 | 32.59 | 39.94 | 39.94 | 12.49 | 30.70 | 28.75 | 38.50 |
| Soups/Canned Foods | 8.16 | 8.16 | 20.30 | 29.21 | 2.35 | 15.49 | 51.50 | 68.83 | 6.77 | 18.65 | 20.91 | 19.72 |
| **Total** | **6.72** | **6.96** | **23.00** | **36.53** | **3.77** | **24.13** | **27.63** | **44.26** | **6.75** | **18.22** | **21.69** | **24.79** |

Notes:

In panel A of the table, we report the average weekly frequency of price changes $\underline{f}$ (in %), for each one of the 11 product categories included in our data, for the three stores. For each category, we computed the $\underline{f}$ as the ratio of the total number of price changes per week in the category, to the number of products in the category (Levy et al., 1997, Table 1, p. 797, Gorodnichenko and Talavera 2017). The average weekly frequency of a price change at each store is calculated for the transaction price, the regular price (as classified and presented by the store), the filtered price (the prices after removing temporary price reductions as identified by Nakamura and Steinsson's (2008) sales filter A), and the reference prices. We use Chahrour's (2008) algorithm with a 13-week rolling window to derive the reference prices. The "total" row gives the average weekly frequency computed over all goods, in each store. In panel B of the table, we report the implied average duration of the prices in weeks. The average duration is calculated as $-\left[ln\left(1-\underline{f}\right)\right]^{-1}$.

TABLE 6

STATISTICAL SIGNIFICANCE OF THE PAIRWISE DIFFERENCES IN THE MEAN PRICE CHANGES FREQUENCY

|  | EDLP (Loblaw's) | HYB (Super-C) |
|---|---|---|
| **Hi-Lo** (Provigo) | Transaction: (23.29%, 13.83%) $\chi^2 = 152.39$*** <br> Regular: (4.06%, 13.38%) $\chi^2 = 281.09$*** <br> Filtered: (3.55%, 4.25%) $\chi^2 = 3.35$* <br> Reference: (2.23%, 2.70%) $\chi^2 = 2.33$ | Transaction: (23.29%, 13.76%) $\chi^2 = 162.89$*** <br> Regular: (4.06%, 5.34%) $\chi^2 = 9.80$*** <br> Filtered: (3.55%, 4.50%) $\chi^2 = 6.24$** <br> Reference: (2.23%, 3.95%) $\chi^2 = 26.17$*** |
| **EDLP** (Loblaw's) |  | Transaction: (13.83%, 13.76%) $\chi^2 = 0.01$ <br> Regular: (13.38%, 5.22%) $\chi^2 = 208.19$*** <br> Filtered: (4.25%, 4.58%) $\chi^2 = 0.40$ <br> Reference: (2.70%, 3.95%) $\chi^2 = 13.01$*** |

Notes:

The figures in the parentheses, in the format "(row, column)," are the average weekly price change frequencies at the corresponding pairs of stores.

* $p < 0.10$, *** $p < 0.01$

TABLE 7

THE HAZARD OF A PRICE CHANGE

|  | (1) Transaction Prices | (2) Regular Prices | (3) Filtered Prices | (4) Reference Prices |
|---|---|---|---|---|
| EDLP Store | 0.66*** | 3.62*** | 1.26 | 1.20 |
|  | (0.063) | (0.410) | (0.159) | (0.146) |
| HYB Store | 0.64*** | 1.41*** | 1.28*** | 1.70*** |
|  | (0.052) | (0.144) | (0.124) | (0.162) |
| Price Level | 0.99 | 1.02* | 1.02 | 1.02 |
|  | (0.011) | (0.009) | (0.014) | (0.136) |
| Private Label | 0.70*** | 0.85 | 0.84 | 1.13 |
|  | (0.082) | (0.102) | (0.097) | (0.140) |
| January Dummy | 1.29** | 1.38** | 2.78*** | 2.61*** |
|  | (0.129) | (0.176) | (0.366) | (0.306) |
| Christmas Dummy | 2.20*** | 1.53 | 6.85*** | 2.45*** |
|  | (0.468) | (0.499) | (3.821) | (0.802) |
| $\chi^2$ | 93.69*** | 259.7*** | 86.3*** | 117.7*** |
| N | 2,951 | 1,479 | 945 | 782 |

Notes:

The results of estimating hazard function regressions of the hazard of a price change. The hazard functions allow the hazard for different categories to be non-proportional. Column (1) gives the results for transaction price changes. Column (2) gives the results for regular price changes. Column (3) gives the results for filtered price changes, using Nakamura and Steinsson's (2008) sales filter A, to remove temporary price reductions. Column (4) gives the results for reference price changes, using Chahrour's (2011) algorithm to identify the reference prices. The numbers in the table show the hazard ratios. EDLP Store is a dummy for goods offered at the EDLP store. HYB Store is a dummy for goods offered at the HYB store (base group: Hi-Lo store). Price Level is the average transaction prices over the 52-week sample period. Private Label is a dummy for private label goods. January Dummy is a dummy for price changes that occur in January. Christmas Dummy is a dummy for price changes that occur on the week of December 25. The regressions also include fixed effects for the product location in the store, for the aisle (back/front/middle) and for the shelf position (bottom/top/middle). Robust standard errors, clustered at the good-store level, are reported in parentheses.

* $p < 10\%$. ** $p < 5\%$. *** $p < 1\%$.

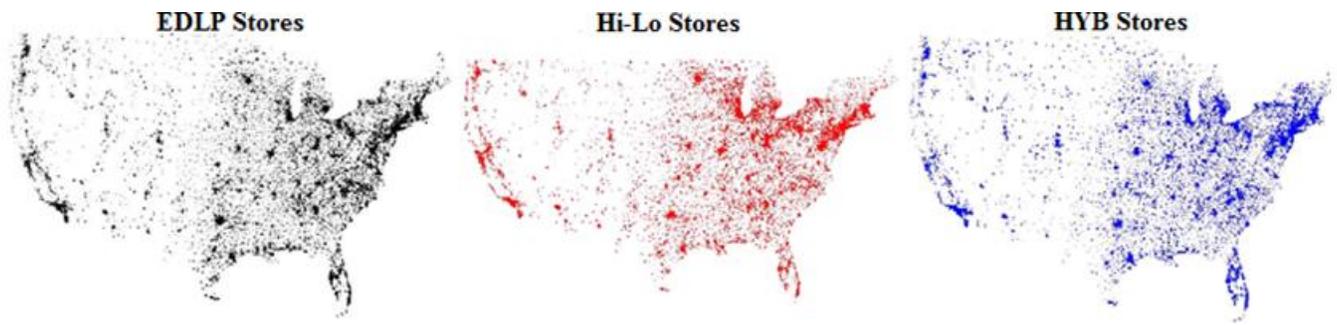

Figure 1. Distribution of Food Stores by Pricing Format across the US (Source: Ellickson and Misra, 2008)

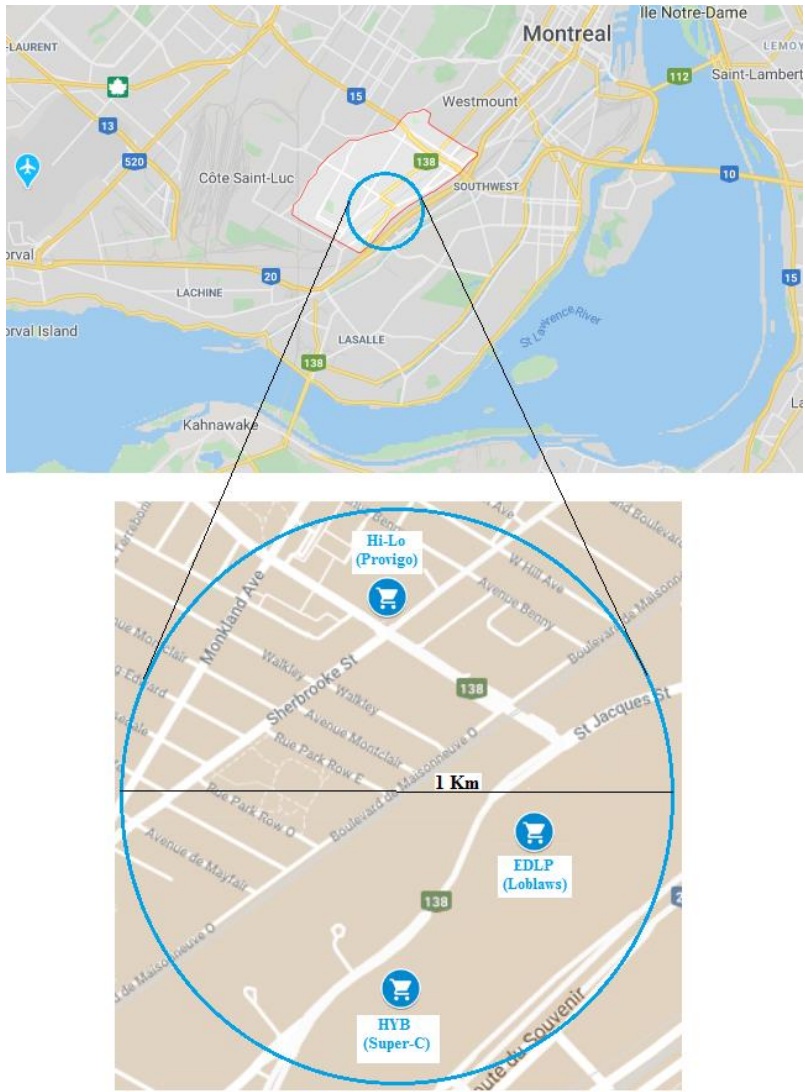

Figure 2. Notre-Dame-de-Grâce District in Montreal, Canada, (the Red Polygon, Top Map), and the Location of the Three Stores in the District (Magnified Blue Circle, Lower Map)

Notes: The exact addresses of the stores are: **Provigo (Hi-Lo)**: 6485 Sherbrooke Street, W., Montreal; **Loblaw's (EDLP)**: 6600 St Jacques Street, Montreal; and **Super-C (HYB)**: 6900 St Jacques Street, Montreal

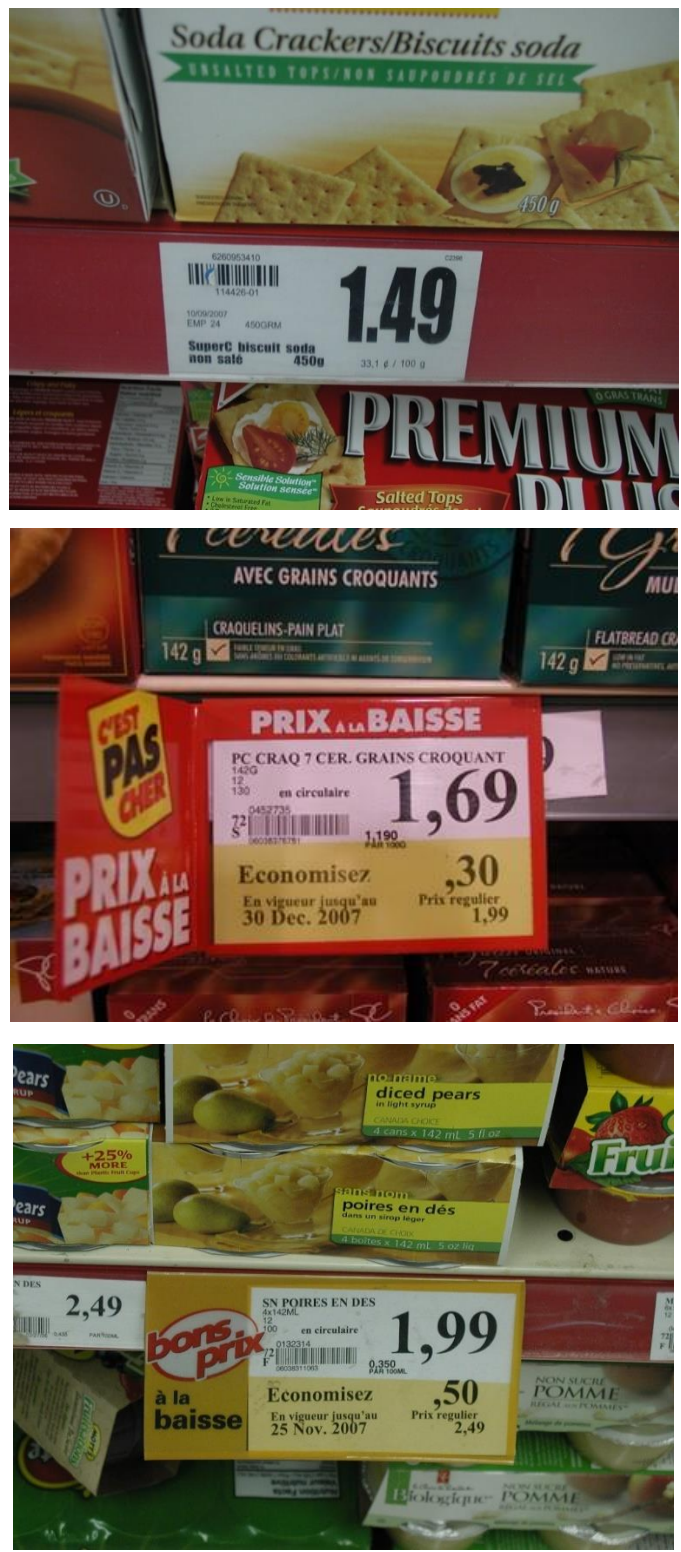

Figure 3. Price Tag Examples with Transaction and Regular Prices

Notes: The top image shows the transaction price (C$1.49), which is also the regular price, of Biscuit Soda at Super-C (HYB). The middle image shows the transaction price (C$1.69) and the regular price (C$1.99) of Grains Croquant at Loblaw's (EDLP). The bottom image shows the transaction price (C$1.99) and the regular price (C$2.49) of Poires en Dés at Provigo (Hi-Low).

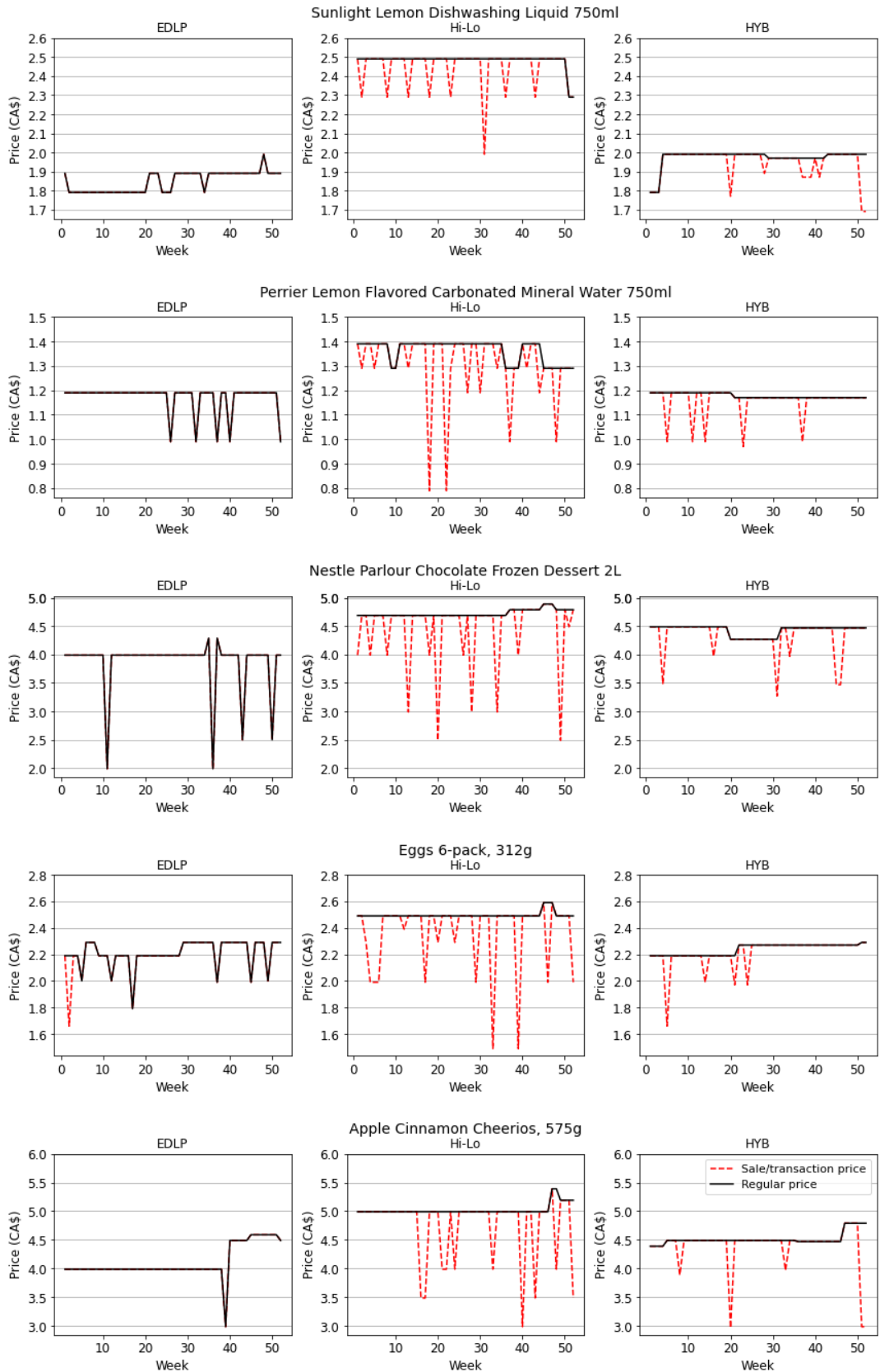

Figure 4. Examples of Weekly Regular Prices (Black Solid Line) and Weekly Transaction Prices (Red Dotted Line) for Five National Brand Goods at the Three Stores (EDLP, Hi-Lo, and HYB), July 30, 2003–July 23, 2004

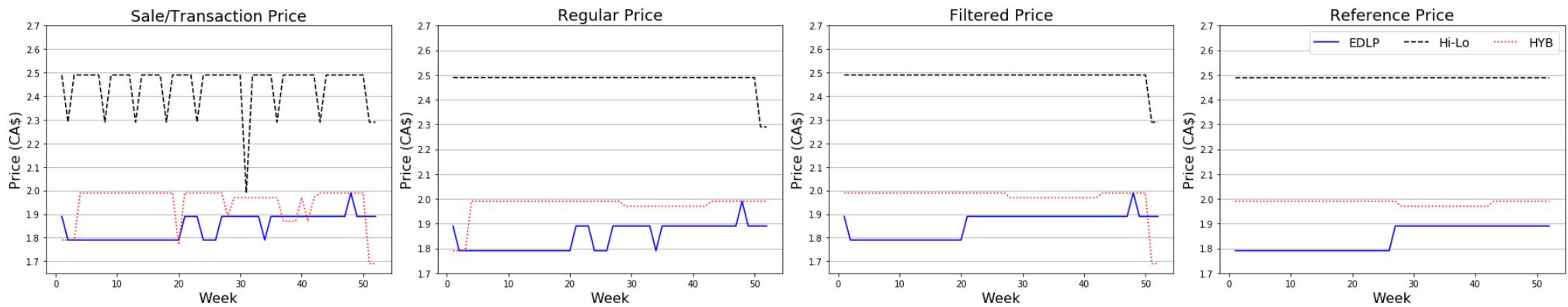

Figure 5. Four Weekly Price Series (Transaction Price, Regular Price, Filtered Price, and Reference Price) of Sunlight Lemon Dishwashing Liquid 750ml. at the Three Stores: EDLP (blue solid line), Hi-Lo (black dashed line), and HYB (red dotted line), July 30, 2003–July 23, 2004

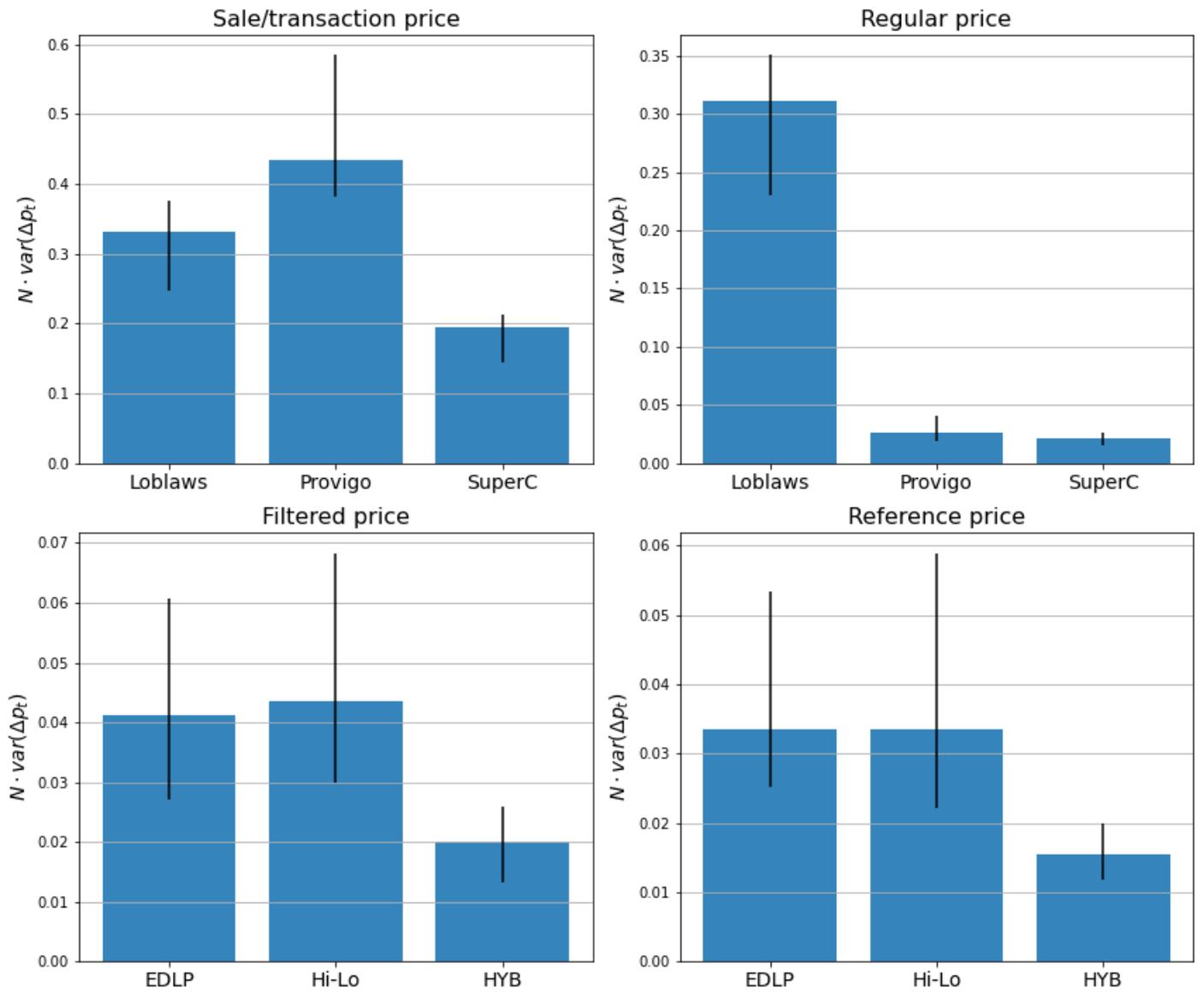

Figure 6. The product of the average number of price changes per year and the variance of the size of price changes. The vertical lines depict 95% confidence intervals. The *y*-axis scales vary across figures.

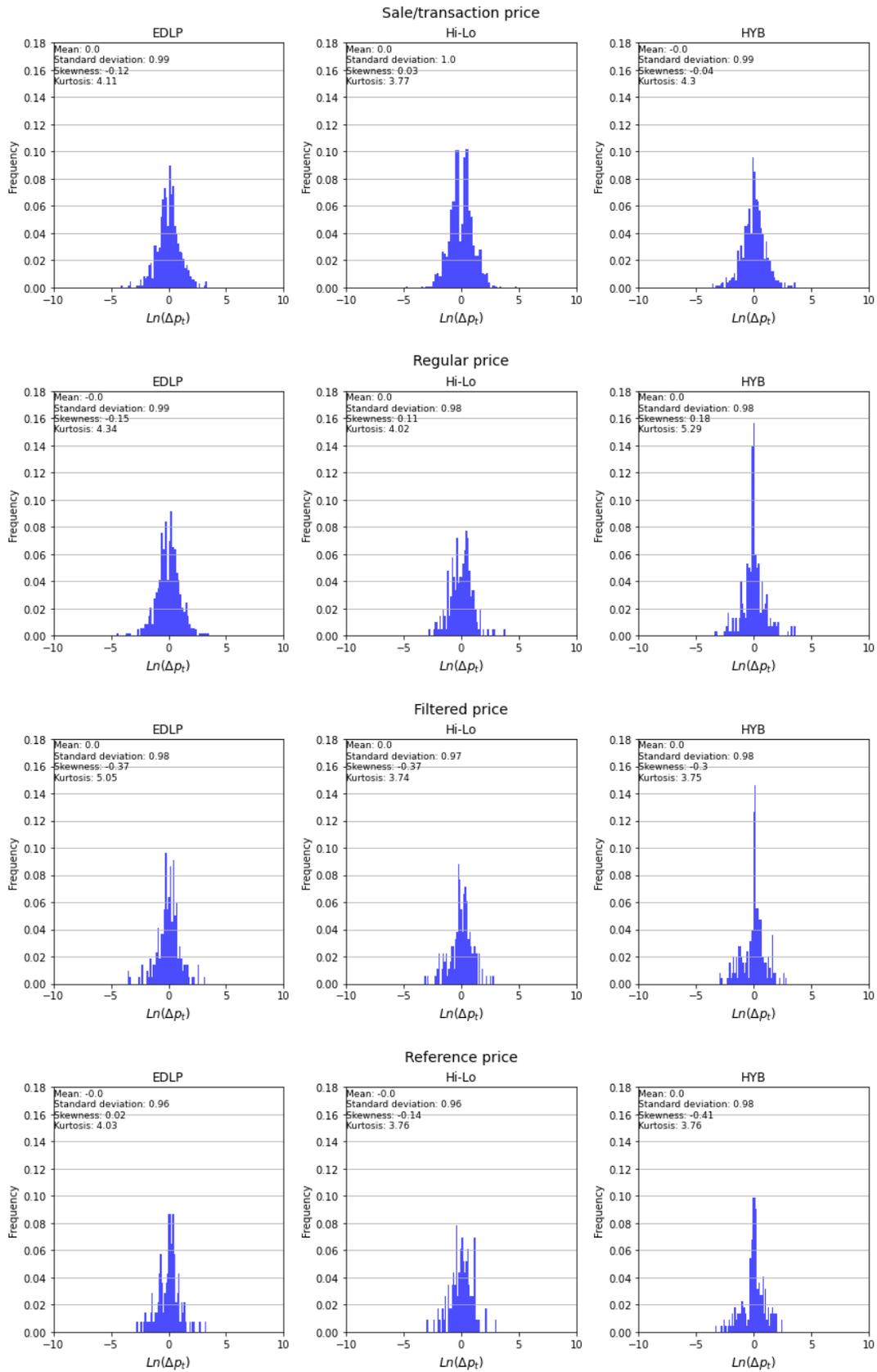

Figure 7. Histograms of the standardized log price differences, conditional on a price change. We standardize the log price differences by product category and store.

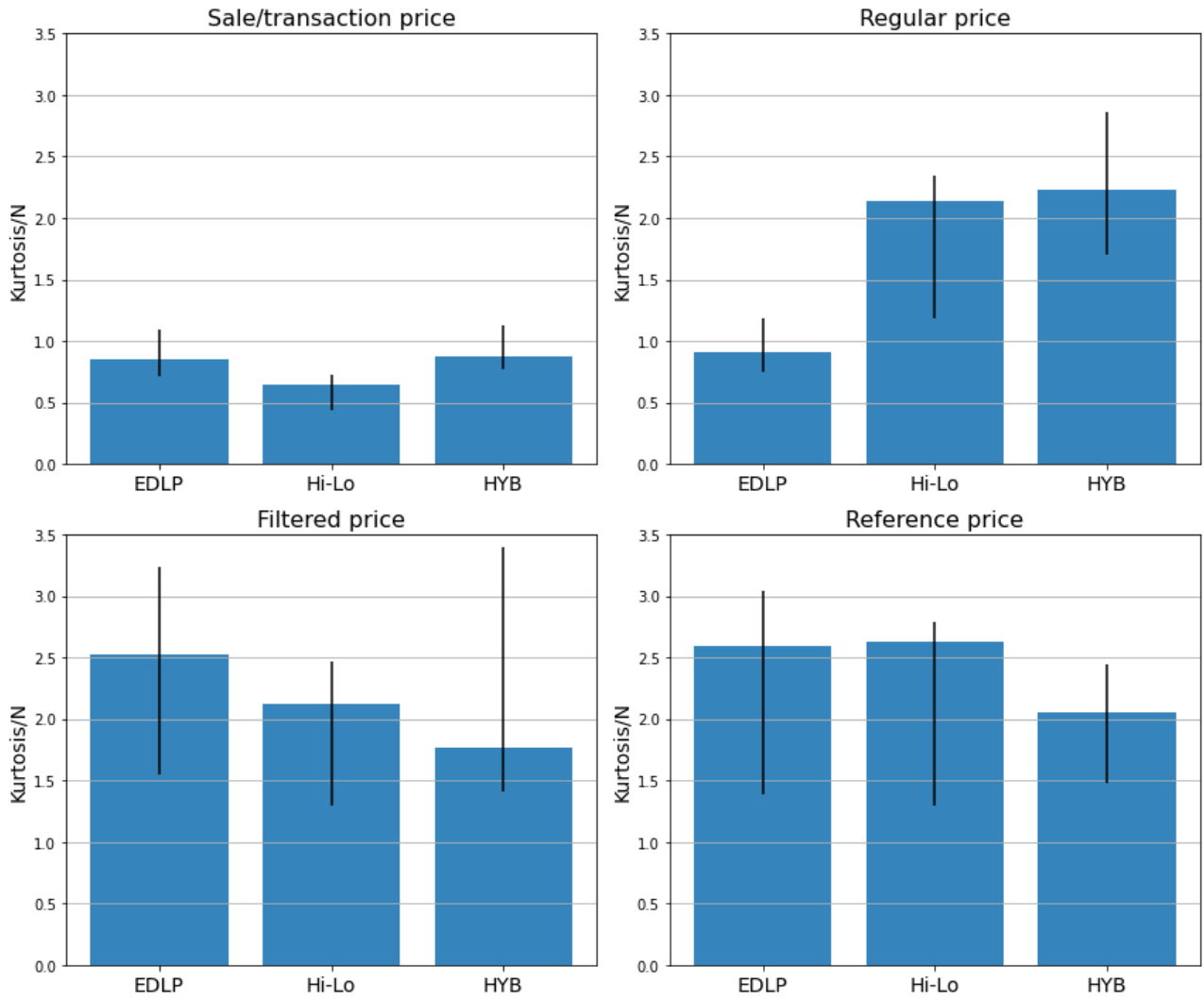

Figure 8. The ratio of the Kurtosis of the size of price changes to the average number of price changes per year. The vertical lines depict 95% confidence intervals.

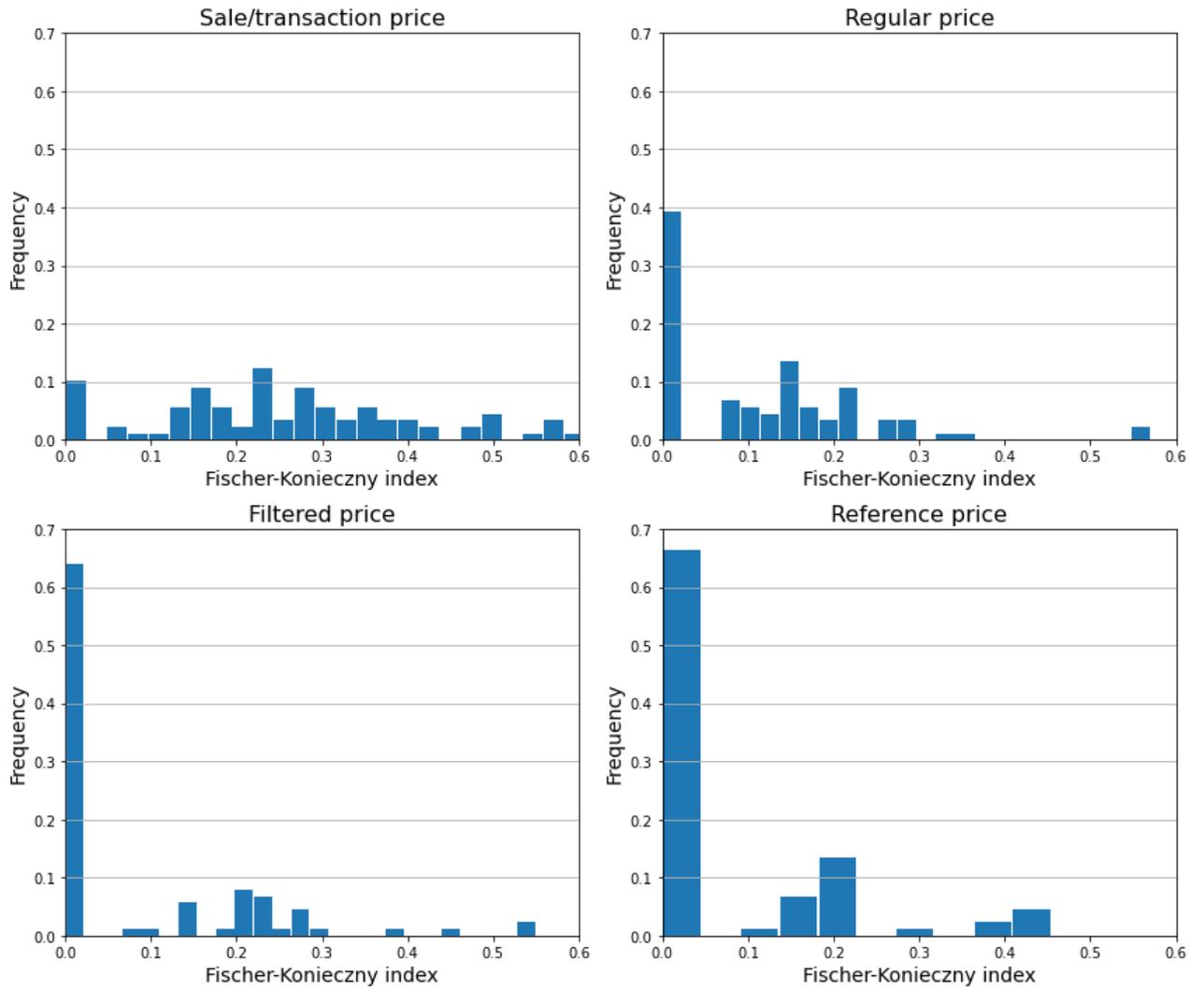

Figure 9. Fischer-Konieczny's (2000) index of synchronization for the 89 national brand products in the sample.

# Supplementary Online Web Appendix

# Retail Pricing Format and Rigidity of Regular Prices

Sourav Ray
Department of Marketing and Consumer Science, G.S. Lang School of Business and Economics
University of Guelph
Guelph, ON N1G-1M8, Canada
s_ray@uoguelph.ca

Avichai Snir
Department of Economics, Bar-Ilan University
Ramat-Gan 5290002, Israel
Avichai.Snir@gmail.com

Daniel Levy
Department of Economics, Bar-Ilan University
Ramat-Gan 5290002, Israel,
Department of Economics, Emory University
Atlanta, GA 30322, USA,
International Centre for Economic Analysis,
ISET at Tbilisi State University
Tbilisi 0108, Georgia, and
Rimini Center for Economic Analysis
Daniel.Levy@biu.ac.il

June 29, 2023

**TABLE OF CONTENTS**

# APPENDIX A. ROBUSTNESS CHECK: CATEGORY-LEVEL PRICES AT THE THREE STORES

In the paper, we show that when looking at the aggregate, store-level data, the High-Low (*Hi-Lo*) store has the highest regular and transaction average prices. We also show that the price level at the Hybrid (*HYB*) store is somewhat lower than at the Every Day Low Price (*EDLP*) store. Below, we show that the same pattern holds when we focus on prices at the category level as well.

Table A1 reports the average regular and transaction prices in each of the three stores. Panel A reports the average regular prices and panel B reports the average transaction prices.

The results are similar for regular and transaction prices, and therefore we discuss only the regular prices. Comparing the EDLP store with the Hi-Lo store, we find that in all the categories, the average prices at the EDLP store are lower than at the Hi-Lo store. In 9 of the 11 categories, the differences are statistically significant. In one additional category, the differences are marginally significant. Thus, the prices at the EDLP store are lower than at the Hi-Lo store not only at the aggregate level. They are lower also when we consider individual categories.

Comparing the EDLP store with the HYB store, we find that in 5 categories, the average prices at the EDLP store are lower than at the HYB store. In 2 categories, the differences are statistically significant. In 6 categories, the average prices at the HYB store are lower than at the EDLP store. In 5 of these categories, the differences are statistically significant. Thus, it seems that in some categories, prices at the EDLP store are below those at the HYB store, in some categories the prices in the two stores are quite similar, and in some categories, prices at the HYB store are lower than at the EDLP store. However, there are more categories in which prices are lower at the HYB store than categories in which the prices are lower at the EDLP store. Overall, therefore, the average price at the HYB store is below the average price at the EDLP store.

Comparing the Hi-Lo store with the HYB store, we find that in all the categories, the average prices at the HYB store are lower than at the Hi-Lo store. In 9 of the 11 categories, the differences are statistically significant. Thus, the prices at the HYB store are lower than at the Hi-Lo store not only at the aggregate level. They are lower also at the level of individual categories.

Table A1. Category Level Summary Statistics on Average Prices

**A. Regular Prices**

| Product Category | EDLP (Loblaw's) | Hi-Lo (Provigo) | HYB (Super-C) | EDLP vs Hi-Lo *Wilcoxon* | EDLP vs HYB *Wilcoxon* | Hi-Lo vs HYB *Wilcoxon* |
|---|---|---|---|---|---|---|
| Baby Products & Foods | 1.96 (1.129) | 2.30 (1.099) | 2.05 (1.103) | 5.46*** | 2.90*** | 5.40*** |
| Beverages | 6.54 (8.126) | 7.00 (8.565) | 5.95 (7.915) | 4.16*** | 5.23*** | 8.31*** |
| Breakfast/Cereals | 3.94 (1.060) | 4.37 (0.981) | 4.10 (1.001) | 13.56*** | 0.84 | 9.79*** |
| Condiments, Sauces & Spread | 2.53 (0.908) | 2.91 (1.160) | 2.63 (1.0125) | 4.68*** | 0.66 | 3.42*** |
| Dairy Products | 3.79 (1.657) | 3.96 (1.743) | 3.84 (1.562) | 1.17 | 0.08 | 0.78 |
| Frozen Food | 4.47 (2.279) | 5.11 (2.621) | 4.44 (2.298) | 5.12*** | 0.72 | 5.01*** |
| Health & Beauty Aid | 3.28 (1.268) | 3.61 (1.182) | 3.50 (1.220) | 4.05*** | 3.30*** | 1.27 |
| Household | 5.40 (1.995) | 6.32 (2.246) | 4.86 (2.015) | 8.96*** | 4.81*** | 14.00*** |
| Juices | 2.93 (1.179) | 3.03 (1.140) | 2.60 (1.156) | 1.84* | 4.58*** | 6.09*** |
| Paper Towel, Tissue & Pet Supplies | 6.58 (4.695) | 7.32 (5.141) | 5.86 (3.952) | 4.84*** | 3.71*** | 6.66*** |
| Soup / Canned Foods | 1.61 (0.665) | 1.78 (0.685) | 1.33 (0.609) | 4.89*** | 8.89*** | 12.32*** |
| Overall | 4.12 (3.500) | 4.58 (3.764) | 3.98 (3.356) | 8.66*** | 3.16*** | 11.42*** |

**B. Transaction Prices**

| Product Category | EDLP (Loblaw's) | Hi-Lo (Provigo) | HYB (Super-C) | EDLP vs Hi-Lo *Wilcoxon* | EDLP vs HYB *Wilcoxon* | Hi-Lo vs HYB *Wilcoxon* |
|---|---|---|---|---|---|---|
| Baby Products & Foods | 1.96 (1.129) | 2.29 (1.108) | 2.04 (1.100) | 5.22*** | 2.84*** | 5.02*** |
| Beverages | 6.54 (8.127) | 6.77 (8.480) | 5.90 (7.875) | 2.35*** | 5.76*** | 6.89*** |
| Breakfast/Cereals | 3.94 (1.060) | 4.22 (1.019) | 4.04 (1.015) | 9.57*** | 0.75 | 7.48*** |
| Condiments, Sauces & Spread | 2.53 (0.908) | 2.80 (1.127) | 2.61 (1.013) | 3.10*** | 0.31 | 2.26** |
| Dairy Products | 3.78 (1.650) | 3.91 (1.709) | 3.79 (1.531) | 0.65 | 0.39 | 0.66 |
| Frozen Food | 4.47 (2.281) | 4.96 (2.588) | 4.39 (2.313) | 4.11*** | 1.02 | 4.23*** |
| Health & Beauty Aid | 3.21 (1.238) | 3.59 (1.191) | 3.47 (1.219) | 4.64*** | 3.37*** | 1.50 |
| Household | 5.40 (1.995) | 6.20 (2.247) | 4.80 (1.990) | 7.90*** | 5.73*** | 13.51*** |
| Juices | 2.93 (1.179) | 2.93 (1.153) | 2.56 (1.138) | 0.10 | 5.18*** | 5.02*** |
| Paper Towel, Tissue & Pet Supplies | 6.58 (4.695) | 7.27 (5.134) | 5.85 (3.947) | 4.45*** | 3.84*** | 6.49*** |
| Soup / Canned Foods | 1.61 (0.665) | 1.71 (0.706) | 1.31 (0.608) | 2.88*** | 9.49*** | 10.57*** |
| Overall | 4.11 (3.501) | 4.47 (3.728) | 3.94 (3.340) | 6.60*** | 3.99*** | 10.18*** |

Notes: The table reports the category-level average prices. The prices are in Canadian Dollars (C$). The EDLP column gives the average prices at the EDLP store. The Hi-Lo column gives the average prices at the Hi-Lo store. The HYB column gives the average prices at the HYB store. The "EDLP vs Hi-Lo" column gives the values of Wilcoxon rank sum $z$-test statistics for comparing the EDLP and Hi-Lo store prices. The "EDLP vs HYB" column gives the values of Wilcoxon rank sum $z$-test statistics for comparing the EDLP and HYB store prices. The "Hi-Lo vs HYB" column gives the values of Wilcoxon rank sum $z$-test statistics for comparing the Hi-Lo and HYB stores. * $p < 0.10$, ** $p < 0.05$, *** $p < 0.01$

# APPENDIX B. ROBUSTNESS CHECK: COMPARISON OF THE WEEKLY FREQUENCY OF PRICE CHANGES ACROSS STORES, AT THE CATEGORY LEVEL

In the paper, we compare the weekly frequencies of price changes at the store level. In this appendix, we show that the results remain unchanged if we conduct the comparisons at the category level.[1]

In Table B1, we report Pearson $\chi^2$ test statistics for comparing the frequencies of price changes at the EDLP and Hi-Lo stores. Column 1 reports the results of comparing the frequencies of the transaction price changes, Column 2 reports the results of comparing the frequencies of the regular price changes (as defined and classified by the store), Column 3 reports the results of comparing the frequencies of the filtered price changes, and Column 4 reports the results of comparing the frequencies of the reference price changes.[2]

In each cell, the name of the store indicates the name of the store that has the higher frequency of price changes. In the transaction prices column, we find that in 10 categories, the Hi-Lo store has a higher frequency of price changes than the EDLP store. In 8 of these 10 categories, the differences are statistically significant.

When we consider the regular prices, we find that in all 11 categories, the frequency of price changes is higher at the EDLP store. In 9 of the 11 categories, the differences are statistically significant.

When we consider filtered prices, we find that the frequency of price changes is higher at the EDLP store in 9 of the 11 categories, but only in one category is the difference statistically significant, and in one additional category, it is marginally significant.

---

[1] The weekly frequency of price changes is given by the ratio of the total number of price changes per week in the category, to the number of products in the category (Levy et al., 1997, Table 1, p. 797, Gorodnichenko and Talavera 2017).

[2] We obtain the filtered series by using the Nakamura and Steinsson's (2008) sales filter A to remove temporary price reductions from the series of transaction prices. We apply Chahrour's (2011) sales filter to the series of transaction prices to obtain the reference prices.

When we consider reference prices, we find that the frequency of price changes is higher at the EDLP store than at the Hi-Lo store in 7 of the 11 categories. Only one of the differences is statistically significant.

Thus, when we look at the category level, we find the same pattern as at the overall store level, as discussed in the paper. If we focus on transaction prices, the Hi-Lo store has a higher frequency of price changes. When we focus on regular prices, in all categories the EDLP store has a higher frequency of price changes. When we focus on filtered prices, the EDLP store has the higher frequency of price changes in 8 out of 11 categories, with one of the differences being statistically significant and another one being marginally significant. When we look at the reference prices, only one of the category level differences is statistically significant.

In Table B2, we report the Pearson $\chi^2$-test statistics for comparing the frequencies of price changes at the EDLP and HYB stores. In each cell, we note the name of the store that has the higher frequency of price changes.

In the transaction prices column, we find that in 4 of the 11 categories, the EDLP store has a higher frequency of price changes than the HYB store. One of the differences is significant statistically and one is marginally significant. In 7 categories, the HYB store has a higher frequency of price changes, where in one case the difference is statistically significant, and in another, the difference is marginally significant.

When we consider regular prices, we find that in all categories, the frequency of price changes is higher at the EDLP store. In 9 of the categories, the differences are statistically significant.

In the filtered prices column, the frequency of price changes is higher at the EDLP store in 4 categories. None of these differences is statistically significant. At the HYB store, the frequency of price changes is higher in 7 categories. One of the differences is statistically significant.

When we look at the column of reference prices, we find that the frequency of price changes is higher at the HYB store than at the EDLP store in 10 categories. Three of the differences are statistically significant, and one additional difference is marginally significant.

Thus, our findings at the category level are similar to our findings at the store level. When we look at the transaction and filtered prices, in some categories the EDLP store has a higher frequency of price changes than the HYB store, but the differences are at best marginally significant. When we look at the regular prices, in all categories the EDLP store has the higher frequency of price changes. When we look at the reference prices, in 10 of the 11 categories, the frequency of price changes is higher at the HYB store than at the EDLP store.

In Table B3, we report the $\chi^2$ test statistics for comparing the average prices at the Hi-Lo and HYB stores. In each cell, the name of the store indicates the name of the store that has the higher frequency of price changes.

In the transaction prices column, we find that in 10 of the 11 categories, the Hi-Lo store has a higher frequency of price changes than the HYB store. In 8 categories, the differences are statistically significant, and in one additional category, it is marginally significant.

When we study the regular prices, we find that in 10 of the 11 categories, the frequency of price changes is higher at the HYB store. In two of the categories, the differences are statistically significant, and in two additional categories, the differences are marginally significant.

In the filtered prices column, the frequency of price changes is higher at the Hi-Lo store in 2 categories. One of the differences is statistically significant. The frequency of price changes is higher at the HYB store in 9 categories. In 2 categories, the differences are statistically significant and in 2 additional categories, the differences are marginally significant.

When we look at the column of reference prices, we find that the frequency of price changes is higher at the HYB store than at the Hi-Lo store in 10 categories. In 5 categories, the differences are statistically significant, and in one additional category, the difference is marginally significant.

Thus, when we look at the category level, we find the same pattern as when we look at the store level. When we consider transaction prices, in 10 of 11 categories the Hi-Lo store has a higher frequency of price changes. When we look at the regular prices, in 10 of the 11 categories, the HYB store has a higher frequency of price changes. When we look at the filtered prices, the HYB store has a higher frequency of price changes in 9 categories. When we look at the reference prices, the HYB store has a higher frequency of price changes in 10 categories.

Table B1. Comparing the Frequency of Price Changes at the EDLP and the Hi-Lo Stores

| Product Category | Transaction Prices | Regular Prices | Filtered Prices | Reference Prices |
|---|---|---|---|---|
| Baby Products & Foods | 0.00 | EDLP 12.64*** | EDLP 2.70 | EDLP 2.01 |
| Beverage | Hi-Lo 19.60*** | EDLP 76.07*** | EDLP 0.58 | EDLP 1.02 |
| Breakfast/Cereals | Hi-Lo 20.62*** | EDLP 56.35*** | EDLP 0.10 | EDLP 0.68 |
| Condiments, Sauces & Spread | Hi-Lo 14.83*** | EDLP 59.17*** | EDLP 0.09 | Hi-Lo 0.55 |
| Dairy Products | Hi-Lo 0.12 | EDLP 23.80*** | EDLP 1.61* | EDLP 0.63 |
| Frozen Food | Hi-Lo 4.48** | EDLP 26.66*** | 0.00 | EDLP 0.54 |
| Health & Beauty Aid | Hi-Lo 1.19 | EDLP 1.75 | Hi-Lo 0.07 | Hi-Lo 0.02 |
| Households | Hi-Lo 33.80*** | EDLP 35.10*** | EDLP 4.05** | EDLP 4.26** |
| Juices | Hi-Lo 39.88*** | EDLP 25.29*** | EDLP 0.55 | Hi-Lo 0.25 |
| Paper Towel, Tissue & Pet Supplies | Hi-Lo 40.14*** | EDLP 0.37 | EDLP 0.05 | Hi-Lo 1.51 |
| Soups/Canned Foods | Hi-Lo 31.20*** | EDLP 4.38** | EDLP 2.66 | EDLP 1.64 |
| Total | Hi-Lo 151.26*** | EDLP 284.01*** | EDLP 3.50* | EDLP 1.94 |

Notes: The table gives the $\chi^2$-test statistics for comparing the average frequencies of weekly price changes in the EDLP and Hi-Lo stores. The transaction price column gives the $\chi^2$-test statistics for comparing the average frequency of weekly transaction price changes. The regular price column gives the $\chi^2$-test statistics for comparing the average frequency of weekly regular price changes. The filtered price column gives the $\chi^2$-test statistics for comparing the average frequency of weekly filtered price changes. The reference price column gives the $\chi^2$-test statistics for comparing the average frequency of weekly reference price changes. The name of the store indicates that the average frequency of price changes at that store is higher than the average frequency at the other store. * $p < 10\%$, ** $p < 5\%$, *** $p < 1\%$

Table B2. Comparing the Frequency of Price Changes at the EDLP and the HYB Stores

| Product Category | Transaction Prices | Regular Prices | Filtered Prices | Reference Prices |
|---|---|---|---|---|
| Baby Products & Foods | HYB 0.13 | EDLP 0.04 | EDLP 0.34 | HYB 3.67* |
| Beverage | EDLP 4.63** | EDLP 74.49*** | HYB 6.57** | HYB 0.01 |
| Breakfast/Cereals | HYB 2.14 | EDLP 32.71*** | HYB 2.08 | HYB 5.41** |
| Condiments, Sauces & Spread | EDLP 3.81* | EDLP 47.30*** | EDLP 0.06 | HYB 3.99** |
| Dairy Products | HYB 0.07 | EDLP 21.12*** | EDLP 0.14 | EDLP 0.24 |
| Frozen Food | EDLP 0.02 | EDLP 16.12*** | HYB 1.31 | HYB 1.05 |
| Health & Beauty Aid | EDLP 0.46 | EDLP 8.77*** | EDLP 0.73 | HYB 0.02 |
| Households | HYB 2.83* | EDLP 10.82*** | HYB 1.83 | HYB 1.18 |
| Juices | HYB 0.04 | EDLP 16.47*** | HYB 2.42 | HYB 8.04*** |
| Paper Towel, Tissue & Pet Supplies | HYB 5.36** | EDLP 0.25 | HYB 0.31 | HYB 0.81 |
| Soups/Canned Foods | HYB 0.57 | EDLP 7.60*** | HYB 0.01 | HYB 0.79 |
| Total | EDLP 0.01 | EDLP 216.06*** | HYB 0.44 | HYB 13.01*** |

Notes: The table gives the $\chi^2$-test statistics for comparing the average frequencies of weekly price changes in the EDLP and HYB stores. The transaction price column gives the $\chi^2$-test statistics for comparing the average frequency of weekly transaction price changes. The regular price column gives the $\chi^2$-test statistics for comparing the average frequency of weekly regular price changes. The filtered price column gives the $\chi^2$-test statistics for comparing the average frequency of weekly filtered price changes. The reference price column gives the $\chi^2$-test statistics for comparing the average frequency of weekly reference price changes. The name of the store indicates that the average frequency of price changes at that store is higher than the average frequency at the other store. * $p < 10\%$, ** $p < 5\%$, *** $p < 1\%$

Table B3. Comparing the Frequency of Price Changes at the Hi-Lo and the HYB Stores

| Product Category | Transaction Prices | Regular Prices | Filtered Prices | Reference Prices |
|---|---|---|---|---|
| Baby Products & Foods | HYB 0.13 | HYB 11.60*** | HYB 4.57** | HYB 8.12*** |
| Beverage | Hi-Lo 44.27*** | HYB 0.28 | Hi-Lo 3.22* | HYB 1.24 |
| Breakfast/Cereals | Hi-Lo 9.02*** | HYB 3.57* | HYB 3.08* | HYB 9.52*** |
| Condiments, Sauces & Spread | Hi-Lo 30.30*** | HYB 0.09 | HYB 0.00 | HYB 1.68 |
| Dairy Products | Hi-Lo 0.01 | HYB 0.39 | HYB 0.90 | HYB 0.12 |
| Frozen Food | Hi-Lo 5.81** | HYB 2.94* | HYB 1.31 | HYB 3.09* |
| Health & Beauty Aid | Hi-Lo 3.11* | Hi-Lo 2.77* | Hi-Lo 1.25 | HYB 0.00 |
| Households | Hi-Lo 19.42*** | HYB 9.41*** | HYB 11.12*** | HYB 9.63*** |
| Juices | Hi-Lo 37.89*** | HYB 1.15 | HYB 0.67 | HYB 5.66** |
| Paper Towel, Tissue & Pet Supplies | Hi-Lo 22.61*** | HYB 0.02 | HYB 0.63 | Hi-Lo 0.16 |
| Soups/Canned Foods | Hi-Lo 34.39*** | Hi-Lo 0.08 | HYB 2.82* | HYB 4.59** |
| Total | Hi-Lo 161.69*** | HYB 8.68*** | HYB 6.59** | HYB 24.86*** |

Notes: The table gives the $\chi^2$-test statistics for comparing the average frequencies of weekly price changes in the Hi-Lo and HYB stores. The transaction price column gives the $\chi^2$-test statistics for comparing the average frequency of weekly transaction price changes. The regular price column gives the $\chi^2$-test statistics for comparing the average frequency of weekly regular price changes. The filtered price column gives the $\chi^2$-test statistics for comparing the average frequency of weekly filtered price changes. The reference price column gives the $\chi^2$-test statistics for comparing the average frequency of weekly reference price changes. Positive values indicate that the average frequency of price changes at the HYB store is higher than the average frequency at the Hi-Lo store. * $p < 10\%$, ** $p < 5\%$, *** $p < 1\%$

# APPENDIX C. DISTRIBUTION OF THE PRICE ENDINGS: LAST DIGIT AND LAST TWO DIGITS

In Figure C1, we present the distribution of the last digit of the prices in our data. According to the figure, digit 9 is the dominant price ending, which is in line with the common retail price-setting practice. See Levy et al. (2011), Anderson et al. (2015), and Snir and Levy (2021), and the studies cited therein.

In our data, 9-ending prices comprise more than 90% of the prices at the EDLP and Hi-Lo stores, similar to the price-ending distribution patterns Anderson et al. (2015) find in their data. At the HYB store, we find that prices ending with "7" are also common, which is in line with the practice of discount stores, often reported in trade publications. See, for example, Risley (2020).

In Figure C2, we present the distribution of the last two digits of the prices in our data. According to the figure, 99-ending prices are a dominant price feature in our data, also in line with the findings reported in the literature. See, for example, Levy et al. (2011).

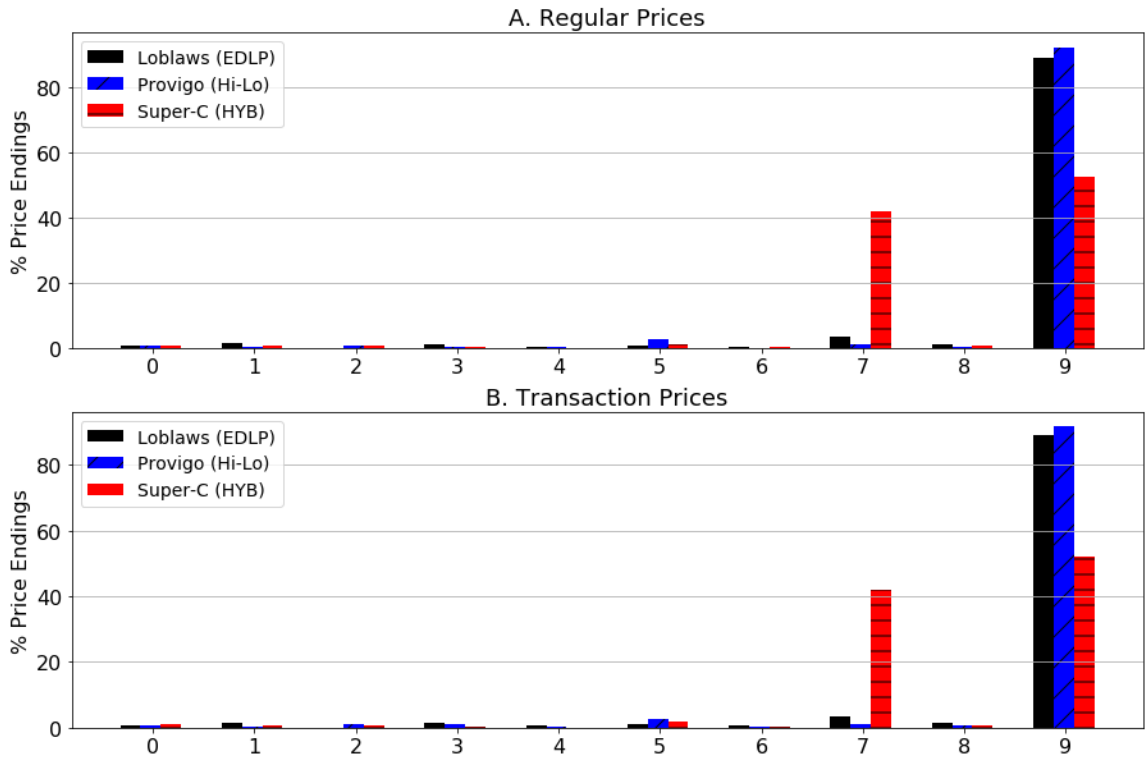

Figure C1. The Distribution of the Right-Most Digits by Store Format

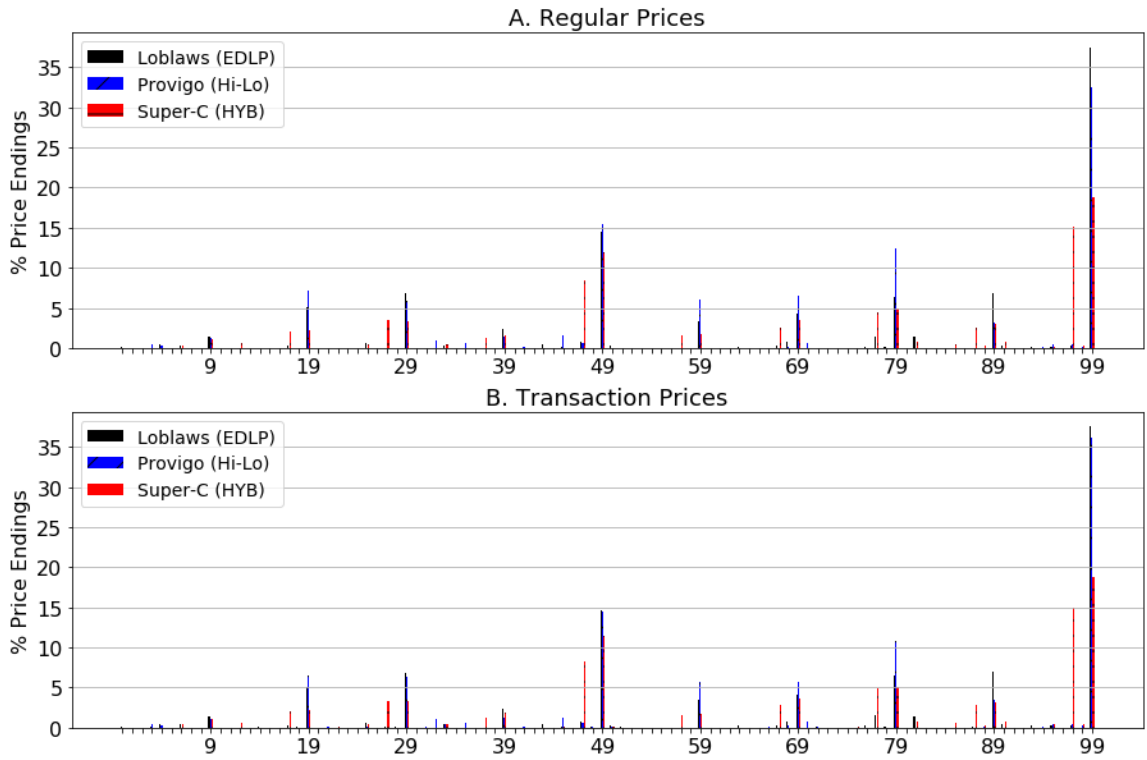

Figure C2. Distribution of the Two Right-Most Digits by Store Format

# APPENDIX D. DETAILED LIST OF PRODUCTS SAMPLED AND THE CORRESPONDING REGULAR AND TRANSACTION PRICES

Table D1. Detailed List of the Products Sampled, by Product Category, by Brand (NB, PL), and by Store Pricing Format, and the Corresponding Average Regular and Transaction Prices

**A. National Brand Products**

| Product Category | Product | EDLP (Loblaw's) | | Hi-Lo (Provigo) | | HYB (Super-C) | |
|---|---|---|---|---|---|---|---|
| | | Regular Price | Transaction price | Regular Price | Transaction price | Regular Price | Transaction price |
| Baby Products and Foods | Dove Baby Soap | 1.94 (0.164) | 1.94 (0.164) | 2.19 (0.000) | 2.16 (0.136) | 1.98 (0.009) | 1.98 (0.009) |
| Baby Products and Foods | Farley's Biscuits 300g | 3.75 (0.159) | 3.75 (0.159) | 3.99 (0.002) | 3.99 (0.002) | 3.68 (0.010) | 3.68 (0.010) |
| Baby Products and Foods | Heinz Blueberry 213ml | 0.81 (0.000) | 0.81 (0.000) | 0.99 (0.000) | 0.97 (0.065) | 0.87 (0.000) | 0.87 (0.000) |
| Baby Products and Foods | Heinz Mixed Cereal 227g | 2.52 (0.205) | 2.52 (0.205) | 2.99 (0.000) | 2.99 (0.000) | 2.83 (0.150) | 2.80 (0.161) |
| Baby Products and Foods | Pablum Soya Cereal 454g | 0.77 (0.000) | 0.77 (0.000) | 1.32 (0.000) | 1.32 (0.000) | 0.87 (0.005) | 0.87 (0.005) |
| Beverage | Bleue Dry 12x341ml | 15.36 (0.433) | 15.36 (0.433) | 15.80 (0.246) | 15.17 (1.256) | 15.34 (0.217) | 15.34 (0.222) |
| Beverage | Coca-Cola Classic | 1.28 (0.061) | 1.28 (0.061) | 1.77 (0.089) | 1.42 (0.228) | 1.29 (0.054) | 1.28 (0.082) |
| Beverage | Molson Dry Beer 12x341ml | 15.32 (0.541) | 15.32 (0.541) | 15.75 (0.252) | 15.03 (1.260) | 15.34 (0.217) | 15.34 (0.222) |
| Beverage | Molson Dry Beer 24x341ml | 24.33 (1.079) | 24.33 (1.079) | 26.33 (0.236) | 26.26 (0.519) | 24.14 (0.933) | 23.89 (1.060) |
| Beverage | Montclair 1L | 0.99 (0.052) | 0.99 (0.052) | 0.99 (0.000) | 0.96 (0.057) | 0.98 (0.010) | 0.92 (0.091) |
| Beverage | Pepsi Diet 12x355ml | 4.00 (0.321) | 4.00 (0.321) | 4.66 (0.105) | 4.19 (0.336) | 4.02 (0.093) | 3.89 (0.333) |
| Beverage | Perrier Lemon 750ml | 1.17 (0.056) | 1.17 (0.060) | 1.36 (0.045) | 1.31 (0.138) | 1.18 (0.010) | 1.16 (0.058) |
| Beverage | Sprite 1L | 1.26 (0.217) | 1.26 (0.217) | 1.54 (0.050) | 1.52 (0.152) | 0.69 (0.000) | 0.69 (0.026) |
| Breakfast/Cereals | Alpha Bits 400g | 3.58 (0.353) | 3.58 (0.353) | 4.01 (0.078) | 3.83 (0.443) | 3.49 (0.000) | 3.45 (0.124) |
| Breakfast/Cereals | Cheerios Apple 575g | 4.11 (0.290) | 4.11 (0.290) | 5.02 (0.092) | 4.76 (0.551) | 4.51 (0.105) | 4.39 (0.380) |
| Breakfast/Cereals | Cheerios Multi-Grain 450g | 4.13 (0.247) | 4.13 (0.247) | 4.77 (0.205) | 4.58 (0.479) | 4.52 (0.104) | 4.41 (0.376) |
| Breakfast/Cereals | Chex Honey Nut 430g | 3.85 (0.413) | 3.85 (0.413) | 4.03 (0.080) | 3.95 (0.405) | 4.01 (0.066) | 3.99 (0.132) |
| Breakfast/Cereals | Corn Flakes 750g | 3.79 (0.329) | 3.79 (0.329) | 4.17 (0.147) | 4.00 (0.485) | 3.61 (0.178) | 3.55 (0.258) |
| Breakfast/Cereals | Life 730g | 3.89 (0.241) | 3.89 (0.241) | 3.99 (0.000) | 3.89 (0.251) | 3.94 (0.045) | 3.75 (0.384) |
| Breakfast/Cereals | Nesquick Cereal 775g | 6.86 (0.633) | 6.86 (0.633) | 7.01 (0.094) | 6.73 (0.819) | 6.97 (0.091) | 6.91 (0.184) |

| Category | Product | | | | | | |
|---|---|---|---|---|---|---|---|
| Breakfast/Cereals | Pops Corn 375g | 3.96 (0.323) | 3.94 (0.361) | 4.99 (0.000) | 4.71 (0.667) | 3.73 (0.287) | 3.68 (0.347) |
| Breakfast/Cereals | Shreddies Cereal 620g | 3.90 (0.190) | 3.90 (0.190) | 4.19 (0.000) | 4.07 (0.343) | 3.75 (0.242) | 3.73 (0.283) |
| Breakfast/Cereals | Special K Red berries 350g | 4.42 (0.175) | 4.42 (0.175) | 4.49 (0.000) | 4.41 (0.182) | 4.49 (0.007) | 4.37 (0.223) |
| Breakfast/Cereals | Sugar Crisp 400g | 3.58 (0.353) | 3.58 (0.353) | 4.18 (0.073) | 3.96 (0.504) | 3.49 (0.000) | 3.45 (0.124) |
| Condiments, Sauces and Spread | Canton Vegetable Delight 990ml | 3.19 (0.261) | 3.19 (0.261) | 3.42 (0.193) | 3.36 (0.234) | 3.34 (0.091) | 3.33 (0.118) |
| Condiments, Sauces and Spread | Classics Dressing 250ml | 1.83 (0.177) | 1.83 (0.177) | 1.97 (0.040) | 1.87 (0.259) | 1.87 (0.059) | 1.81 (0.179) |
| Condiments, Sauces and Spread | French's Yellow Mustard 400ml | 2.04 (0.153) | 2.04 (0.153) | 1.99 (0.000) | 1.95 (0.160) | 2.16 (0.068) | 2.15 (0.087) |
| Condiments, Sauces and Spread | HEINZ Tomato KETCHUP 1L | 3.04 (0.250) | 3.04 (0.250) | 3.68 (0.175) | 3.51 (0.360) | 3.20 (0.236) | 3.17 (0.250) |
| Condiments, Sauces and Spread | Hellmann's Mayonaise 1L | 3.95 (0.243) | 3.95 (0.243) | 4.79 (0.000) | 4.54 (0.589) | 4.02 (0.116) | 3.98 (0.231) |
| Condiments, Sauces and Spread | Miracle Whip Dressing Sauce 1L | 3.90 (0.257) | 3.90 (0.257) | 4.79 (0.000) | 4.47 (0.658) | 3.98 (0.010) | 3.94 (0.237) |
| Condiments, Sauces and Spread | Regular Sugar 2kg | 2.57 (0.064) | 2.57 (0.064) | 2.61 (0.054) | 2.53 (0.215) | 2.58 (0.010) | 2.58 (0.042) |
| Condiments, Sauces and Spread | Sifto Table Salt 1kg | 1.10 (0.051) | 1.10 (0.051) | 1.40 (0.101) | 1.40 (.101) | 1.07 (0.036) | 1.07 (0.038) |
| Condiments, Sauces and Spread | VH Soya Sauce 450ml | 1.47 (0.047) | 1.47 (0.047) | 1.64 (0.067) | 1.59 (0.125) | 1.48 (0.024) | 1.47 (0.040) |
| Dairy Products | Natrel 1% Partly Skimmed Milk 2L | 2.85 (0.071) | 2.85 (0.071) | 2.84 (0.063) | 2.84 (0.063) | 2.84 (0.052) | 2.84 (0.052) |
| Dairy Products | Extra Large Eggs 12un | 2.46 (0.115) | 2.46 (0.115) | 2.52 (0.082) | 2.52 (0.082) | 2.48 (0.028) | 2.47 (0.049) |
| Dairy Products | Lactantia 2% Skimmed Milk 2L | 3.01 (0.030) | 3.01 (0.030) | 3.00 (0.056) | 3.00 (0.056) | 2.94 (0.041) | 2.94 (0.041) |
| Dairy Products | Lactantia Butter 454g | 3.89 (0.243) | 3.89 (0.243) | 4.15 (0.168) | 4.07 (0.287) | 3.95 (0.073) | 3.85 (0.219) |
| Dairy Products | Large Eggs 12un | 1.92 (0.302) | 1.92 (0.302) | 2.40 (0.100) | 2.35 (0.236) | 1.96 (0.181) | 1.89 (0.298) |
| Dairy Products | Omega Eggs 12un | 3.20 (0.063) | 3.20 (0.063) | 3.23 (0.081) | 3.22 (0.100) | 3.15 (0.050) | 3.12 (0.071) |
| Dairy Products | P'tit Quebec Cheese 600g | 6.32 (0.943) | 6.27 (0.977) | 7.04 (0.134) | 6.74 (0.825) | 6.59 (0.143) | 6.37 (0.762) |
| Dairy Products | Quebon 3.25% Bottle Milk 2L | 3.02 (0.033) | 3.02 (0.033) | 3.00 (0.030) | 3.00 (0.030) | 3.02 (0.034) | 3.02 (0.034) |
| Dairy Products | Saputo Cheese 700g | 7.28 (0.279) | 7.28 (0.279) | 7.51 (0.071) | 7.46 (0.174) | 7.12 (0.166) | 6.97 (0.337) |
| Dairy Products | Soya 1.89L | 3.93 (0.080) | 3.93 (0.080) | 3.95 (0.053) | 3.87 (0.191) | 3.98 (0.096) | 3.96 (0.122) |
| Frozen Food | Arctic Garden California Style 2kg | 6.13 (0.151) | 6.13 (0.151) | 6.91 (0.160) | 6.91 (0.160) | 6.26 (0.185) | 6.26 (0.185) |
| Frozen Food | Arctic Garden Thai Style 1.75kg | 6.77 (0.288) | 6.77 (0.288) | 7.23 (0.124) | 7.23 (0.124) | 6.76 (0.295) | 6.76 (0.295) |
| Frozen Food | Delissio Pizza 840g | 7.12 (0.813) | 7.12 (0.813) | 8.81 (0.060) | 8.19 (1.270) | 7.49 (0.008) | 7.41 (0.518) |
| Frozen Food | 6 Eggs 312g | 2.21 (0.106) | 2.20 (0.130) | 2.50 (0.024) | 2.37 (0.251) | 2.24 (0.040) | 2.21 (0.107) |
| Frozen Food | Minis Ice Cream 100ml | 0.67 (0.044) | 0.67 (0.044) | 0.70 (0.008) | 0.68 (0.052) | 0.80 (0.037) | 0.78 (0.070) |

| Category | Product | | | | | |
|---|---|---|---|---|---|---|---|
| Frozen Food | Nestle Parlour 2L | 3.87 (0.482) | 3.87 (0.482) | 4.73 (0.060) | 4.45 (.605) | 4.43 (0.090) | 4.33 (0.294) |
| Frozen Food | Quebon Classic 2L | 4.56 (0.489) | 4.56 (0.489) | 4.92 (0.047) | 4.87 (0.340) | 3.96 (0.172) | 3.68 (0.459) |
| Health & Beauty Aid | Alberto Hairspray 300ml | 2.66 (0.240) | 2.57 (0.310) | 2.79 (0.368) | 2.79 (0.370) | 2.89 (0.008) | 2.89 (0.008) |
| Health & Beauty Aid | Colgate Total 75ml | 1.49 (0.002) | 1.49 (0.002) | 1.77 (0.058) | 1.76 (0.077) | 1.57 (0.034) | 1.54 (0.084) |
| Health & Beauty Aid | Dove All Day 354ml | 4.85 (0.124) | 4.81 (0.285) | 4.92 (0.304) | 4.92 (0.304) | 4.90 (0.186) | 4.85 (0.200) |
| Health & Beauty Aid | Dove Soap 2x100g | 1.98 (0.198) | 1.98 (0.198) | 2.37 (0.144) | 2.29 (0.213) | 1.97 (0.050) | 1.96 (0.060) |
| Health & Beauty Aid | Finesse Extra Body Shampoo 300ml | 1.89 (0.111) | 1.89 (0.111) | 2.90 (0.258) | 2.90 (0.258) | 2.86 (0.247) | 2.79 (0.360) |
| Health & Beauty Aid | Fructis Style 300ml | 3.06 (0.251) | 2.95 (0.429) | 3.16 (0.200) | 3.11 (0.172) | 3.35 (0.250) | 3.31 (0.280) |
| Health & Beauty Aid | Gillette Shaving Cream 60g | 3.20 (0.120) | 3.20 (0.120) | 3.91 (0.181) | 3.87 (0.235) | 3.29 (0.008) | 3.25 (0.153) |
| Health & Beauty Aid | Head & Shoulder 400ml | 5.30 (0.078) | 5.06 (0.547) | 5.48 (0.254) | 5.45 (0.268) | 5.42 (0.086) | 5.38 (0.142) |
| Health & Beauty Aid | Pantene Shampoo 400ml | 4.59 (0.265) | 4.39 (0.489) | 4.80 (0.408) | 4.80 (0.412) | 4.91 (0.054) | 4.86 (0.150) |
| Health & Beauty Aid | Scope Mouthwash Original Mint 1L | 3.73 (0.156) | 3.73 (0.156) | 4.00 (0.419) | 3.99 (0.441) | 3.86 (0.099) | 3.86 (0.099) |
| Households | Arctic Power 3.3kg | 6.66 (0.556) | 6.66 (0.556) | 8.49 (0.000) | 8.34 (0.590) | 6.85 (0.530) | 6.80 (0.665) |
| Households | Canola Harvest Oil 1.89L | 4.80 (0.523) | 4.80 (0.523) | 5.45 (0.250) | 5.32 (0.491) | 4.60 (0.498) | 4.58 (0.502) |
| Households | Downy April Fresh 3L | 5.58 (0.157) | 5.58 (0.157) | 6.74 (0.115) | 6.64 (0.279) | 5.68 (.010) | 5.67 (0.038) |
| Households | Five Rose Flour 2.5kg | 3.97 (0.139) | 3.97 (0.139) | 4.00 (0.058) | 3.93 (0.188) | 4.32 (0.056) | 4.32 (0.057) |
| Households | Fleecy Fresh Air 5L | 4.97 (0.539) | 4.97 (0.539) | 6.01 (0.065) | 5.89 (0.479) | 5.17 (0.138) | 5.14 (0.161) |
| Households | Mazola Corn Cooking Oil 2L | 5.01 (0.349) | 5.01 (0.349) | 6.08 (0.191) | 5.91 (0.518) | 5.71 (0.239) | 5.38 (0.459) |
| Households | Palmolive Dishwashing Liquid 625L | 2.00 (0.047) | 2.00 (0.047) | 2.62 (0.045) | 2.54 (0.236) | 1.98 (0.010) | 1.98 (0.010) |
| Households | Purex 3.78L | 5.83 (0.165) | 5.83 (0.165) | 8.02 (0.069) | 7.98 (0.211) | 5.78 (0.010) | 5.78 (0.010) |
| Households | Robin Hood Flour 10kg | 7.99 (0.000) | 7.99 (0.000) | 8.16 (0.484) | 8.05 (0.784) | 6.59 (0.689) | 6.26 (0.827) |
| Households | Sunlight Detergent with bleach 3.3kg | 7.79 (0.417) | 7.79 (0.417) | 8.99 (0.000) | 8.78 (0.800) | 6.91 (0.193) | 6.90 (0.202) |
| Households | Sunlight Dishwashing Liquid 750ml | 1.85 (0.054) | 1.85 (0.054) | 2.48 (0.039) | 2.45 (0.100) | 1.97 (0.047) | 1.95 (0.080) |
| Households | Tide Detergent Power 3.4kg | 8.41 (0.281) | 8.41 (0.281) | 9.99 (0.000) | 9.72 (0.792) | 8.53 (0.113) | 8.46 (0.379) |
| Juices | Del Monte 1L | 1.12 (0.119) | 1.12 (0.119) | 1.21 (0.086) | 1.15 (0.161) | 1.16 (0.062) | 1.14 (0.078) |
| Juices | Oasis Classic 960ml | 1.23 (0.168) | 1.23 (0.168) | 1.41 (0.141) | 1.29 (0.241) | 1.25 (0.083) | 1.19 (0.137) |
| Juices | Ocean Spray Cocktail 1.89L | 3.69 (0.028) | 3.69 (0.028) | 3.79 (0.028) | 3.73 (0.158) | 3.68 (0.010) | 3.62 (0.193) |
| Juices | Rougemont 1.89L | 2.57 (0.092) | 2.57 (0.092) | 2.59 (0.000) | 2.57 (0.153) | 2.39 (0.023) | 2.39 (0.023) |

| Product Category | Product | EDLP (Loblaw's) Regular Price | EDLP (Loblaw's) Transaction price | Hi-Lo (Provigo) Regular Price | Hi-Lo (Provigo) Transaction price | HYB (Super-C) Regular Price | HYB (Super-C) Transaction price |
|---|---|---|---|---|---|---|---|
| Juices | Tropicana Orange Juice 1.89L | 3.40 (0.283) | 3.40 (0.283) | 3.49 (0.028) | 3.38 (0.256) | 3.44 (0.107) | 3.39 (0.165) |
| Juices | Welch's Fruit 1.82L | 4.52 (0.076) | 4.52 (0.076) | 4.59 (0.059) | 4.50 (0.210) | 4.50 (0.068) | 4.39 (0.219) |
| Paper Towel, Tissue & Pet Supplies | Cat Chow 4kg | 10.09 (0.198) | 10.09 (0.198) | 10.95 (0.135) | 10.95 (0.135) | 9.98 (0.010) | 9.96 (0.098) |
| Paper Towel, Tissue & Pet Supplies | Cottonolle Paper Towel 30RL | 13.95 (0.277) | 13.95 (0.277) | 13.76 (0.645) | 13.76 (0.645) | 9.02 (0.118) | 8.96 (0.339) |
| Paper Towel, Tissue & Pet Supplies | Dog Chow 2kg | 4.99 (0.000) | 4.99 (0.000) | 5.38 (0.191) | 5.28 (0.278) | 4.98 (0.010) | 4.98 (0.010) |
| Paper Towel, Tissue & Pet Supplies | Kleenex Tissue 230FE | 2.45 (0.146) | 2.45 (0.146) | 2.79 (0.000) | 2.77 (0.065) | 2.58 (0.010) | 2.57 (0.046) |
| Paper Towel, Tissue & Pet Supplies | Puffs Plus Lotion 144FE | 2.55 (0.102) | 2.55 (0.102) | 2.89 (0.000) | 2.86 (0.097) | 2.48 (0.010) | 2.45 (0.087) |
| Paper Towel, Tissue & Pet Supplies | Puppy Chow 8kg | 11.06 (0.172) | 11.06 (0.172) | 14.26 (0.598) | 14.10 (0.813) | 11.47 (0.009) | 11.45 (0.097) |
| Paper Towel, Tissue & Pet Supplies | Scotties Tissue 150 FE | 0.99 (0.014) | 0.99 (0.014) | 1.24 (0.050) | 1.20 (0.111) | 0.99 (0.000) | 0.99 (0.000) |
| Soup / Canned Foods | Aylmer Whole Tomato 796ml | 1.29 (0.000) | 1.29 (0.000) | 1.42 (0.062) | 1.31 (0.218) | 1.28 (0.010) | 1.26 (0.090) |
| Soup / Canned Foods | Del Monte Fruit Cocktail 796ml | 2.72 (0.184) | 2.72 (0.184) | 2.93 (0.093) | 2.86 (0.211) | 2.75 (0.153) | 2.73 (0.164) |
| Soup / Canned Foods | Green Giant Beans 398ml | 1.03 (0.164) | 1.03 (0.164) | 1.17 (0.054) | 1.13 (0.140) | 1.15 (0.076) | 1.09 (0.123) |
| Soup / Canned Foods | Pastene Diced Tomato 796ml | 1.42 (0.054) | 1.42 (0.054) | 1.59 (0.000) | 1.55 (0.119) | 1.35 (0.082) | 1.31 (.110) |

**B. Private Label Products**

| Product Category | Product | EDLP (Loblaw's) Regular Price | EDLP (Loblaw's) Transaction price | Hi-Lo (Provigo) Regular Price | Hi-Lo (Provigo) Transaction price | HYB (Super-C) Regular Price | HYB (Super-C) Transaction price |
|---|---|---|---|---|---|---|---|
| Beverage | PC Natural Spring Water 1.5L | 0.76 (0.067) | 0.76 (0.067) | 0.77 (0.044) | 0.75 (0.054) | | |
| Beverage | PC Cola 2L | 0.96 (0.082) | 0.95 (0.105) | 1.07 (0.036) | 1.04 (0.070) | | |
| Beverage | Super C Natural Spring Water 1.5L | | | | | 0.68 (0.024) | 0.68 (0.024) |
| Beverage | Super C Cola 2L | | | | | 0.99 (0.017) | 0.93 (0.100) |
| Beverage | Super C Mineral Water 1L | | | | | 0.79 (0.007) | 0.79 (0.007) |
| Breakfast/Cereals | PC Corn Flakes 750g | 2.94 (0.179) | 2.94 (0.179) | 3.28 (0.249) | 3.26 (0.269) | | |
| Breakfast/Cereals | PC Crispy Rice 525g | 2.20 (0.346) | 2.20 (0.346) | 2.74 (0.109) | 2.69 (0.105) | | |
| Breakfast/Cereals | Super C Corn Flakes 675g | | | | | 2.75 (0.088) | 2.74 (0.103) |
| Condiments, Sauces and Spread | PC Ketchup 1 L | 2.09 (0.149) | 2.09 (0.149) | 2.30 (0.197) | 2.20 (0.219) | | |

| Category | Product | | | | | | |
|---|---|---|---|---|---|---|---|
| Condiments, Sauces and Spread | PC Original Whipped Salad 950ml | 2.60 (0.212) | 2.60 (0.212) | 3.38 (0.650) | 3.35 (0.661) | | |
| Dairy Products | Super C Butter 454g | | | | | 2.95 (0.080) | 2.95 (.100) |
| Dairy Products | Super C Cheddar Cheese 600g | | | | | 5.09 (0.220) | 5.08 (0.215) |
| Frozen Food | Super C Buttermilk Pancake 310kg | | | | | 1.73 (0.098) | 1.73 (0.098) |
| Frozen Food | Super C Pizza Lunch 1.2kg | | | | | 6.34 (0.230) | 6.33 (0.234) |
| Households | PC Fabric Softener 3L | 3.99 (0.000) | 3.99 (0.000) | 4.52 (0.118) | 4.49 (0.171) | | |
| Households | PC Laundry Detergent 3.4kg | 6.82 (0.382) | 6.82 (0.382) | 6.99 (0.000) | 6.85 (0.348) | | |
| Households | Super C Dishwashing 850ml | | | | | 1.77 (0.030) | 1.76 (0.037) |
| Households | Super C Fabric Softener 3.6L | | | | | 1.98 (0.010) | 1.98 (0.010) |
| Households | Super C Laundry Detergent 3.6kg | | | | | 5.94 (0.196) | 5.80 (0.383) |
| Households | Super C Maize Oil 2L | | | | | 3.98 (0.009) | 3.97 (0.036) |
| Juices | PC Juice Cocktail 1.89L | 2.91 (0.451) | 2.91 (0.451) | 3.22 (0.257) | 2.99 (0.458) | | |
| Juices | PC White Grape Juice 1.82 L | 3.99 (0.000) | 3.99 (0.000) | 3.96 (0.073) | 3.82 (0.283) | | |
| Juices | Super C Fruit Punch Drink 2L | | | | | 1.54 (.043) | 1.54 (.043) |
| Juices | Super C Orange Juice 1.89L | | | | | 2.85 (0.087) | 2.85 (0.087) |
| Paper Towel, Tissue and Pet Supplies | Super C Bathroom Double Tissue 24un | | | | | 9.82 (0.084) | 9.80 (0.140) |
| Paper Towel, Tissue and Pet Supplies | Super C Facial Tissue 250un | | | | | 1.46 (0.037) | 1.44 (0.102) |
| Soup / Canned Foods | Super C Mais 398ml | | | | | 0.85 (0.032) | 0.85 (0.032) |
| Soup / Canned Foods | Super C Small Peas 398ml | | | | | 0.95 (0.061) | 0.95 (0.061) |
| Soup / Canned Foods | Super C Tomatoes 796ml | | | | | 0.98 (0.010) | 0.98 (0.027) |

# APPENDIX E. ALTERNATIVE CALCULATIONS OF THE AVERAGE PRICE DURATION

Carvalho (2006) shows that because of Jensen's inequality, calibrating sticky price models using the information on average frequencies, as we do in the paper, underestimates the stickiness of prices. In this appendix, therefore, we calculate an alternative measure of price durations:

(E1) $\quad -\frac{1}{N}\sum_{i \in C} \ [ln(1-f_i)]^{-1}$,

where $f_i$ is the weekly price change frequency of product $i$ in category $C$, and $N$ is the number of products in category $C$.

However, the use of equation E1 with our data has a significant drawback. If for a given price measure and a given store, a product has no price changes, then we are forced to drop it from the calculation, biasing the estimates downwards. Our estimates are therefore a lower bound of the price durations. This downward bias is less severe for transaction prices, which are relatively volatile, but it is likely to be important for reference prices and perhaps also for the filtered and regular prices. The results are summarized in Table E1. Panel A presents the implied average durations based on the average frequencies, as we do in the paper.

Panel B presents the results based on Equation E1. The expected durations of transaction prices at the EDLP, Hi-Lo, and HYB stores are 10.70 weeks, 8.94 weeks, and 10.55 weeks, respectively. These durations are 59.2%, 137.1%, and 56.3% longer than the corresponding durations in Panel A.

For regular prices, the expected duration figures in panel B are 10.94 weeks, 27.66 weeks, and 21.96 weeks for the EDLP, Hi-Lo and HYB stores, respectively. These durations are 57.2%, 14.6%, and 20.5% longer than the corresponding durations in Panel A. For Nakamura and Steinsson's (2008) filtered prices, the expected durations in panel B are 26.01 weeks, 29.53 weeks, and 24.44 weeks for the EDLP, Hi-Lo and HYB stores, respectively. These durations are 13.1%, 6.9%, and 12.7% longer than the corresponding durations in Panel A. For Chahrour's (2011) reference price, the estimated durations in panel B are often shorter than in Panel A, indicating that for reference prices, our lower bound perhaps is not a good measure of price stickiness.

To obtain a better measure of the underestimation of price stickiness implied by using the average frequencies, we calculate the average price durations using only the observations that we used to calculate Equation E1. I.e., we calculate $-\left[ln\,(1-\underline{f})\right]^{-1}$, where $\underline{f}$ is the ratio of the total number of price changes per week in the category to the number of products in the category, using information only on products that have at least one price change. In other words, the sample that we use is the same as the sample that we used to calculate Panel B of Table E1, making the results comparable. The results are reported in Table E2.

Focusing on the bottom rows, we find that for transaction prices, the durations reported in Panel B of Table E1 are 72.3% (EDLP), 142.9% (Hi-Lo), and 59.4% (HYB) greater than in Table E2. For regular prices, as defined by the store, the durations reported in Panel B of Table E1 are 70.1% (EDLP), 44.4% (Hi-Lo), and 26.6% (HYB) greater than in Table E2. For Nakamura and Steinsson (2008) filtered prices, the durations reported in Panel B of Table E1 are 38.9% (EDLP), 36.3% (Hi-Lo), and 19.5% (HYB) greater than in Table E2. For Chahrour's (2011) reference prices, the durations reported in Panel B of Table E1 are 32.4% (EDLP), 25.4% (Hi-Lo), and 22.2% (HYB) greater than in Table E2.

It therefore seems that in comparison to Equation (E1), the downward bias generated by using the average frequency to calculate price stickiness is most pronounced when prices are flexible. The bias is also affected by the variance in the frequency of price changes across products. Consequently, the greatest differences between Table E2 and Panel B of Table E1 are for the transaction prices of the Hi-Lo and EDLP stores. The differences are smallest for Nakamura and Steinsson (2008) filtered prices and Chahrour's (2011) reference prices of the HYB store.

Table E1. Implied Price Duration

| | A. Implied Average Price Duration in Weeks based on average frequencies ||||||||||||
|---|---|---|---|---|---|---|---|---|---|---|---|---|
| **Product Category** | EDLP (Loblaw's) | | | | Hi-Lo (Provigo) | | | | HYB (Super-C) | | | |
| | Transaction | Regular | Filtered | Reference | Transaction | Regular | Filtered | Reference | Transaction | Regular | Filtered | Reference |
| Baby Products & Foods | 16.83 | 16.83 | 51.50 | 129.50 | 16.83 | 259.50 | 259.50 | N/A | 14.79 | 18.07 | 36.64 | 32.00 |
| Beverages | 3.81 | 3.89 | 13.94 | 34.16 | 2.27 | 22.10 | 16.83 | 51.50 | 5.09 | 17.37 | 28.10 | 33.14 |
| Breakfast/Cereals | 6.38 | 6.53 | 33.30 | 47.78 | 3.60 | 35.08 | 35.08 | 67.10 | 5.21 | 20.30 | 21.01 | 22.61 |
| Condiments, Sauces & Spread | 4.59 | 4.59 | 22.38 | 43.50 | 2.88 | 20.68 | 24.37 | 33.14 | 6.17 | 19.00 | 24.13 | 21.78 |
| Dairy Products | 6.24 | 6.42 | 16.83 | 22.10 | 5.91 | 18.07 | 24.26 | 28.39 | 5.99 | 15.49 | 19.62 | 25.50 |
| Frozen Food | 6.10 | 6.35 | 25.50 | 35.90 | 4.21 | 27.50 | 25.50 | 51.50 | 6.27 | 15.63 | 18.22 | 24.13 |
| Health & Beauty Aid | 7.61 | 9.69 | 17.43 | 23.13 | 6.15 | 12.83 | 16.83 | 22.10 | 8.61 | 18.75 | 22.10 | 22.10 |
| Households | 10.04 | 10.04 | 26.46 | 35.90 | 4.40 | 45.00 | 48.03 | 80.39 | 7.73 | 17.98 | 18.40 | 26.34 |
| Juices | 5.99 | 6.20 | 31.50 | 58.93 | 2.37 | 21.39 | 23.97 | 45.72 | 5.79 | 15.49 | 18.40 | 18.40 |
| Paper Towel, Tissue & Pet Supplies | 25.50 | 25.50 | 35.90 | 60.17 | 4.84 | 32.59 | 39.94 | 39.94 | 12.49 | 30.70 | 28.75 | 38.50 |
| Soups/Canned Foods | 8.16 | 8.16 | 20.30 | 29.21 | 2.35 | 15.49 | 51.50 | 68.83 | 6.77 | 18.65 | 20.91 | 19.72 |
| **Total** | **6.72** | **6.96** | **23.00** | **36.53** | **3.77** | **24.13** | **27.63** | **44.26** | **6.75** | **18.22** | **21.69** | **24.79** |
| | B. Expected Price Duration in Weeks ||||||||||||
| Baby Products & Foods | 15.67 | 15.67 | 34.16 | 25.50 | 26.94 | 51.50 | 51.50 | | 32.99 | 33.29 | 36.33 | 35.25 |
| Beverages | 4.77 | 4.85 | 19.55 | 37.96 | 5.98 | 28.77 | 27.46 | 47.16 | 9.79 | 24.19 | 27.90 | 35.03 |
| Breakfast/Cereals | 7.93 | 7.99 | 30.70 | 42.83 | 4.66 | 32.24 | 30.31 | 48.61 | 6.12 | 20.73 | 23.07 | 22.03 |
| Condiments, Sauces & Spread | 6.01 | 6.01 | 24.47 | 30.55 | 7.78 | 21.44 | 25.39 | 27.35 | 7.98 | 24.40 | 28.87 | 28.39 |
| Dairy Products | 10.52 | 10.60 | 24.19 | 29.14 | 13.82 | 24.80 | 28.10 | 27.42 | 8.95 | 18.71 | 21.88 | 30.19 |
| Frozen Food | 8.32 | 8.38 | 31.07 | 40.66 | 15.40 | 31.69 | 30.45 | 47.16 | 9.08 | 17.79 | 20.20 | 27.88 |
| Health & Beauty Aid | 12.29 | 14.15 | 24.37 | 29.06 | 11.05 | 17.31 | 21.51 | 26.80 | 11.20 | 19.86 | 22.90 | 22.90 |
| Households | 12.07 | 12.07 | 24.63 | 26.04 | 5.48 | 39.80 | 42.04 | 41.39 | 11.15 | 20.73 | 21.38 | 31.73 |
| Juices | 15.16 | 15.25 | 29.83 | 36.33 | 5.08 | 27.35 | 27.97 | 28.75 | 7.94 | 16.29 | 22.25 | 19.54 |
| Paper Towel, Tissue & Pet Supplies | 26.22 | 26.22 | 28.96 | 38.50 | 13.65 | 27.23 | 27.66 | 27.66 | 15.63 | 30.91 | 29.83 | 35.40 |
| Soups/Canned Foods | 8.57 | 8.57 | 17.41 | 22.61 | 2.76 | 14.97 | 25.50 | 38.50 | 9.31 | 23.89 | 24.88 | 23.64 |
| **Total** | **10.70** | **10.94** | **26.01** | **33.52** | **8.94** | **27.66** | **29.53** | **36.23** | **10.55** | **21.96** | **24.44** | **28.30** |

Notes: In panel A of the table, we report the implied average duration of the prices in weeks. The average duration is calculated as $-\left[\ln(1-f)\right]^{-1}$, for each one of the 11 product categories included in our data, for the three stores. For each category, we computed the $f$ as the ratio of the total number of price changes per week in the category, to the number of products in the category (Levy et al., 1997, Table 1, p.

797, Gorodnichenko and Talavera 2017). The average weekly frequency of a price change at each store is calculated for the transaction price, the regular price (as classified and presented by the store), the filtered price (the prices after removing temporary price reductions as identified by Nakamura and Steinsson's (2008) sales filter A), and the reference prices. We use Chahrour's (2008) algorithm with a 13-week rolling window to derive the reference prices. The "total" row gives the average weekly frequency computed over all goods, in each store. In panel B, we calculate the expected durations as: $-\frac{1}{N}\sum_{i \in C} [ln(1-f_i)]^{-1}$ where $f_i$ is the frequency of price changes of product $i$ in category $C$, and $N$ is the total number of products in the category.

Table E2. Implied Price Duration Using Only Products with at least One Price Change.

| | A. Implied Average Price Duration in Weeks based on average frequencies | | | | | | | | | | | |
|---|---|---|---|---|---|---|---|---|---|---|---|---|
| **Product Category** | EDLP (Loblaw's) | | | | Hi-Lo (Provigo) | | | | HYB (Super-C) | | | |
| | Transaction | Regular | Filtered | Reference | Transaction | Regular | Filtered | Reference | Transaction | Regular | Filtered | Reference |
| Baby Products & Foods | 9.89 | 9.89 | 30.70 | 25.50 | 9.89 | 51.50 | 51.50 | N/A | 11.73 | 14.35 | 29.21 | 25.50 |
| Beverages | 3.81 | 3.89 | 12.49 | 27.23 | 2.27 | 19.84 | 15.09 | 41.10 | 5.09 | 15.74 | 22.90 | 30.09 |
| Breakfast/Cereals | 6.38 | 6.53 | 25.50 | 36.64 | 3.60 | 24.13 | 24.13 | 46.30 | 5.21 | 16.83 | 17.43 | 18.75 |
| Condiments, Sauces & Spread | 4.59 | 4.59 | 16.13 | 23.50 | 2.88 | 14.90 | 17.58 | 20.91 | 6.17 | 19.00 | 24.13 | 21.78 |
| Dairy Products | 6.24 | 6.42 | 16.83 | 22.10 | 5.91 | 18.07 | 24.26 | 25.50 | 5.99 | 15.49 | 19.62 | 25.50 |
| Frozen Food | 6.10 | 6.35 | 25.50 | 30.70 | 4.21 | 27.50 | 25.50 | 44.07 | 6.27 | 15.63 | 18.22 | 21.39 |
| Health & Beauty Aid | 7.61 | 9.69 | 17.43 | 20.77 | 6.15 | 12.83 | 16.83 | 22.10 | 8.61 | 18.75 | 22.10 | 22.10 |
| Households | 8.53 | 8.53 | 18.75 | 20.30 | 4.40 | 32.00 | 37.63 | 34.16 | 7.73 | 17.98 | 18.40 | 26.34 |
| Juices | 5.17 | 5.36 | 23.50 | 29.21 | 2.37 | 18.65 | 20.91 | 22.61 | 5.79 | 15.49 | 18.40 | 18.40 |
| Paper Towel, Tissue & Pet Supplies | 21.78 | 21.78 | 25.50 | 34.16 | 4.84 | 23.13 | 22.61 | 22.61 | 11.05 | 27.23 | 25.50 | 29.83 |
| Soups/Canned Foods | 5.99 | 5.99 | 15.09 | 21.78 | 2.35 | 11.49 | 25.50 | 34.16 | 6.77 | 18.65 | 20.91 | 19.72 |
| **Total** | **6.21** | **6.43** | **18.73** | **25.31** | **3.68** | **19.15** | **21.66** | **28.89** | **6.62** | **17.35** | **20.46** | **23.15** |

Notes: In panel A of the table, we report the implied average duration of the prices in weeks. The average duration is calculated as $-\left[ln\left(1-\underline{f}\right)\right]^{-1}$, for each one of the 11 product categories included in our data, for the three stores. For each category, we computed the $\underline{f}$ as the ratio of the total number of price changes per week in the category, to the number of products in the category (Levy et al., 1997, Table 1, p. 797, Gorodnichenko and Talavera 2017). We use only observations on products that had at least 1 price change. The average weekly frequency of a price change at each store is calculated for the transaction price, the regular price (as classified and presented by the store), the filtered price (the prices after removing temporary price reductions as identified by Nakamura and Steinsson's (2008) sales filter A), and the reference prices. We use Chahrour's (2008) algorithm with a 13-week rolling window to derive the reference prices. The "total" row gives the average weekly frequency computed over all goods, in each store.

# APPENDIX F. HISTOGRAMS OF PRICE CHANGES

Figure F1 depicts the histograms of the size of price changes. We find that there is a large variation in the kurtoses, both across stores and across price measures. If we look at the transaction price, we find that the kurtoses are between 3.48 at the Hi-Lo store and 4.63 at the HYB store. When we focus on regular prices, the kurtosis at the EDLP store remains almost unchanged (4.29), but the removal of sales, which are usually large in percentage terms, leads to an increase in the kurtoses at the Hi-Lo (8.52) and HYB (5.78) stores. For the filtered prices, the kurtoses are more similar across the three stores: 8.64 at the EDLP store, 7.52 at the Hi-Lo store, and 7.48 at the HYB store. There is also a large variation in the kurtoses of the reference prices: 5.17 at the EDLP store, 7.32 at the Hi-Lo store, and 4.24 at the HYB store.

Figure F1. Histograms of the Size of Price Changes

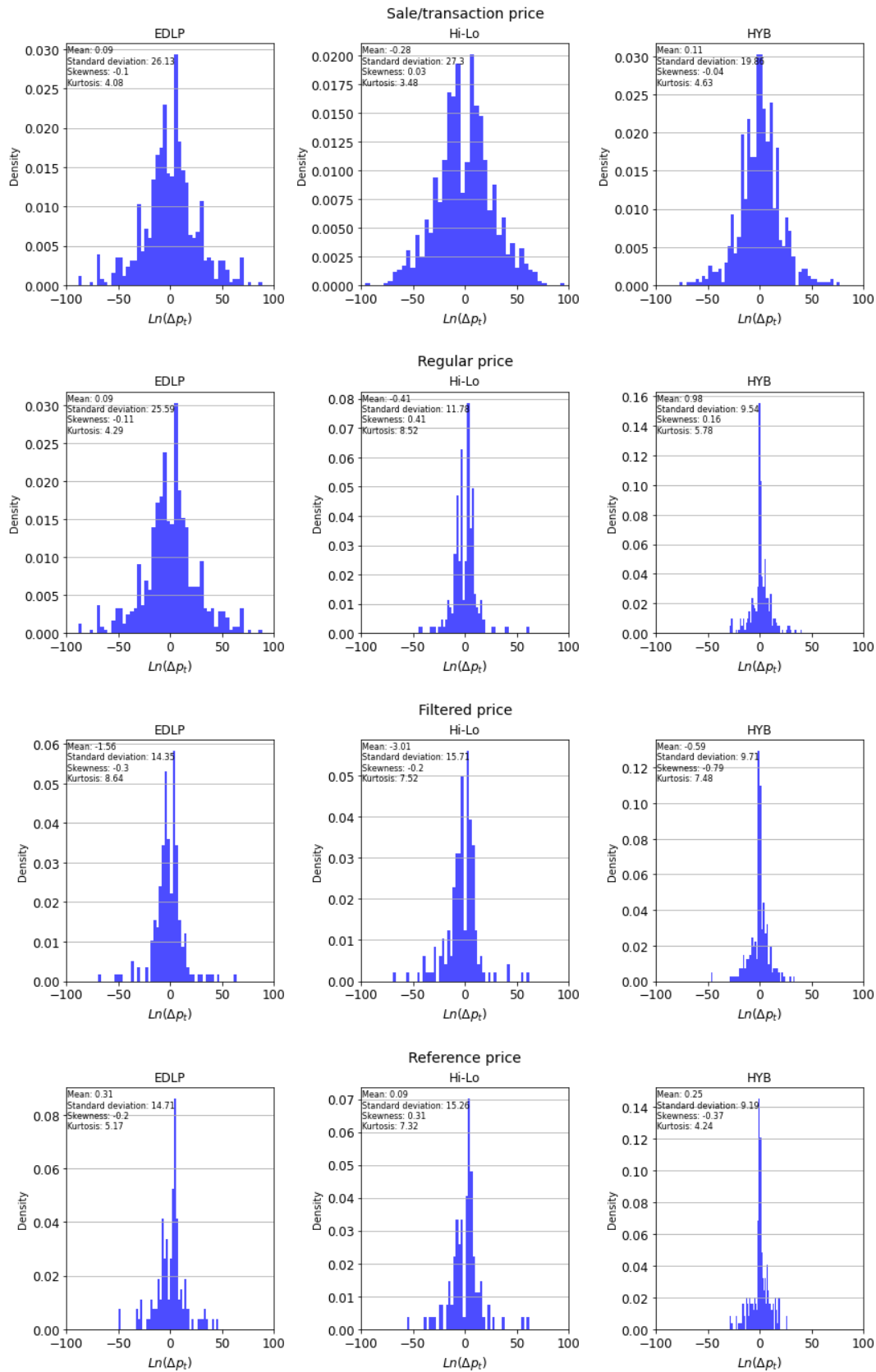

Notes: The figure shows the histograms of the size of price changes calculated as $100 \times [Ln(p_{s,i,t}) - Ln(p_{s,i,t-1})]$, where $p$ is the relevant price measure of product $i$ offered in store $s$ on week $t$. The scale of the y-axis varies across the figures.

# APPENDIX G. AVERAGE FREQUENCIES OF PRICE CHANGES AND IMPLIED DURATIONS: Nakamura and Steinsson's (2008) Sales Filter

One disadvantage of sales filters such as Nakamura and Steinsson's (2008) is that they can be less precise near the endpoints. For example, Figure 5 in the paper illustrates that if a price-cut takes place near the end of the sample period, the filter is unable to determine whether it is temporary or not.

This problem is likely to be more important in a short (time series) dataset, such as ours, than in longer datasets, where endpoints compose a smaller share of the observations. To estimate the effect of endpoints on the precision of our estimates for the frequency of price changes, we run the Nakamura and Steinsson's (2008) sales filter again, this time assuming that if the transaction price decreased less than 6 weeks away from the endpoint, without bouncing back up again, then we excluded these observations. We chose the value of 6 weeks because we calibrated the filter such that the maximum length of a sale is 6 weeks.

Panel A of Table G1 presents the average frequencies in each of the 11 product categories, in each of the 3 stores. We find that overall, excluding the price changes close to the end of the sample reduces the frequency of price changes relative to the values reported in Table 5 in the paper. In Table G1, the average frequencies of the EDLP, Hi-Lo, and HYB stores are 3.69%, 2.91%, and 4.20%, respectively. The corresponding values in Table 5 are 4.25%, 3.55%, and 4.50%. The average frequency decreases, therefore, by 13.2%, 18.0%, and 9.33%, respectively.

Panel B of Table G1 presents the average durations implied by the average frequencies of price changes. As in Panel B of Table 5, we calculate the average durations as $-\left[ln\,(1-\underline{f})\right]^{-1}$ where $\underline{f}$ is the average frequency of price changes. We find that the average durations are 28.20, 33.15 and 25.76 weeks at the EDLP, Hi-Lo and HYB stores, respectively. The corresponding values in Table 5 are 23.00, 27.63 and 21.69 weeks. Assuming no price changes close to the end points, therefore, increases the estimated durations by 18.8%–22.6%.

However, as we discuss in Appendix E, using the average frequencies might bias the average durations downwards. We therefore recalculate the expected durations following the same procedure as in Appendix E. The results are reported in Panel C.

We find that the expected durations are 28.20, 33.15 and 25.76 weeks for the EDLP, Hi-Lo, and HYB stores, respectively. This is compared to 26.01, 29.53, and 24.44 weeks, respectively, reported in Table E1. When we focus on the expected durations, therefore, the effects of assuming no price changes close to the end points are more modest: 8.4% for the EDLP store, 12.3% for the Hi-Lo store, and 5.4% for the HYB store.

Table G1. Average Frequencies of Price Changes and Implied Durations: Nakamura and Steinsson's (2008) Sales Filter

| A. Average Weekly Frequency of Price Changes | | | |
|---|---|---|---|
| | Transaction | Regular | Filtered |
| Baby Products & Foods | 1.54% | 0.00% | 2.31% |
| Beverages | 6.35% | 5.38% | 3.15% |
| Breakfast/Cereals | 2.37% | 1.92% | 4.17% |
| Condiments, Sauces & Spread | 3.67% | 3.32% | 3.85% |
| Dairy Products | 5.58% | 3.85% | 4.65% |
| Frozen Food | 3.85% | 3.02% | 4.70% |
| Health & Beauty Aid | 3.65% | 4.23% | 4.23% |
| Households | 3.43% | 1.79% | 5.05% |
| Juices | 2.64% | 3.37% | 5.05% |
| Paper Towel, Tissue & Pet Supplies | 2.47% | 1.92% | 3.21% |
| Soups/Canned Foods | 4.33% | 1.44% | 4.67% |
| **Total** | **3.69%** | **2.91%** | **4.20%** |
| B. Implied Average Price Duration in Weeks | | | |
| Baby Products & Foods | 64.50 | N/A | 42.83 |
| Beverages | 15.25 | 18.07 | 31.28 |
| Breakfast/Cereals | 41.75 | 51.50 | 23.50 |
| Condiments, Sauces & Spread | 26.73 | 29.60 | 25.50 |
| Dairy Products | 17.43 | 25.50 | 21.01 |
| Frozen Food | 25.50 | 32.59 | 20.77 |
| Health & Beauty Aid | 26.87 | 23.13 | 23.13 |
| Households | 28.62 | 55.50 | 19.31 |
| Juices | 37.32 | 29.21 | 19.31 |
| Paper Towel, Tissue & Pet Supplies | 39.94 | 51.50 | 30.70 |
| Soups/Canned Foods | 22.61 | 68.83 | 20.91 |
| **Total** | **26.59** | **33.82** | **23.29** |
| C. Expected Price Duration in Weeks | | | |
| Baby Products & Foods | 25.50 | N/A | 38.50 |
| Beverages | 15.74 | 33.24 | 26.04 |
| Breakfast/Cereals | 39.36 | 39.94 | 26.97 |
| Condiments, Sauces & Spread | 22.84 | 30.37 | 31.76 |
| Dairy Products | 24.63 | 28.96 | 23.33 |
| Frozen Food | 31.07 | 36.64 | 20.08 |
| Health & Beauty Aid | 33.39 | 22.25 | 23.76 |
| Households | 25.50 | 39.94 | 23.55 |
| Juices | 32.72 | 36.64 | 22.79 |
| Paper Towel, Tissue & Pet Supplies | 34.16 | 22.61 | 30.91 |
| Soups/Canned Foods | 18.27 | 38.50 | 24.88 |
| **Total** | **28.20** | **33.15** | **25.76** |

# APPENDIX H. COMPARISON WITH A RETAIL FOOD STORE IN ISRAEL

One weakness of our dataset is that it covers only 52 weeks. This has two effects on our estimates of price rigidity. First, when we look at the filtered and reference price series, we find many products with no price changes, biasing our estimates of the duration of prices downwards.

Second, Nakamura and Steinsson's (2008) sales filter, which we use to calculate the filtered price series, as well as Chahrour's (2011) algorithm which we use to calculate the reference price series, are less accurate near the endpoints. It is possible, therefore, that our estimates of the rigidity of the filtered and reference prices are affected by this inaccuracy. To address this concern, in Appendix G, we provided estimates for Nakamura and Steinsson's (2008) sales filter assuming that all price changes close to the end points are temporary.

In the current appendix, we try to gauge the significance of having a short price series, by using a longer dataset. We use data made available by the Israeli retail "price transparency" law. Since 2015, all major Israeli retailers are required to post their prices online. Prices are posted for both online and brick-and-mortar stores. Prices of all products in each store are posted online once every day. If prices are updated during the day, the internet site should be updated within one hour of the price change. See Bonomo et al. (2023) for more details about the price transparency law.

We have data for one store which belongs to the largest supermarket chain, "Shufersal Deal-Extra." The chain positions itself as a discount store, a form of HYB format, offering relatively low prices along with temporary price cuts. The particular store we sampled is located in the city of Nesher, in the north of Israel. By Israeli standards, it is a large store, carrying over 9,800 different products. We have weekly data on 2,256 products for the period January 7, 2018–April 11, 2021 (171 weeks). For each product, we have both the transaction and regular prices, as posted online by the chain.

To make our results comparable to the results we report in the paper, we use data only for products with no more than 3 missing observations. This leaves us with 447 products.

The average price is 13.44 NIS with a standard deviation of 9.89 NIS, and the average regular price is 14.56 NIS with a standard deviation of 10.59 NIS.[3]

In addition to the transaction and regular price series that we have, for each product we generate a series of filtered prices using Nakamura and Steinsson's (2008) sales filter, and a series of reference prices using Chahrour's (2011) algorithm. For each product, we therefore have four price measures: transaction, regular, filtered and reference.

For each price measure, we calculate the average frequency of price changes, the implied average durations based on the average frequencies, i.e., $-\left[ln\left(1-\underline{f}\right)\right]^{-1}$, where $\underline{f}$ is the ratio of the total number of price changes per week in the category to the number of products in the category, and the expected implied durations, $-\frac{1}{N}\sum_{i=1}^{N} [ln(1-f_i)]^{-1}$, where $f_i$ is the weekly price change frequency of product $i$, and $N$ is the number of products.

In Panel A of Table H1, we present the results when we use all the observations. We find that for the transaction prices, regular prices, and the reference prices, the frequencies of price changes in the Israeli store are similar to the frequencies we find at the Canadian HYB store. The likelihood that the transaction price changes in each week is 13.62%, the likelihood that the regular price changes in each week is 4.68%, and the likelihood that the reference price changes is 4.47%. The finding that prices of a Hi-Lo store in Israel changes at a similar rate to a Canadian HYB store is consistent with Dhyne et al. (2006) that show that there are fewer temporary price cuts in Europe than in the US.

For the filtered prices, we find that the frequency of price changes is higher than for the regular prices as advertised by the store. It turns out that this happens because the store occasionally sets a high regular price, which is kept unchanged for a long period, and a lower transaction price. In other words, the store advertises certain products as being "on sale" for long periods. On such occasions, when the transaction price is changed (i.e., the size of the "discount" on the product is changed), the Nakamura and Steinsson (2008) sales filter identifies it as a filtered price change.

---

[3] The average US Dollar–NIS exchange rate during that period was 3.53 NIS for 1 US Dollar with standard deviation of 0.114 NIS.

Looking at the implied average durations, we find that the results for Israel are again quite similar to the results for the Canadian HYB store. The implied average duration for the transaction price in Israel (Canadian HYB) is 6.82 (6.75), for the regular price it is 20.85 (18.22), for the filtered prices it is 16.78 (21.69), and for the reference prices it is 21.87 (24.79).

It therefore seems that using a short series had only a modest effect on the implied average duration of prices. However, when we calculate the expected duration of prices, the effect of omitting products with no price changes seems to have had a significant effect on the results. The expected duration, in weeks, of the transaction prices in the Israeli (Canadian HYB) data is 18.34 (10.55), of the regular prices, 69.95 (21.96), of the filtered prices, 43.97 (24.44), and of the reference prices, 47.85 (28.30).

Thus, the short data series that we use in the paper likely leads to a significant underestimation of the expected duration of prices. There is a need in larger dataset to draw stronger conclusions.

The inclusion of endpoints, on the other hand, seems to have had only a modest effect on the estimates of duration and average/expected duration. This can be seen in Panel B, which shows the results when for each product we remove observations that are up to 6 weeks from the first or the last observations. We remove observations near the endpoints since Nakamura and Steinsson's (2008) and Chahrour's (2011) algorithms are likely to be less precise near the endpoints.

The results are almost unaffected compared to Panel A. Thus, imprecision around the endpoints does not seem to be a significant problem for price rigidity estimates, although the problem is likely to be more severe when the dataset is short.

Table H1. Frequency of Price Changes and Implied Durations, Israeli Dataset

| | Transaction price | Regular price | Filtered price | Reference price |
|---|---|---|---|---|
| **A. All observations** | | | | |
| Frequency of price changes | 13.62% | 4.68% | 5.79% | 4.47% |
| Implied average duration (weeks) | 6.82 | 20.85 | 16.78 | 21.87 |
| Expected duration (weeks) | 18.34 | 69.95 | 43.97 | 47.85 |
| **B. Excluding end points** | | | | |
| Frequency of price changes | 13.45% | 4.57% | 5.76% | 4.56% |
| Implied average duration (weeks) | 6.92 | 21.40 | 16.84 | 21.43 |
| Expected duration (weeks) | 17.16 | 66.60 | 42.33 | 46.43 |

Notes: Results for Israeli store "Shufersal," store number 71, located in Nesher. Weekly data for 447 products, over the period January 7, 2018–April 11, 2021. The frequency of price changes is the average weekly frequency of price changes $\underline{f}$ (in %). We compute $\underline{f}$ as the ratio of the total number of price changes per week in the category, to the number of products in the category. The implied average duration is calculated as $-\left[\ln(1-\underline{f})\right]^{-1}$. The expected duration is calculated as $-\frac{1}{N}\sum_{i=1}^{N} [\ln(1-f_i)]^{-1}$, where $f_i$ is the weekly price change frequency of product $i$, and $N$ is the number of products. Panel A uses all observations. In Panel B, for each product we exclude observations that are less than 6 weeks from the first or last observation.

# APPENDIX I. RETAIL SUPERMARKET LANDSCAPE IN CANADA

Retail sales of Canadian food stores amounted to about C$ 144 billion in 2021. The top Canadian food retailer is Loblaw Companies Ltd. With 28% market share, followed by Sobeys with 20%. Other leading food retailers include Metro Inc., Costco, and Walmart. Figure I1 shows the market share of top-10 retail food chain store operators in Canada.

Of the nearly 27,000 food stores in Canada, over one third were Ontario. Loblaw Companies Ltd., with over 2,400 stores nationwide, had the largest number of stores among grocery retailers in Canada and generated about 37 billion Canadian dollars in food sales in 2021. Sobeys Inc. followed with more than 1,400 stores and sales reaching just over 28 billion dollars in the same year. Revenues of Costco, Walmart, and Metro, were not far behind with 27, 22, and 18 billion Canadian dollars, respectively. Figure I2 shows the number of grocery stores in Canada by regions.

Figure I1. Top Grocery Retailers in Canada by Market Share, 2021

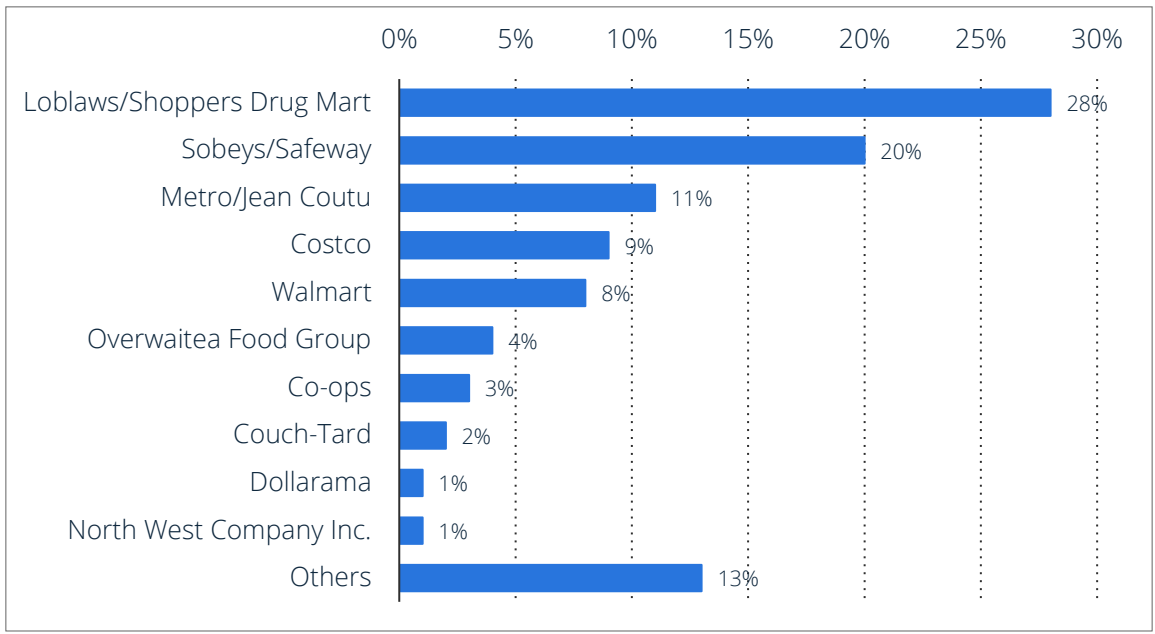

Source: Statista, https://www.statista.com/

Figure I2. Number of Grocery Stores in Canada by Regions, 2022

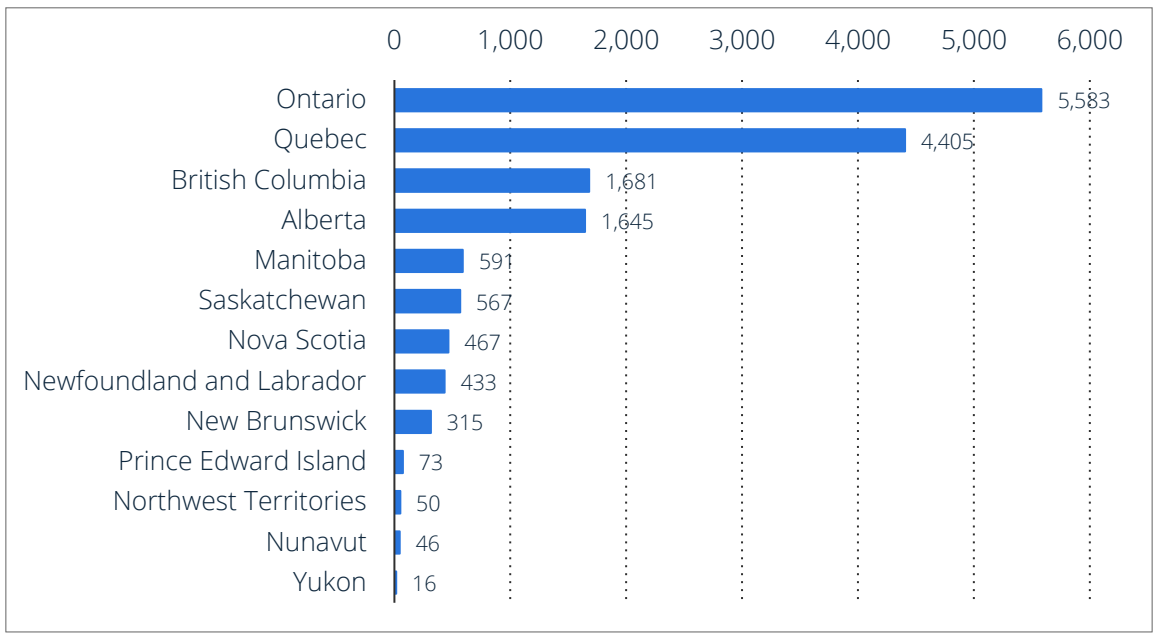

Source: Statista, https://www.statista.com/

# APPENDIX J. RETAIL SUPERMARKET LANDSCAPE IN THE US

In Table J1, we list the 15 largest retail food chains in the US, and their store pricing format distribution. According to the table, some chains have a dominant pricing format. For example, H.E. Butt employs the EDLP format at 96% of its stores, Food-Lion at 86% of its stores, and Walmart at 73% of its stores. Thus, at these chains, EDLP is the most common format. Most chains, however, use all three formats. For example, Kroger employs Hi-Lo at 47% of the stores, HYB at 40% of its stores, and EDLP at 13% of the stores. Stop & Shop, employs EDLP, Hi-Lo and HYB pricing formats, at 7%, 50% and 43% of its stores, respectively.

The chains with a high proportion (say, 30% or more) of Hi-Lo stores include A&P, Safeway, Stop & Shop, Kroger, Pathmark, and Lucky, each employing the Hi-Lo format at 35%–55% of their stores. The chains with a high proportion of EDLP stores include H.E. Butt, Food-Lion, Walmart, Winn-Dixie, Albertson's, Cub-Food, and Pathmark, each employing the EDLP format at 33%–96% of their stores. The chains operating a high proportion of HYB stores include Publix, Fred-Meyer, Giant, Stop & Shop, Safeway, Albertson's, Kroger, Lucky, Cub-Food, A&P, and Winn-Dixie, each employing a HYB format at 30%–71% of their stores. In sum, in the US retail food market, all three pricing formats are common and widespread.

The above figures suggest that the pricing format is not a chain-level variable. It turns out, however, there may be substantial variability in the pricing formats used by a retail chain even at the level of local markets. Consider, for example, Pathmark stores located in New Jersey, in the "small" area around the Raritan River, between Madison and Raritan Bay, as shown in Figure J1. In the magnified area of the figure, there are 37 Pathmark stores and as the Figure shows, they follow very different pricing formats, despite their close proximity to each other.

The variability in the pricing format is not limited to a particular chain. According to Ellickson and Misra (2008), this is characteristic of the entire retail food industry, irrespective of chain/store size, and irrespective of whether or not the stores are vertically integrated or not.

In Figure 1 in the paper, we show the spatial distribution of the pricing format across the US. As the figure shows, there are no clear differences between the spatial distributions of the three pricing formats.

However, if we look at the actual shares of each pricing format across the US regions, then we find some differences. In Table J2, we present the pricing format distribution across 8 regions of the US. According to the table, EDLP format stores are particularly popular in the South, South-East, Southern Central, and the South-West regions of the US. Hi-Lo format stores are particularly popular in the Great Lakes, Southern Central, North-East, and West Coast regions. HYB format stores are particularly popular in the North-West, South-West, West Cost, North-East, and South-East regions of the US. Thus, there is a regional variation in the prevalence of the different pricing formats, although all three formats are present in all parts of the US.

Table J1. Store Pricing Format Distribution for the 15 Largest Supermarket Retail Chains in the US

| Supermarket Chain | Number of Stores | Percentage of | | |
|---|---|---|---|---|
| | | EDLP Stores | Hi-Lo Stores | HYB Stores |
| Kroger | 1,399 | 13 | 47 | 40 |
| Food-Lion | 1,186 | 86 | 2 | 12 |
| Winn-Dixie | 1,174 | 67 | 3 | 30 |
| Safeway | 1,165 | 5 | 52 | 43 |
| Albertson | 922 | 48 | 11 | 41 |
| Fred-Meyer | 821 | 18 | 22 | 60 |
| Lucky | 813 | 27 | 35 | 38 |
| Giant | 711 | 11 | 29 | 60 |
| A&P | 698 | 15 | 55 | 30 |
| Publix | 581 | 16 | 13 | 71 |
| Walmart | 487 | 73 | 1 | 26 |
| Cub-Foods | 375 | 40 | 26 | 34 |
| H.E. Butt | 250 | 96 | 1 | 3 |
| Stop & Shop | 189 | 7 | 50 | 43 |
| Pathmark | 135 | 33 | 42 | 25 |

Source: Ellickson and Misra (2008)

Table J2. Distribution of Store Pricing Formats by Regions

| US Region | Percentage of | | |
|---|---|---|---|
| | EDLP Stores | Hi-Lo Stores | HYB Stores |
| West Coast | 22 | 39 | 39 |
| North-West | 17 | 32 | 51 |
| South-West | 32 | 20 | 48 |
| South | 43 | 32 | 25 |
| Southern Central | 28 | 45 | 27 |
| Great Lakes | 17 | 54 | 29 |
| North-East | 23 | 40 | 37 |
| South-East | 40 | 23 | 37 |

Notes: The figures in the table are the averages for 17,388 stores in the US, with annual revenues of at least $2 million.

Source: Ellickson and Misra (2008).

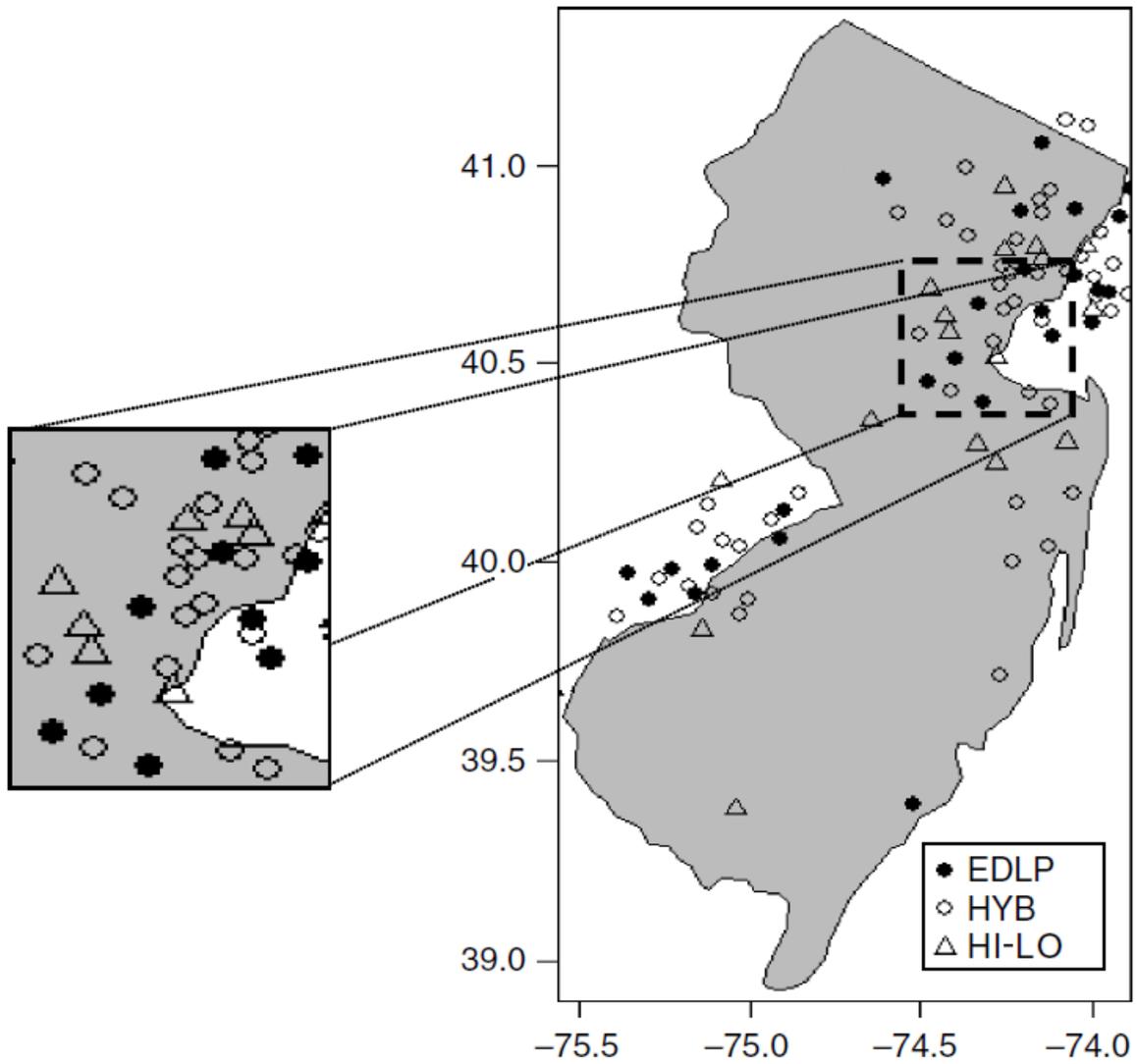

Figure J1. Local Variability in the Pricing Format of Pathmark Stores in New Jersey, Zooming-In the Area Around the Raritan River, between Madison and Raritan Bay (Source: Ellickson and Misra 2008)

# REFERENCES

Anderson, E., N. Jaimovich, and D. Simester (2015), "Price Stickiness: Empirical Evidence of the Menu Cost Channel," **Review of Economics and Statistics** 97(4), 813–826.

Bonomo, M, C. Carvalho, O. Kryvstov, S. Ribon, and R. Rigato (2023), "Multi-Product Pricing: Theory and Evidence from Large Retailers," **Economic Journal** 133(651), 905–927.

Carvalho, C. (2006), "Heterogeneity in Price Stickiness and the Real Effects of Monetary Shocks," **Frontiers of Macroeconomics** 2(1), 1–56.

Chahrour, R.A. (2011), "Sales and Price Spikes in Retail Price Data," **Economics Letters** 110, 143–146.

Dhyne, E., L.J Álvarez, H. Le Bihan, G. Veronese, D. Dias, J. Hoffmann, N. Jonker, P. Lünnemann, F. Rumler an J. Vilmunen (2006), "Price Changes in the Euro Area and the United States: Some Facts from Individual Consumer Price Data," **Journal of Economic Perspectives** 20(2), 171–192.

Ellickson, P., and S. Misra (2008), "Supermarket Pricing Strategies," **Marketing Science** 27(5), 811–828.

Gorodnichenko, Y. and O. Talavera (2017), "Price Setting in Online Markets: Basic Facts, International Comparisons, and Cross-Border Integration," **American Economic Review** 107(1), 249–282.

Levy, D., M. Bergen, S. Dutta, and R. Venable (1997), "The Magnitude of Menu Costs: Direct Evidence from Large U.S. Supermarket Chains," **Quarterly Journal of Economics** 112(3), 791–825.

Levy, D., D. Lee, H.A. Chen, R. Kauffman, and M. Bergen (2011), "Price Points and Price Rigidity," **Review of Economics and Statistics** 93(4), 1417–1431.

Nakamura, E., and J. Steinsson (2008), "Five Facts about Prices: a Reevaluation of Menu Cost Models," **Quarterly Journal of Economics** 123(4), 1415–1464.

Risley, D. (2020), "Does Ending the Price in 7 Really Matter?" *Guide to Pricing Your Products (6-Part Series)*, September 3, 2020, https://www.blogmarketingacademy.com/prices-ending-in-7/, accessed December 28, 2020.

Snir, A., and D. Levy (2021), "If You Think 9-Ending Prices Are Low, Think Again," **Journal of the Association for Consumer Research** 6(1), 33–47.